\documentclass[twocolumn,twocolappendix]{aastex631} 
\usepackage{multirow}
\usepackage{gensymb}
\usepackage{float}
\usepackage{subfigure}
\usepackage{graphicx}
\usepackage{threeparttable}
\usepackage{amsmath}
\usepackage{color}
\usepackage{mathrsfs}
\usepackage{soul}
\setstcolor{red}
\newcommand{\NII}{[N\,\textsc{ii}]}

\newcommand{\OIII}{[O\,\textsc{iii}]}
\newcommand{\Ha}{\mbox{H$\alpha$}}
\newcommand{\Hb}{\mbox{H$\beta$}}
\newcommand{\Ms}{{\mbox{$M_*$}}}
\newcommand{\Md}{{\mbox{$M_\mathrm{d}$}}}
\newcommand{\lbol}{{\mbox{$L_\mathrm{bol}$}}}
\newcommand{\Msun}{{\mbox{$M_\odot$}}}
\newcommand{\um}{{\mbox{$\mu$m}}}
\newcommand{\usurf}{{\mbox{$\mathrm{mag\,arcsec^{-2}}$}}}
\newcommand{\sersic}{{S\'ersic}}

\def\degree{${}^{\circ}$}

\shortauthors{Li et al.}

\begin{document}
\title{Panchromatic Photometry of Low-redshift, Massive Galaxies Selected from SDSS Stripe~82}
\author[0000-0002-3309-8433]{Yang A. Li}
\affiliation{Kavli Institute for Astronomy and Astrophysics, Peking University, Beijing 100871, China}
\affiliation{Department of Astronomy, School of Physics, Peking University, Beijing 100871, China}

\author[0000-0001-6947-5846]{Luis C. Ho}
\affiliation{Kavli Institute for Astronomy and Astrophysics, Peking University, Beijing 100871, China}
\affiliation{Department of Astronomy, School of Physics, Peking University, Beijing 100871, China}

\author[0000-0002-4569-9009]{Jinyi Shangguan}
\email{shangguan@mpe.mpg.de}
\affil{Max-Planck-Institut f\"{u}r extraterrestrische Physik, Gie{\ss}enbachstr. 1, D-85748 Garching, Germany}

\author[0000-0001-5105-2837]{Ming-Yang Zhuang}
\affil{Kavli Institute for Astronomy and Astrophysics, Peking University, Beijing 100871, China}
\affil{Department of Astronomy, School of Physics, Peking University, Beijing 100871, China}
\affil{Department of Astronomy, University of Illinois at Urbana-Champaign, Urbana, IL 61801, USA}

\author[0000-0001-8496-4162]{Ruancun Li}
\affil{Kavli Institute for Astronomy and Astrophysics, Peking University, Beijing 100871, China}
\affil{Department of Astronomy, School of Physics, Peking University, Beijing 100871, China}

\begin{abstract}
The broadband spectral energy distribution of a galaxy encodes valuable information on its stellar mass, star formation rate (SFR), dust content, and possible fractional energy contribution from nonstellar sources. We present a comprehensive catalog of panchromatic photometry, covering 17 bands from the far-ultraviolet to 500~$\mu$m, for 2685 low-redshift ($z=0.01-0.11$), massive ($M_* > 10^{10}\,M_\odot$) galaxies selected from the Stripe~82 region of the Sloan Digital Sky Survey, one of the largest areas with relatively deep, uniform observations over a wide range of wavelengths. Taking advantage of the deep optical coadded images, we develop a hybrid approach for matched-aperture photometry of the multi-band data. We derive robust uncertainties and upper limits for undetected galaxies, deblend interacting/merging galaxies and sources in crowded regions, and treat contamination by foreground stars. We perform spectral energy distribution fitting to derive the stellar mass, SFR, and dust mass, critically assessing the influence of flux upper limits for undetected photometric bands  and applying corrections for systematic uncertainties based on extensive mock tests. Comparison of our measurements with those of commonly used published catalogs reveals good agreement for the stellar masses. While the SFRs of galaxies on the star-forming main sequence show reasonable consistency, galaxies in and below the green valley show considerable disagreement between different sets of measurements. Our analysis suggests that one should incorporate the most accurate and inclusive photometry into the spectral energy distribution analysis, and that care should be exercised in interpreting the SFRs of galaxies with moderate to weak star formation activity.
\end{abstract}


\section{Introduction}

The galaxy population displays a bewildering variety of morphology, internal structure, kinematics, gas content, star formation rate and efficiency, and level of central black hole accretion. At the same time, the diversity of galaxy properties is interwoven by a number of empirical scaling relations that link them to each other, and many physical parameters are strongly coupled to the galaxy mass (e.g., \citealt{Kennicutt2012, KH2013, Cappellari2016, Saintonge2022}). Environments also play an important role (e.g., \citealt{Peng2010}), as does, of course, cosmic epoch (e.g., \citealt{Shapley2011, Madau2014}). Within this backdrop, low-redshift galaxies serve as a valuable laboratory to explore the myriad outstanding complexities of galaxy evolution. Apart from providing witness to the latest stage in the cosmic lifecycle of galaxies, the local Universe offers the obvious advantage of proximity. Nearby galaxies are bright and well-resolved. In the last two decades, the advent of large-area sky surveys has furnished a rich inventory of multiwavelength data suitable for probing the statistical physical properties of nearby galaxies, whose panchromatic coverage includes, among others, the ultraviolet (UV) by the Galaxy Evolution Explorer (GALEX; \citealt{Martin2005GALEX}), the optical by the Sloan Digital Sky Survey (SDSS; \citealt{York2000}), the near-infrared (NIR) by the Two Micron All Sky Survey (2MASS; \citealt{Skrutskie2006}), the mid-infrared (MIR) by the Wide-field Infrared Survey Explorer (WISE; \citealt{Wright_WISE2010}, and, over much more limited but still sizable areas, the far-infrared (FIR) by Herschel \citep{Pilbratt2010}. Despite their relatively shallow depth and moderate resolution, these large-area databases constitute an invaluable resource for probing, at the very minimum, several fundamental properties of nearby galaxies on global scales, through analysis of their integrated, broadband spectral energy distribution (SED). Broadband SED fitting furnishes key physical parameters such as stellar mass (\Ms), star formation rate (SFR), dust mass ($M_d$), and even an assessment of the contribution by active galactic nuclei (AGNs) to the total luminosity (e.g., \citealt{daCunha2008, Conroy2013, Boquien2019CIGALE}). Coupled with optical imaging and spectroscopy (e.g., from SDSS), one can explore the rich interplay between star formation, black hole accretion, and the interstellar medium on the one hand, and galaxy structure and environment (e.g., tidal interactions, mergers, environment) on the other hand.

The SDSS Stripe~82 region is one of the largest areas observed with relatively high and uniform sensitivity observed across a wide range of wavelengths. Scanned repeatedly by the SDSS survey for $\sim 80$ times, the coadded optical images are about 2 magnitudes deeper than the SDSS legacy survey \citep{Abazajian2009, Jiang2014}, rendering them especially sensitive to low-surface brightness features in galaxy outskirts (e.g., \citealt{Fliri2016}). Apart from the standard all-sky coverage provided by GALEX, 2MASS, and WISE, numerous dedicated multiwavelength surveys have been conducted in the Stripe~82 region, among them the VISTA-CFHT survey in the NIR \citep{Geach2017}, the Spitzer-IRAC Equatorial Survey in the MIR \citep{Timlin2016}, high-resolution imaging with the Very Large Array in the radio \citep{Hodge2011}, and Chandra observations in the X-rays \citep{LaMassa2016}. Of most relevance for securing wavelength coverage that impacts the derivation of SFRs and dust masses, a portion of Stripe~82 was observed in the FIR by the Herschel Stripe~82 Survey (HerS; \citealt{VieroHerS2014}) and the Herschel Multi-tiered Extragalactic Survey (HerMES; \citealt{Oliver2012}), which makes Stripe~82 the largest area scanned by Herschel to a uniform depth \citep{Lutz2014,Shirley2021HELP}. These Herschel observations have much better sensitivity than the all-sky FIR surveys executed by the Infrared Astronomical Satellite \citep{Neugebauer1984} or the AKARI Far Infrared Surveyor \citep{Kawada2007}. The above-summarized databases distinguish Stripe~82 as one of the best regions to study the comprehensive SED of a sizable population of representative low-redshift galaxies. 

Stripe~82 has been the target of diverse investigations. Most previous studies of low-redshift galaxies in this survey area largely capitalized on the advantages afforded by the deep optical and NIR imaging to probe their low-surface brightness features, morphology, and local environment. Examples include the construction of a large, mass-limited sample of massive galaxies \citep{Bundy2015}, searching for tidal features in early-type systems \citep{Kaviraj2010, Yoon2020}, surveying for ultra-diffuse galaxies \citep{Zaritsky2021}, investigating the host galaxies of nearby AGNs \citep{Matsuoka2014} and their incidence of mergers \citep{Karhunen2014, Hong2015}, and measuring internal galactic substructures, such as bulges and disks \citep{Bottrell2019, Sachdeva2020}, disk truncations and extended halos \citep{Peters2017}, and asymmetry \citep{Yesuf2021}. \citet{Ellison2016} used Herschel observations of a subset of Stripe~82 galaxies as a training set for artificial neural network prediction of total IR luminosity for SDSS galaxies. In a similar spirit, \citet{Rosario2016} assessed the systematics of optical SFRs with the aid of IR-based SFRs computed from Herschel and WISE observations of low-redshift Stripe~82 galaxies. Nevertheless, the valuable repository of multiwavelength data for Stripe~82 has yet to be fully exploited to characterize systematically the comprehensive (UV to FIR) SEDs of its constituent galaxies to derive their basic physical properties. Value-added databases of SDSS galaxies, such as that maintained by MPA-JHU \citep{Kauffmann2003,Brinchmann2004,Salim2007} and the GALEX-SDSS-WISE Legacy Catalog 1 and 2 \citep{Salim2016,Salim2018}, certainly encompass the Stripe~82 field. However, the SED analysis that underpins these catalogs did not incorporate the broadest or most inclusive wavelength coverage in the photometry used, in large part because most SDSS galaxies lack Herschel observations.

A more subtle, often overlooked limitation of previous efforts is that they usually amalgamate measurements from extant primary source catalogs, which employed different methods of photometry and uncertainty estimation. This can introduce hard-to-quantify systematic uncertainties and bias the SED fitting results \citep[e.g.,][]{Hill2011GAMA, Clark2018}. Such systematic uncertainties are especially pernicious for low-redshift galaxies, as many catalogs performed photometric measurements assuming that the sources are point-like [using point-spread function (PSF) fitting photometry], while, in actuality, the galaxies are resolved or marginally resolved \citep{Wright2016LAMBDAR, CalderonCastillo2019, Miller2020}. One can measure galaxy fluxes consistently across different bands by modeling its extended surface brightness distribution. For instance, \citet{Lang2016Tractor} measured fluxes in the four WISE MIR bands (3.4 to 22~$\mu$m) using the code {\tt Tractor} \citep{Lang2016TractorCode} by constraining the galaxy with the SDSS $r$-band ``exponential + de~Vaucouleurs'' model. Alternatively, the {\tt GALAPAGOS} code \citep{Haussler2013} can fit the light distribution of the galaxy with a parametric model to yield its total flux simultaneously in multiple bands \citep{Vika2013SHAPEGALFIT, Psychogyios2020}. The latest version of the code \citep{Haussler2022} even permits simultaneous multi-band decomposition of the bulge and disk components. {\tt ProFuse} \citep{Robotham2022} offers similar functionalities. While effective and efficient, the necessarily idealized and oversimplified fits usually invoked in parametric methods might introduce systematic uncertainties depending on the degree to which the models reproduce true light distribution of the galaxy.

The complications of parametric models can be obviated by adopting a model-independent strategy for aperture photometry. The aperture size, shape, and orientation can be chosen to optimally capture the desired boundaries of the source, and they can be fixed for consistency across different bands. As an example, \cite{Wright2016LAMBDAR} conducted forced-aperture photometry of galaxy images in 21 bands (UV to FIR) for the Galaxy and Mass Assembly sample (GAMA; \citealt{Baldry2010, Driver2011, Driver2016}) by using as prior an aperture defined from a high-resolution optical image and then calculating the working aperture for other bands, after convolving the prior aperture with the PSF of the corresponding bands. \cite{Clark2018} adopted a similar approach for the extensive panchromatic (UV to submillimeter) data for the DustPedia sample of nearby galaxies. Developed in the same spirit, the source detection and photometry code \texttt{ProFound} \citep{Robotham2018profound} starts with a segmentation map created from the stacked optical and NIR images and then enlarges it to generate the source aperture according to the surface brightness depth and resolution of the image.

We present a comprehensive catalog of panchromatic photometry for 2781 low-redshift ($z = 0.01-0.11$) galaxies selected from the HerS region of Stripe~82, covering 17 bands from the far-UV (FUV) to 500~$\mu$m. Of these, 2668 sources belong to a primary sample of active and inactive galaxies with stellar masses $M_* > 10^{10}\, M_{\odot}$. We perform SED fitting to derive stellar masses, SFRs, dust masses, and AGN luminosity fractions, which will serve as the foundation for a series of forthcoming works. This paper describes the technical details of our approach to securing robust and physically consistent measurements of total galaxy fluxes and associated uncertainties across many bands, and, where necessary, proper upper limits. As most of the galaxies in our sample are marginally to fully resolved from the UV to the MIR, but not in the FIR, we develop a hybrid approach of matched-aperture photometry, which can be used effectively for the 14 shorter wavelength bands (FUV to 4.6~\um), along with profile-fitting photometry for the longer wavelength bands. Interacting galaxies and sources in crowded regions, such as the cores of galaxy groups and clusters, are decomposed properly using multi-band, simultaneous two-dimensional (2D) image decomposition. Special care is devoted to estimating the full error budget, which includes uncertainties associated with background variance and contamination by foreground stars, which we treat using a new method. We design a series of mock tests to evaluate the impact of including certain critical bands and their respective upper limits in the SED analysis. Comparing our new measurements with those of previous works reveals that SED fitting-based SFRs of galaxies with moderate to low star formation activity should be used with considerable caution.

The paper is structured as follows. Section~\ref{sec_sample} introduces the sample definition and the multiwavelength data used in this work. Our methodology is introduced in Section~\ref{sec_method}, which includes our techniques for image preprocessing and deblending, aperture-matched photometry, model-fitting photometry for heavily blended galaxies, and estimating uncertainties and upper limits. Section~\ref{sec_SEDfit} presents the method for SED fitting, the results obtained therefrom, the effect of AGNs, the influence of nondetections, and parameter uncertainties. Section~\ref{sec_comparison} compares the stellar masses and SFRs derived in this work with those of other widely used catalogs. The implications of this work are discussed in Section~\ref{sec_discussion}, and the main conclusions are summarized in Section~\ref{sec_summary}. Some technical details are relegated to Appendices A--C. We adopt a $\Lambda$CDM cosmology with $H_0=70\,\rm km\,s^{-1}\,Mpc^{-1}$ and $\Omega_{\Lambda}=0.7$. Stellar masses and SFRs reported in this study are based on the \citet{Chabrier2003} stellar initial mass function.

\begin{figure}
\centering
\includegraphics[width=0.5\textwidth]{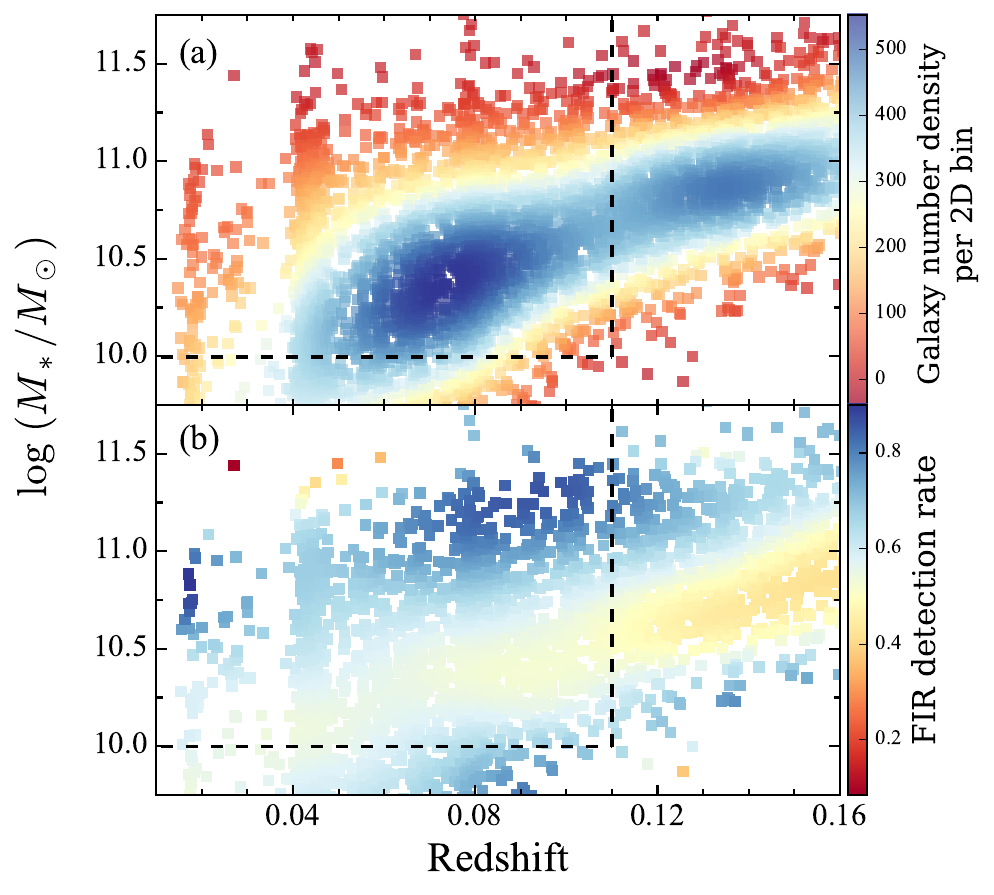}
\caption{The distribution of stellar mass \citep{Salim2018} and redshift for galaxies in Stripe~82. We code the targets by (a) the number density of SDSS galaxies in each stellar mass-redshift bin and (b) the detection rate in Herschel/SPIRE 250\,$\mu$m (HELP catalog; \citealt{Shirley2021HELP}). Galaxies with $\mathrm{S/N}>2$ are considered detections. We bin stellar mass in steps of 0.5\,dex and redshift in steps of 0.01. The detection fraction is defined as the ratio between the number of detections over the number of galaxies in each bin. The cuts of redshift and stellar mass for our sample are shown as black dash lines (see text for more details).}
\label{fig:sample}
\end{figure}

\begin{deluxetable*}{r  c  R  c  c c c c c r r c c c c}
    \tablecaption{Sample Information} 
    \tablewidth{12pc}
    \label{tab_info}
    \tablehead{                                                                    
    \colhead{ID} &   
    \colhead{R. A.} &                    
    \colhead{Decl.}&
    \colhead{$z$}&
    \colhead{Method}&
    \colhead{Interaction}&
    \colhead{Companion}&
    \colhead{$a_{25}$}&
    \colhead{$b_{25}$}&
    \colhead{$\rm PA_{25}$}&
    \colhead{$R_{50}$}&
    \colhead{$R_{90}$}&
    \colhead{$R_e$}&
    \colhead{$n$} &
    \colhead{Spectral}
    \\
    \colhead{ } & 
    \colhead{(\degree) } & 
    \colhead{(\degree) } & 
    \colhead{ } &
    \colhead{ Flag } &
    \colhead{ Flag } &
    \colhead{ ID } &
    \colhead{ (\arcsec) } & 
    \colhead{ (\arcsec) } &
    \colhead{ (\degree) } &
    \colhead{ (\arcsec) } &
    \colhead{ (\arcsec) } &
    \colhead{ (\arcsec) } &
    \colhead{ } &
    \colhead{ Class} \\
    \colhead{(1) } & 
    \colhead{(2) } & 
    \colhead{(3) } & 
    \colhead{(4) } & 
    \colhead{(5) } & 
    \colhead{(6) } & 
    \colhead{(7) } & 
    \colhead{(8) } & 
    \colhead{(9) } & 
    \colhead{(10) } & 
    \colhead{(11) } & 
    \colhead{(12) } & 
    \colhead{(13) } & 
    \colhead{(14) } & 
    \colhead{(15) } 
    }
    \startdata
    1 & 22.89043 & 0.55986 & 0.0781 & 2 & 1 & 2486 & 24.93 & 17.20 & 35.35 & 1.34 & 1.34 & 5.07 & 6.68 & Inactive \\
    2 & 28.69295 & 1.05575 & 0.0817 & 1 & 0 & 0 & 13.38 & 4.52 & 93.62 & 4.54 & 4.54 & 7.51 & 5.09 & Inactive \\
    3 & 30.35632 & 1.09832 & 0.0779 & 2 & 1 & 2473 & 39.13 & 33.07 & 169.75 & 4.16 & 4.16 & 7.03 & 3.63 & AGN \\
    4 & 31.79894 & -0.83094 & 0.0702 & 1 & 0 & 0 & 10.30 & 8.92 & 20.71 & 0.91 & 0.91 & 0.69 & 2.21 & Inactive \\
    5 & 28.71352 & 0.70223 & 0.0791 & 1 & 0 & 0 & 10.30 & 9.03 & 96.06 & 1.08 & 1.08 & \nodata & \nodata & Inactive \\
    6 & 31.34923 & 0.13916 & 0.0767 & 1 & 0 & 0 & 12.43 & 10.75 & 83.72 & 1.06 & 1.06 & \nodata & \nodata & Inactive \\
    7 & 26.96477 & -0.28846 & 0.0792 & 1 & 0 & 0 & 12.83 & 8.00 & 132.25 & 0.94 & 0.94 & 0.74 & 2.96 & Inactive \\
    8 & 19.92329 & 1.12774 & 0.0892 & 1 & 1 & 100 & 12.68 & 9.44 & 63.66 & 1.23 & 1.23 & \nodata & \nodata & Inactive \\
    9 & 21.99442 & -1.24683 & 0.0773 & 1 & 0 & 0 & 10.73 & 9.69 & 39.57 & 0.94 & 0.94 & \nodata & \nodata & Inactive \\
    10 & 14.38344 & 0.27412 & 0.0797 & 1 & 0 & 0 & 11.52 & 9.52 & 176.16 & 0.97 & 0.97 & 0.71 & 3.04 & Inactive \\
    \enddata
\tablecomments{
Col. (1): Index number.
Col. (2): Right ascension (J2000).
Col. (3): Declination (J2000).
Col. (4): Spectroscopic redshift from SDSS.
Col. (5): Flag for the photometric method: 0 = no photometry because of serious contamination from saturated stars or projected background galaxies, optical isophotal size is unavailable, or strongly affected by distortion effects from the edge of the GALEX field; 1 = aperture photometry; 2 = photometry by 2D decomposition; 3 = aperture photometry for supplementary object; 4 =  photometry by 2D decomposition for supplementary object.
Col. (6): Flag to indicate whether the target is in an interacting system: 0 = isolated; 1 = merger pair system; 2 = cluster central region; 3 = uncertain (a merger system or an isolated galaxy with subtructures); 4 = merger system consisting of $>2$ galaxies.
Col. (7): Index number of the merger companion if interaction flag = 1; if the merger system includes more than two galaxies, the companion is the closest galaxy.
Col. (8): Semi-major axis of the aperture based on $\mu_r = 25\,\rm mag\,arcsec^{-2}$; see Section~\ref{sec_method} for details.
Col. (9): Semi-minor axis of the aperture based on $\mu_r = 25\,\rm mag\,arcsec^{-2}$.
Col. (10): Position angle of the aperture based on $\mu_r = 25\,\rm mag\,arcsec^{-2}$.
Col. (11): Half-light radius from SDSS database.
Col. (12): Petrosian $R_{90}$ from SDSS database.
Col. (13): Effective radius from \cite{Bottrell2019}.
Col. (14): \sersic\ index of the single-\sersic\ model fit from \cite{Bottrell2019}.
Col. (15): Spectroscopic classification based on the optical emission-line diagnostic criteria of \cite{Kewley2001}; objects with weak emission lines are classified here as ``inactive.''
(The full machine-readable table can be found in the online version.)
}
\end{deluxetable*}

\begin{figure}
\centering
\includegraphics[width=0.45\textwidth]{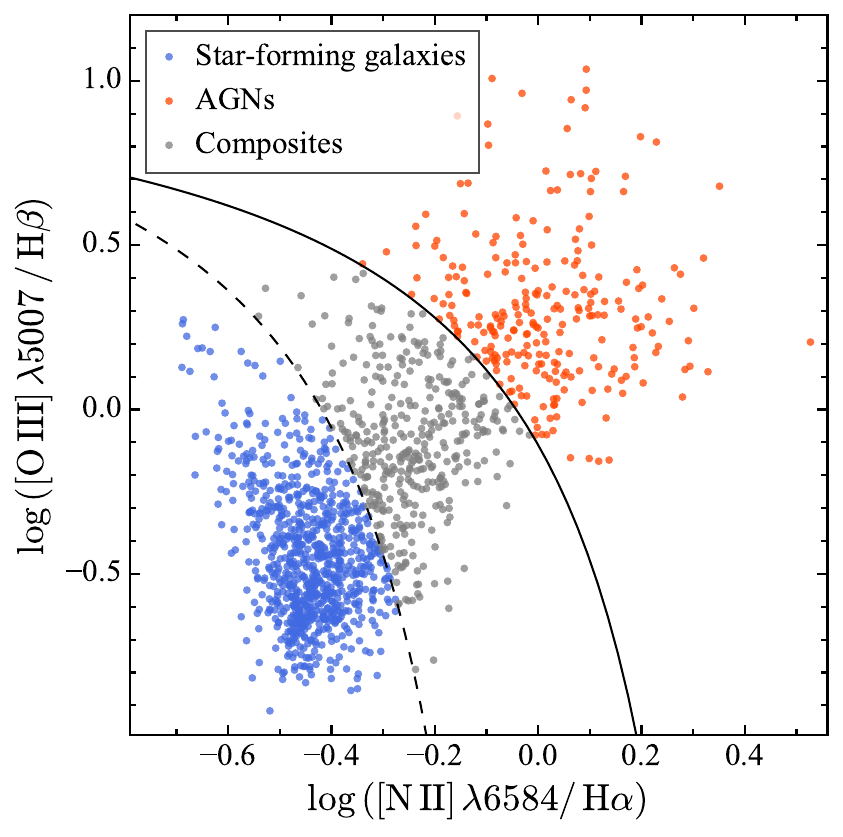}
\caption{
The \OIII/H$\beta$ versus \NII/H$\alpha$ diagnostic diagram for objects with $\mathrm{S/N}>2$ in their optical emission-line fluxes. The extreme starburst boundary (solid line; \citealt{Kewley2001}) and the pure star formation line (dashed line; \citealt{Kauffmann2003}) classify sources into star-forming galaxies (blue), AGNs (red), and composite (star-forming and AGN) systems.}
\label{fig:BPT}
\end{figure}

\section{Observational Material}\label{sec_sample}
\subsection{Sample Definition}

We select galaxies in the $\rm 79\,deg^2$ SDSS Stripe~82 region that is covered by the Herschel Stripe~82 Survey (HerS; \citealt{VieroHerS2014}) with the Herschel/SPIRE instrument. The SDSS DR16 database provides spectroscopic redshift, spectral classification \citep{Bolton2012}, and morphological classification \citep{Lupton2001SDSS} of each target. Among the three spectral types (\texttt{GALAXY}, \texttt{QSO}, and \texttt{STAR}), we only select sources classified as \texttt{GALAXY}. We also require the morphological type of the targets to be \texttt{GALAXY}\footnote{\url{https://www.sdss.org/dr16/algorithms/classify/\#Star/GalaxyClassification}} when selecting targets in CasJobs\footnote{\url{http://skyserver.sdss.org/CasJobs/}}.

Figure~\ref{fig:sample} shows for the Stripe~82 galaxies the distribution of redshift versus stellar mass ($M_*$) derives from the GALEX-SDSS-WISE Legacy Catalog 2 (GSWLC-2; \citealt{Salim2018}). A lower redshift cut of $z>0.01$ is required in the GSWLC-2 to exclude the very nearby galaxies whose distance is poorly indicated by their redshift \citep{Tully2016}. The two notable large-scale structures arise from overdensities at $z\approx 0.07$ and 0.13 (Figure~\ref{fig:sample}a). As one of our primary objectives is to incorporate FIR measurements into the SED analysis in order to obtain more robust SFRs, we require that a significant fraction of the sources be detected at least in the Herschel/SPIRE 250\,$\mu$m band. Figure~\ref{fig:sample}b color codes the data points by the detection rate ($f_{\rm det}$) of SPIRE 250\,$\mu$m as given in the HELP catalog \citep{Shirley2021HELP}, which considers flux densities with $\mathrm{S/N}>2$ as detections. Our sample is defined by the 2681 galaxies with $\Ms>10^{10}\,\Msun$ and $z<0.11$, which have an average 250\,$\mu$m detection rate of $f_{\rm det} \gtrsim 0.55$.  Beyond $z=0.12$, the average detection rate in redshift bins drops to 0.5 or lower.  We exclude 13 galaxies with very low surface brightnesses that lack SDSS isophotal measurements down to a surface brightness level of $\mu_r = 25$\,\usurf, which are needed for our aperture definition (Section~\ref{sec_method}), and an additional two galaxies that lie on the edge of the GALEX field, whose distortion effect strongly impacts the UV data. 

To facilitate scientific analysis using our database, we classify the sample into isolated (interaction flag = 0) and interacting (interaction flag = 1) galaxies. We consider a galaxy as interacting or merging if it contains a physically associated companion, which we define as another galaxy with a spectroscopic redshift difference $\Delta z_\mathrm{spec} < 0.002$ ($< 600\,\rm km\,s^{-1}$) within a projected separation of $d\leq 50$\, kpc \citep[e.g.,][]{Ellison2010, CalderonCastillo2019}. Lacking a reliable spectroscopic redshift, we regard it as interacting or merging if it exhibits obvious tidal features connecting to the main galaxy, and its photometric redshift ($z_{\rm phot}$), considering its uncertainty, is consistent with the $z_{\rm spec}$ of the main galaxy. We additionally require that the two targets differ by less than 3 magnitudes in the $r$ band. This ensures that we do not miss galaxy pairs with mass ratios as low as 10:1, a conventional threshold for minor mergers, while at the same time excluding the multitude of confusing, fainter nearby sources that may be associated with clumpy substructures from the main galaxy. The sample contains 96 galaxies with an interacting companion that has $\Ms \leq 10^{10}\,\Msun$, and hence formally does not satisfy the stellar mass cut of the main sample. We retain these pairs, so long as the companion is not more than 3 magnitudes fainter than the primary galaxy. Since many of these fainter companions only have photometric redshifts, which have relatively large uncertainties, in these instances, if they do not have spectroscopic redshifts, we simply assume that their redshifts are identical to the spectroscopic redshift of the primary companion.

More than half ($\sim 53\%$) of the galaxies have strong emission lines in their optical spectra. We diagnose their photoionization source based on the \OIII\,$\lambda 5007$/\Hb\ versus \NII\,$\lambda 6584$/\Ha\ line-intensity ratio diagram \citep{BPT1981}, adopting the classification criteria of \cite{Kewley2001} and \cite{Kauffmann2003}. As we present in Figure~\ref{fig:BPT}, among the 1472 galaxies for which all four emission lines are detected with $\rm S/N>3$ according to the MPA-JHU catalog, 850 are classified as star-forming galaxies, 247 as active galactic nuclei (AGNs), and 375 as composites (star-forming component mixed with an AGN component). The rest ($\sim 47\%$) of the sample has little to no detectable line emission, consisting mainly of truly quiescent galaxies with minimal star formation and low-ionization nuclear emission-line regions (LINERs; \citealt{Heckman1980}), which are predominantly highly sub-Eddington AGNs \citep{Ho2008, Ho2009}. Since such weak emission-line sources will contribute negligibly to the broadband SED, for the purposes of this study will simply consider them inactive galaxies. The above statistics do not account for 17 sources within the redshift cut that have spectral type \texttt{QSO} in the SDSS database, a generic label for sources that exhibit broad emission lines. Although these objects were excluded from the GSWLC-2 and therefore do not have stellar masses, their stellar masses are expected to exceed $10^{10}\, M_\odot$ and therefore satisfy the selection criteria of our sample. From the catalog of \cite{Liu2019}, these 17 broad-line AGNs have black hole masses $M_{\rm BH} \approx 10^{6.1}-10^{8.2}\,M_\odot$, which, according to the $M_{\rm BH}-M_*$ scaling relation (for all galaxy types) of \cite{Greene2020}, correspond to $M_* \approx 10^{9.7}-10^{11}\,M_\odot$. In summary: the final sample has 2781 galaxies (Table~\ref{tab_info}), consisting of 2668 objects that satisfy the original stellar mass cut ($M_* > 10^{10}\, M_{\odot}$), 96 supplementary objects of lower stellar mass that are companions of members from the original sample, and 17 broad-line (type~1) AGNs included for completeness.

\subsection{Data} \label{sec_data}

Our analysis involves panchromatic photometry of low-redshift, massive galaxies selected from the HerS region of Stripe~82. To this end, we assemble the following sets of multiwavelength images.

\begin{itemize}

\item{\bf Optical:} We extract the $urgiz$ images from Stripe~82\footnote{\url{https://www.sdss.org/dr16/algorithms/magnitudes/}}, whose coadded images are about 2 magnitudes deeper than typical SDSS single-scan images, reaching a limit surface brightness of $\mu_r \approx 28.5\, \usurf$ \citep{Fliri2016}. The median seeing ranges from full width at half-maximum (FWHM) $\sim 1\farcs 31$ in $u$ to $\rm FWHM \approx 1\farcs 04$ in $z$.

\item{\bf UV:} GALEX conducted an all-sky imaging survey in the FUV (central wavelength $\sim1539$\,\AA) and near-UV (NUV; central wavelength $\sim2316$\,\AA) bands with a PSF resolution of $\rm FWHM = 4\farcs2$ and 5\farcs3, respectively \citep{Martin2005GALEX}. Although the General Release 6 and 7 (GR6+7; \citealt{Bianchi2014GALEX, Bianchi2017GALEX}) provide \cite{Kron1980} elliptical aperture magnitudes, the measurements are affected significantly by distortions on the edge of the detector \citep{Morrissey2007GALEX} and source blending \citep{Salim2016}. Instead of applying a statistical correction to the GR6+7 data (e.g., \citealt{Salim2016}), we perform our own aperture photometry after source deblending to measure total galaxy fluxes using a consistent, common aperture across the SED (Section \ref{sec_method}). We download the GALEX image tiles of the Deep, Medium, or All-sky Imaging Surveys ( DIS, MIS and AIS, \citealt{Morrissey2007GALEX}) from CasJob\footnote{\url{https://galex.stsci.edu/casjobs/}}. If a target is covered by different image tiles, we choose the one that has the longest exposure time. To avoid the distortion effect, we do not accept the NUV image if the target is $>0.53\degree$ from the image center.

\item{\bf NIR and MIR:} The NIR data come from images taken in the $J$ (1.24\,\um), $H$ (1.66\,\um), and $K_s$ (2.16\,\um) bands from 2MASS \citep{Skrutskie2006}. Although the spatial resolution of the images is moderate ($\rm FWHM \approx 3\arcsec$), most of the galaxies are still sufficiently resolved that their fluxes are underestimated significantly in the 2MASS Point Source Catalog (PSC; \citealt{Cutri2003}; Appendix~\ref{appdix_fluxcompare}). Meanwhile, the 2MASS Extended Source Catalog (XSC; \citealt{Jarrett2MASS2000}) only contains the measurements of $\sim 60\%$ of our targets. The AllWISE Atlas of the WISE \citep{Wright_WISE2010} all-sky survey provides match-filtered, coadded images in the W1 (3.4\,\um), W2 (4.6\,\um), W3 (12\,\um), and W4 (22\,\um) bands with FWHM resolution 6\farcs1, 6\farcs8, 7\farcs4, and 12\arcsec, respectively. The AllWISE catalog provides fluxes from PSF-fitting as well as aperture photometry, but we find that these measurements underestimate the total flux even for marginally resolved galaxies (Appendix~\ref{appdix_fluxcompare}). The fluxes from the ``unWISE'' catalog \citep{Lang2016Tractor} derived from forced photometry based on the SDSS $r$-band profile of the galaxies closely follow our measurements (Appendix~\ref{appdix_fluxcompare}). However, only $76\%$ of our targets have W4 measurements in the unWISE catalog.

The above considerations compel us to reanalyze all the NIR and MIR data in a uniform, self-consistent manner, following the same precepts applied to the optical and UV data. We download the 2MASS and AllWISE images, including the AllWISE uncertainty maps, from the NASA/IPAC Infrared Science Archive\footnote{\url{https://irsa.ipac.caltech.edu/}}. 

\item{\bf FIR:} The Herschel Stripe~82 Survey (HerS; \citealt{VieroHerS2014}) observed an area of 79\,deg$^2$ in the SDSS Stripe~82 region, covering ($\alpha, \delta)_{\rm J2000}$ = (0h~54m, $-2^{\circ}$) to (2h~24m, $+2^{\circ}$). The SPIRE imaging photometer \citep{Griffin2010} was used to survey the 250, 350, and 500 \um\ bands to an average depth of 13.0, 12.9, and 14.8 mJy~beam$^{-1}$, respectively. Although the rest of the Stripe~82 region is covered by the Herschel Multi-tiered Extragalactic Survey (HerMES; \citealt{Oliver2012}), the achieved depth is significantly shallower, and in this study we only focus on the HerS area. Nearly all of our galaxies are unresolved in SPIRE, which has a PSF FWHM of 18\farcs2, 25\farcs2, and 36\farcs3 at 250, 350, and 500\,\um, respectively. We take advantage of the comprehensive photometry offered by the Herschel Extra-galactic Legacy Project (HELP; \citealt{Shirley2021HELP}), which uses a Bayesian approach to deblend galaxies using prior information from higher resolution measurements at shorter wavelengths. HELP provides a catalog of forced photometry based on the prior information, as well as a ``blind catalog'' that does not use the prior information. We crossmatch our galaxies to the forced-photometry catalog using a search radius of 1\arcsec. Most of the targets that are not found in this catalog are too faint to be detected by Herschel, although an additional $\sim 70$ measurements were located in the blind catalog after enlarging the search radius to $7\farcs4$, which encompasses $>99.7\%$ ($3\,\sigma$) of the observed coordinate separations of the crossmatched Herschel counterparts of the SDSS galaxies. We visually verified that most of them are robust measurements and not contaminated by companion sources (Section~\ref{sec_galfitm}). In total, 96\% of the galaxies have reliable Herschel/SPIRE measurements from the HELP catalog, with $\sim 70\%$ of the targets detected at $250\,\mu$m. For the HELP-unmeasured sources, we estimate flux density upper limits at 250\,\um\ based on the number of observation scans \citep{VieroHerS2014}. We adopt a $2\,\sigma$ upper limit of 20.7\,mJy for regions scanned twice and 18.7\,mJy for those scanned 3 times.

\end{itemize}

\begin{figure*} 
\centering
\includegraphics[width=0.88\textwidth]{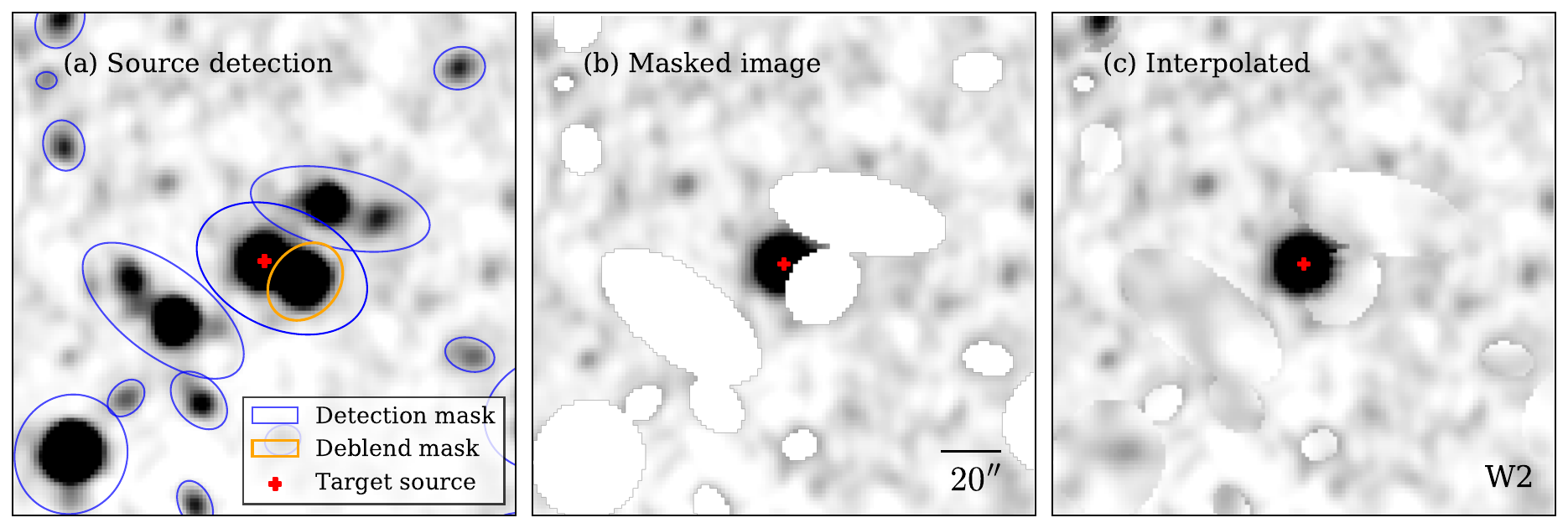}
\caption{Example of preprocessing a W2 image before aperture photometry. (a) Source detection and deblending the target (red cross) from other sources; a nearby contaminating source is masked (orange ellipse). (b) The final mask combines the {\tt{detection mask}} (all sources except the target segment) and the {\tt{deblend mask}} (if a contaminant is deblended). (c) The final image used for aperture photometry, generated using the surface brightness profile model of the target to interpolate the masked pixels with the model and adding noise.}
\label{fig:mask_interpolation}
\end{figure*}

\section{Method}
\label{sec_method}

\begin{figure*}
\centering
\includegraphics[width=1.0\textwidth]{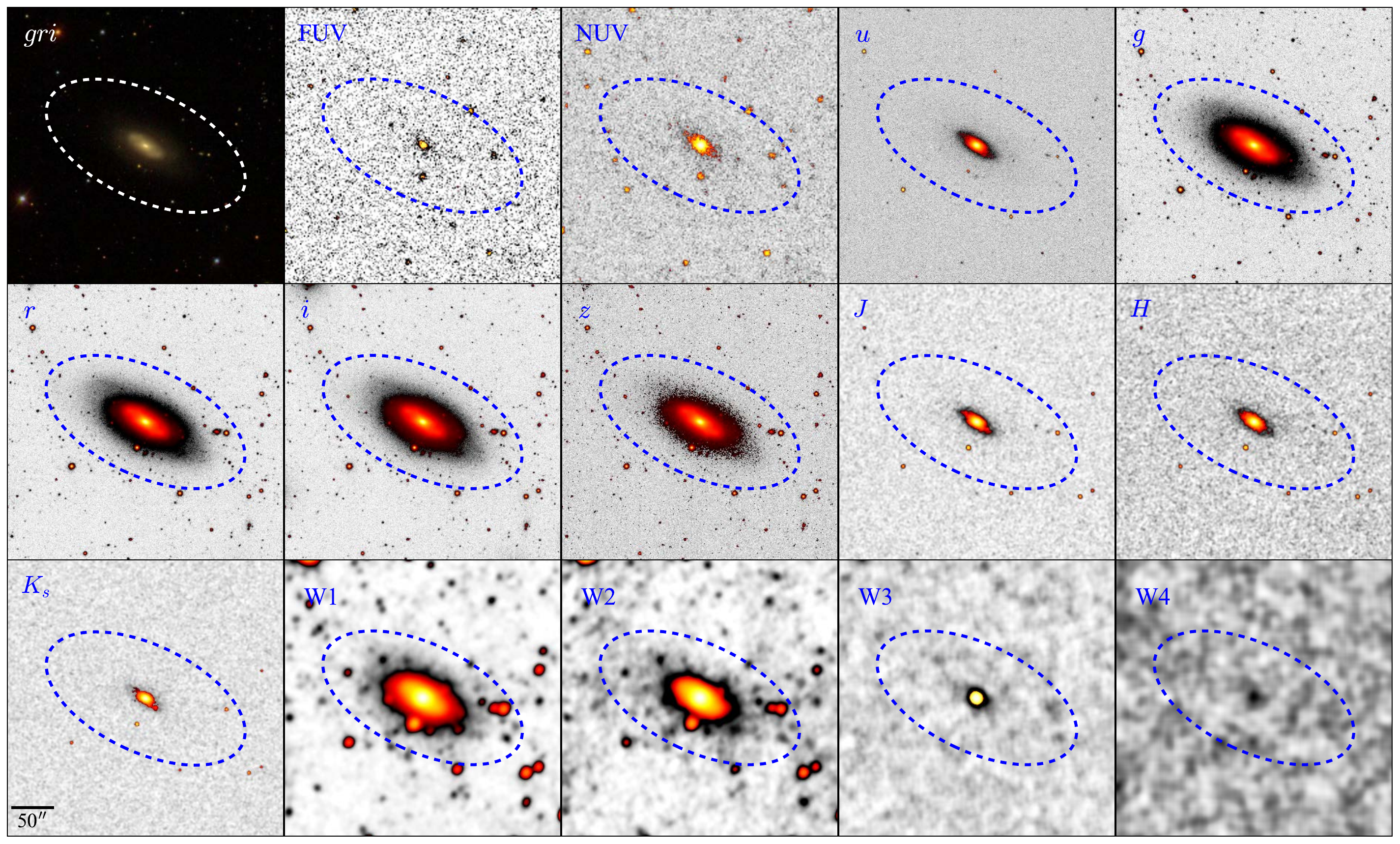}
\caption{Illustration of the aperture (dashed ellipse) for a resolved galaxy in images from FUV to W4, color-coded by the flux intensity in each filter. The aperture is twice the size of the isophote of the galaxy at 25\,\usurf\ in the $r$ band. Optical images come from the coaddition of SDSS Stripe~82 data produced by \citet{Fliri2016}. The first panel shows the SDSS $gri$ color composite image provided by the SDSS SkyServer.}
\label{StellarBandsAptr}
\end{figure*}

Apart from the three Herschel FIR bands, whose fluxes, owing to limitations of spatial resolution, we simply adopt from the HELP catalog (Section~2.2), for all the other 14 bands from FUV to W4 we apply a uniform method to measure the total flux using a common aperture defined by the optical profile of the galaxy. This section describes the steps necessary to preprocess the images, including source masking and detection, background subtraction, deblending, and removal of contaminants. We derive aperture-matched photometry for galaxies that are isolated or that can be deblended adequately. A minority of merging galaxies or galaxies in exceptionally crowded environments require more sophisticated 2D parametric decomposition. The photometric measurements are summarized in Tables~\ref{tab:photometry} and \ref{tab:photometry2}.  All flux and magnitude measurements in our catalog have been corrected for Galactic extinction, based on the extinction curve outlined by \cite{Fitzpatrick2007} and an assumed value of $R_V = 3.1$.

%\vspace{0.5cm}

\subsection{Preprocessing and Deblending}\label{sec_deblend}
We take the aperture of a galaxy (with semi-major axis $a_{25}$ and semi-minor axis $b_{25}$) twice as large as its isophote at 25\,\usurf\ in $r$ band to ensure a large enough aperture size.
We set the dimensions of the image cutouts to $400\arcsec \times 400\arcsec$ if $a_{25} < 50\arcsec$, and to $1000\arcsec \times 1000\arcsec$ if $a_{25} \geq 50\arcsec$. Setting the image size to $\gtrsim 10$ times the galaxy size ensures ample area to compute the background.

The images are preprocessed using the \texttt{Python} package \texttt{photutils} \citep{Bradley2020photutils}. We obtain a preliminary estimate of the background level and its standard deviation using the \texttt{detect\_threshold} function to sigma-clip the background data with a threshold of $2\,\sigma$. After smoothing the image to suppress the noise with a 2D circular Gaussian kernel with a FWHM equal to the image resolution, we run \texttt{detect\_sources} and generate a segmentation map. A source is considered detected if at least 5 connecting pixels have a flux more than 2 times the standard deviation above the background. We then use \texttt{source\_properties} to convert the segmentation map into a source list, which includes information such as the centroid, semi-major axis, semi-minor axis, and position angle (PA) of the galaxy. With this geometric information in hand, we grow the semi-major and semi-minor axes 3 times and derive the {\tt{detection mask}}. An example is shown in Figure~\ref{fig:mask_interpolation}a. To determine and remove the final background, we apply the mask and fit the background pixels with a third-order 2D polynomial model.

The image segmentation of the target may be blended with nearby sources. We use the function \texttt{deblend\_sources} to separate the potentially blended contaminants, setting \texttt{nlevels}\footnote{The number of multi-thresholding levels between the minimum and maximum values within the source segment.} = 32 and \texttt{contrast}\footnote{The fraction of the total (blended) source flux that a local peak must have to be considered a separate object.} = 0.0001 to aggressively break up the original segments into smaller segments based on the saddle point between the flux peaks \citep{Sazonova2021,Zhao2022}. If the companion sources are deblended, we generate the {\tt{deblend mask}} for them. The final mask of the image is a combination of the {\tt{detection mask}} without masking the target and the {\tt{deblend mask}} (e.g., Figure~\ref{fig:mask_interpolation}b).

In order to measure the total flux of the target, we need to interpolate nearby masked pixels. With the mask applied, we first derive the surface brightness profile of the target using the task \texttt{ellipse} in \texttt{Pyraf}\footnote{\url{https://pypi.org/project/pyraf/}} \citep{Tody1993IRAF}. We then use the 2D surface brightness model generated for the target to interpolate the masked pixels that contain significant emissions from the target. We run \texttt{ellipse} twice. In the first run, we fix the center of the ellipse to the centroid of the target according to its SDSS coordinates and fit the ellipticity and PA of the isophotes in logarithmic steps of 0.1 in radius. We choose the isophote with average surface brightness 2 times the standard deviation above the background to represent the shape of the galaxy. In the second run, we fix the ellipticity and PA to the values determined in the first run and fit the isophotes in linear steps of 1 pixel to densely sample the surface brightness profile of the galaxy. The 2D surface brightness model is generated from this output, which is used to replace the masked pixels, after adding Gaussian noise equivalent to the standard deviation of the background (Figure~\ref{fig:mask_interpolation}c). In order to define the region to interpolate the masked pixels, we need to determine the outer boundary where the target emission drops to the level of the background noise fluctuation, which is estimated following the methodology of \cite{LiZY2011}. We define the boundary of the galaxy by the semi-major ($a_\mathrm{profile}$) and semi-minor ($b_\mathrm{profile}$) axis of the isophote whose surface brightness reaches the background noise fluctuation.

Some galaxies suffer from such severe star contamination that the above standard deblending procedure fails to be effective. Under these circumstances, we use a new method developed to remove the contaminating emission from unsaturated stars (Appendix~\ref{sec_star_decontam}). As a brief summary, when the standard deblending in the UV or IR image fails for a star that contaminates the target galaxy, we predict the flux of the star within the aperture in that band and subtract it from the galaxy photometry. We predict of the total magnitude of the star in the UV or IR using a newly developed random forest regression method that uses the star’s optical PSF magnitudes as input. The fraction of the star's flux within the aperture is calculated based on its position relative to that of the target galaxy and the PSF light distribution in the respective band. Isolated stars serve as the training sample, using SDSS PSF magnitudes as input to take advantage of the optica data, which have the highest resolution among all the bands and accurate fluxes from the PSF models. We estimate the magnitude of the star in the GALEX, 2MASS, and WISE (W1$-$W3) bands, which can be predicted based on an $R^2$ score greater than 0.94. A small number (30) of galaxies corrupted by nearby, heavily saturated stars are unsalvageable. We removed them from further consideration, noting that this introduces no bias to the final sample.

\begin{deluxetable*}{rccc cccc cccc}
\tabletypesize{\footnotesize}
\tablecaption{Flux Density Measurements: Ultraviolet, Optical, and Near-infrared \label{tab:photometry}} 
\tablewidth{12pc}
\tablehead{ 
\colhead{ ID } & 
\colhead{$E(B-V)$}&
\colhead{ FUV } & 
\colhead{ NUV } & 
\colhead{ $u$ } & 
\colhead{ $g$ } & 
\colhead{ $r$ } & 
\colhead{ $i$ } & 
\colhead{ $z$ } & 
\colhead{ $J$ } & 
\colhead{ $H$ } & 
\colhead{ $K_s$ } 
\\
\colhead{  } &
\colhead{(mag) } &
\colhead{ ($\mu$Jy) } & 
\colhead{ ($\mu$Jy) } & 
\colhead{ (mJy) } & 
\colhead{ (mJy) } & 
\colhead{ (mJy) } & 
\colhead{ (mJy) } & 
\colhead{ (mJy) } & 
\colhead{ (mJy) } & 
\colhead{ (mJy) } & 
\colhead{ (mJy) } 
\\ 
\colhead{  (1) } &                                   
\colhead{  (2) } &                                   
\colhead{  (3) } &
\colhead{  (4) } &
\colhead{  (5) } &
\colhead{  (6) } &
\colhead{  (7) } &
\colhead{  (8) } &
\colhead{  (9) } &
\colhead{ (10) } &
\colhead{ (11) } &
\colhead{ (12) } 
}
\startdata
1 & 0.0200 & 1.23$\pm$0.15 & 2.37$\pm$0.15 & 0.05$\pm$0.01 & 0.26$\pm$0.02 & 0.60$\pm$0.05 & 1.00$\pm$0.10 & 1.09$\pm$0.04 & 4.48$\pm$1.24 & 0.76$\pm$0.21 & 1.96$\pm$0.29 \\
2 & 0.0257 & 3.10$\pm$0.74 & 10.22$\pm$1.24 & 0.08$\pm$0.02 & 0.25$\pm$0.03 & 0.58$\pm$0.06 & 0.85$\pm$0.12 & 1.06$\pm$0.21 & 4.24$\pm$0.68 & 5.02$\pm$0.87 & 4.87$\pm$0.84 \\
3 & 0.0216 & 52.30$\pm$3.58 & 70.00$\pm$3.75 & 0.30$\pm$0.06 & 1.49$\pm$0.13 & 3.22$\pm$0.24 & 4.83$\pm$0.48 & 5.85$\pm$0.20 & 6.05$\pm$1.69 & 7.35$\pm$2.05 & 6.01$\pm$0.56 \\
4 & 0.0297 & $<1.67$ & $<2.10$ & 0.03$\pm$0.01 & 0.17$\pm$0.02 & 0.37$\pm$0.04 & 0.53$\pm$0.07 & 0.72$\pm$0.14 & 1.12$\pm$0.27 & 1.42$\pm$0.35 & 1.17$\pm$0.50 \\
5 & 0.0284 & $<1.79$ & $<3.15$ & 0.02$\pm$0.01 & 0.13$\pm$0.02 & 0.29$\pm$0.03 & 0.43$\pm$0.06 & 0.56$\pm$0.11 & 0.97$\pm$0.29 & 1.03$\pm$0.50 & $<1.53$ \\
6 & 0.0210 & $<2.75$ & $<3.75$ & 0.04$\pm$0.01 & 0.16$\pm$0.02 & 0.33$\pm$0.04 & 0.46$\pm$0.06 & 0.59$\pm$0.12 & 1.15$\pm$0.32 & $<1.72$ & 1.69$\pm$0.55 \\
7 & 0.0277 & $<2.19$ & $<12.94$ & 0.04$\pm$0.01 & 0.21$\pm$0.02 & 0.47$\pm$0.05 & 0.68$\pm$0.10 & 0.91$\pm$0.18 & 1.05$\pm$0.25 & 1.58$\pm$0.42 & 1.53$\pm$0.53 \\
8 & 0.0231 & 2.51$\pm$1.09 & 5.65$\pm$1.29 & 0.04$\pm$0.01 & 0.21$\pm$0.03 & 0.44$\pm$0.05 & 0.64$\pm$0.09 & 0.82$\pm$0.16 & 1.53$\pm$0.39 & 2.31$\pm$0.54 & 1.39$\pm$0.48 \\
9 & 0.0313 & $<3.95$ & $<4.97$ & 0.03$\pm$0.01 & 0.15$\pm$0.02 & 0.34$\pm$0.04 & 0.48$\pm$0.07 & 0.63$\pm$0.13 & 1.23$\pm$0.29 & 1.17$\pm$0.40 & 1.46$\pm$0.48 \\
10 & 0.0217 & $<3.42$ & $<3.29$ & 0.07$\pm$0.02 & 0.29$\pm$0.03 & 0.60$\pm$0.07 & 0.83$\pm$0.12 & 1.08$\pm$0.22 & 1.59$\pm$0.35 & 1.73$\pm$0.48 & $<1.80$ \\
\enddata
\tablecomments{ 
Col. (1): Index number.
Col. (2): Galactic extinction from \cite{Schlafly_Finkbeiner2011}.
Cols. (3)--(12): Flux densities or upper limits in FUV, NUV, $u$, $g$, $r$, $i$, $z$, $J$, $H$, and $K_s$, corrected for Galactic extinction assuming the extinction curve of \cite{Fitzpatrick2007} and $R_V = 3.1$. The uncertainty is calculated according to Equation~1 in Section~\ref{sec_uncertainty}. If the measured flux density, $F$, is positive but has $\rm S/N <2$, we adopt $2\sigma_{\rm total}+F$ as the flux upper limit; if $F < 0$, we simply adopt $2\sigma_{\rm total}$ as the flux upper limit. See text for details. (The full machine-readable table can be found in the online version.)
}
\end{deluxetable*}

\begin{deluxetable*}{rcccc cccc cc}
\tabletypesize{\footnotesize}
\tablecaption{Flux Density Measurements: Mid-infrared and Far-infrared \label{tab:photometry2}} 
\tablewidth{12pc}
\tablehead{ 
\colhead{ ID}  & 
\colhead{ W1 } & 
\colhead{ W2 } & 
\colhead{ W3 } & 
\colhead{ W3 } & 
\colhead{ W4 } &
\colhead{ W4 } & 
\colhead{ HELP } &
\colhead{ 250~$\mu$m } & 
\colhead{ 350~$\mu$m } & 
\colhead{ 500~$\mu$m } 
\\
\colhead{  }      &
\colhead{ (mJy) } & 
\colhead{ (mJy) } & 
\colhead{  Flag}  &
\colhead{ (mJy) } & 
\colhead{  Flag}  &
\colhead{ (mJy) } & 
\colhead{  Flag}  &
\colhead{ (mJy) } & 
\colhead{ (mJy) } & 
\colhead{ (mJy) }  
\\ 
\colhead{ (1) } &                                   
\colhead{ (2) } &                                   
\colhead{ (3) } &
\colhead{ (4) } &
\colhead{ (5) } &
\colhead{ (6) } &
\colhead{ (7) } &
\colhead{ (8) } &
\colhead{ (9) } &
\colhead{ (10) } &
\colhead{ (11) } 
}
\startdata
1 & 0.99$\pm$0.06 & 0.54$\pm$0.07 & 1 & $<1.16$ & 1 & $<7.74$ & \nodata & $<18.67$ & \nodata & \nodata \\
2 & 2.69$\pm$0.27 & 1.60$\pm$0.17 & 0 & 0.28$\pm$0.06 & 0 & $<2.28$ & force & 2.94$\pm$3.28 & 2.97$\pm$3.17 & 3.55$\pm$4.00 \\
3 & 3.91$\pm$0.21 & 2.75$\pm$0.32 & 1 & $<3.01$ & 1 & $<8.42$ & force & 81.28$\pm$4.42 & 55.53$\pm$6.23 & 15.27$\pm$10.15 \\
4 & 0.54$\pm$0.07 & 0.33$\pm$0.06 & 0 & $<0.45$ & 0 & $<1.99$ & force & 3.19$\pm$3.28 & 3.98$\pm$4.17 & 6.53$\pm$6.41 \\
5 & 0.39$\pm$0.06 & 0.18$\pm$0.06 & 0 & $<0.42$ & 0 & $<1.55$ & force & 1.22$\pm$1.54 & 2.04$\pm$2.21 & 1.88$\pm$2.01 \\
6 & 0.41$\pm$0.05 & 0.15$\pm$0.06 & 0 & $<0.42$ & 0 & $<1.76$ & force & 10.04$\pm$5.95 & 4.27$\pm$4.25 & 4.22$\pm$4.45 \\
7 & 0.70$\pm$0.08 & 0.38$\pm$0.07 & 0 & $<0.35$ & 0 & $<2.41$ & force & 2.40$\pm$2.35 & 3.83$\pm$3.88 & 3.86$\pm$3.94 \\
8 & 0.64$\pm$0.07 & 0.39$\pm$0.07 & 0 & $<0.29$ & 0 & $<2.41$ & force & 2.97$\pm$3.17 & 3.40$\pm$3.39 & 1.96$\pm$2.29 \\
9 & 0.50$\pm$0.07 & 0.23$\pm$0.06 & 0 & $<0.27$ & 0 & $<2.31$ & \nodata & $<18.67$ & \nodata & \nodata \\
10 & 0.72$\pm$0.08 & 0.42$\pm$0.08 & 0 & $<0.68$ & 0 & $<2.01$ & force & 13.97$\pm$5.02 & 12.03$\pm$4.68 & 5.73$\pm$4.67 \\
\enddata
\tablecomments{ 
Col. (1): Index number.
Cols. (2), (3), (5), and (7): Flux density and associated uncertainty for W1, W2, W3, and W4 bands. The deblended flux densities for interacting systems are listed as separate entries.
Cols. (4) and (6): Flag for extended object for W3 and W4, respectively: 0  = unresolved; 1 = resolved according to method of Appendix~\ref{appdix_psffit}; 2 = blended, interacting system.
Col. (8): Flag for cross-matching SDSS coordinates to the HELP catalog: ``forced'' = object located within 1\arcsec\ from an object in the HELP forced-photometry catalog; ``blind'' = object located within 7\farcs4 from an object in the HELP blind-photometry catalog; null otherwise.
Cols. (9)--(11): Flux density and uncertainty of Herschel/SPIRE 250, 350, and 500$\,\mu$m bands.  The value listed is the median flux density of the posterior distribution from the HELP catalog \citep{Shirley2021HELP}. The value is null for objects that do not have a matching counterpart in the HELP catalog. Uncertainty equals half of the difference between the 84th and 16th percentile of the 250$\,\mu$m flux likelihood distribution. (The full machine-readable table can be found in the online version.)
}
\end{deluxetable*}

\subsection{Aperture-matched Photometry}
\label{sec_stellar_photo}

With the decontaminated, deblended images in hand, we perform aperture photometry across all 14 bands (${\rm FUV,\, NUV},\, u,\, g,\, r,\, i,\, z,\, J,\, H,\, K_s,\, {\rm W1,\, W2,\, W3,}$\,and W4) using a common aperture defined by the isophote at the surface brightness level of $\mu_r = 25$\,\usurf. This surface brightness level is trivially satisfied by Stripe~82, which reaches a depth of $\mu_r \approx 28.5\,\usurf$ \citep{Fliri2016}. We set the semi-major axis of the aperture to $a_{25}$ and its semi-minor axis to $b_{25}$ if both axes are larger than 3 times the FWHM of the PSF; otherwise, the minimum aperture radius is set to 3 times the FWHM of the PSF, sufficient to enclose $>90\%$ of the total flux of a point source. This guarantees that most of the flux of compact or edge-on galaxies can be included in the aperture, while obviating the need for aperture correction. Figure~\ref{StellarBandsAptr} shows an example of a galaxy resolved in all the bands. The WISE images are dominated by the confusion noise of the background; source emission that cannot be detected during the preprocessing steps will not bias the aperture photometry.

The W3 and W4 bands require special consideration. The coarse PSFs of these two WISE bands (${\rm FWHM} = 8\farcs4$ and 12\arcsec, respectively) imply that many of the galaxies in our sample are expected to be unresolved \citep{Cluver2014,Rosario2016}. Because the W3 and W4 observations are substantially shallower than those in W1 and W2, our approach of aperture-matched photometry will not be able to detect a significant fraction of the targets. Under these circumstances, PSF-fitting should perform better than aperture photometry. We carried out a series of mock tests to determine conservative empirical criteria to recognize unresolved galaxies under conditions appropriate for our sample selection. As detailed in Appendix~\ref{appdix_psffit}, a galaxy can be safely regarded as unresolved in W3 and W4 if, in the $r$ band, $R_e$ is less than 2\arcsec\ and 4\arcsec, respectively, and $n \ge 3.5$. By these criteria, 1328 sources are unresolved and unblended in W3; the corresponding number in W4 is 1844. We directly adopt the PSF-fitting measurements from the AllWISE catalog for these sources. AllWISE provides the uncertainty of the PSF fitting even if the target is not significantly detected, which enables us to estimate upper limits for nondetections.

To ascertain the degree to which our choice of aperture captures the total flux of the galaxy, we use {\tt{GALFIT}} \citep{Peng2002GALFIT, Peng2010GALFIT} to generate mock multi-band images of a subset of $\sim 300$ relatively isolated galaxies drawn from our parent sample, using as input the best-fit single-component \sersic\ profile parameters from \cite{Bottrell2019}. After convolving the models with the respective PSF, we add the simulated image to a real background image constructed for each specific band to mimic as faithfully as possible actual observations. Application of our aperture photometry procedure reveals that the fraction of the input flux recovered is $>90\%$ in the UV bands,  $>90\%$ in $g$, $r$, and $i$, $>80\%$ in $u$, $>85\%$ in the $z$ and 2MASS bands, $>90\%$ in W1 and W2, and $>80\%$ in W3 and W4. 
For bright objects, such as those with W4 $<6.5$ mag, the input flux recovery levels are close to unity. However, as objects become fainter, the scatter of the flux recovery increases to a standard deviation of $10\%-20\%$  around a median value of $\sim 1$. The underestimation of flux depends on luminosity and is relatively minor for more than 84\% of the galaxies when compared to the typical uncertainties (Section~\ref{sec_SEDfit}) associated with stellar mass (0.05~dex) and SFR (0.18~dex).

\begin{figure*}
\centering
\includegraphics[width=0.9\textwidth]{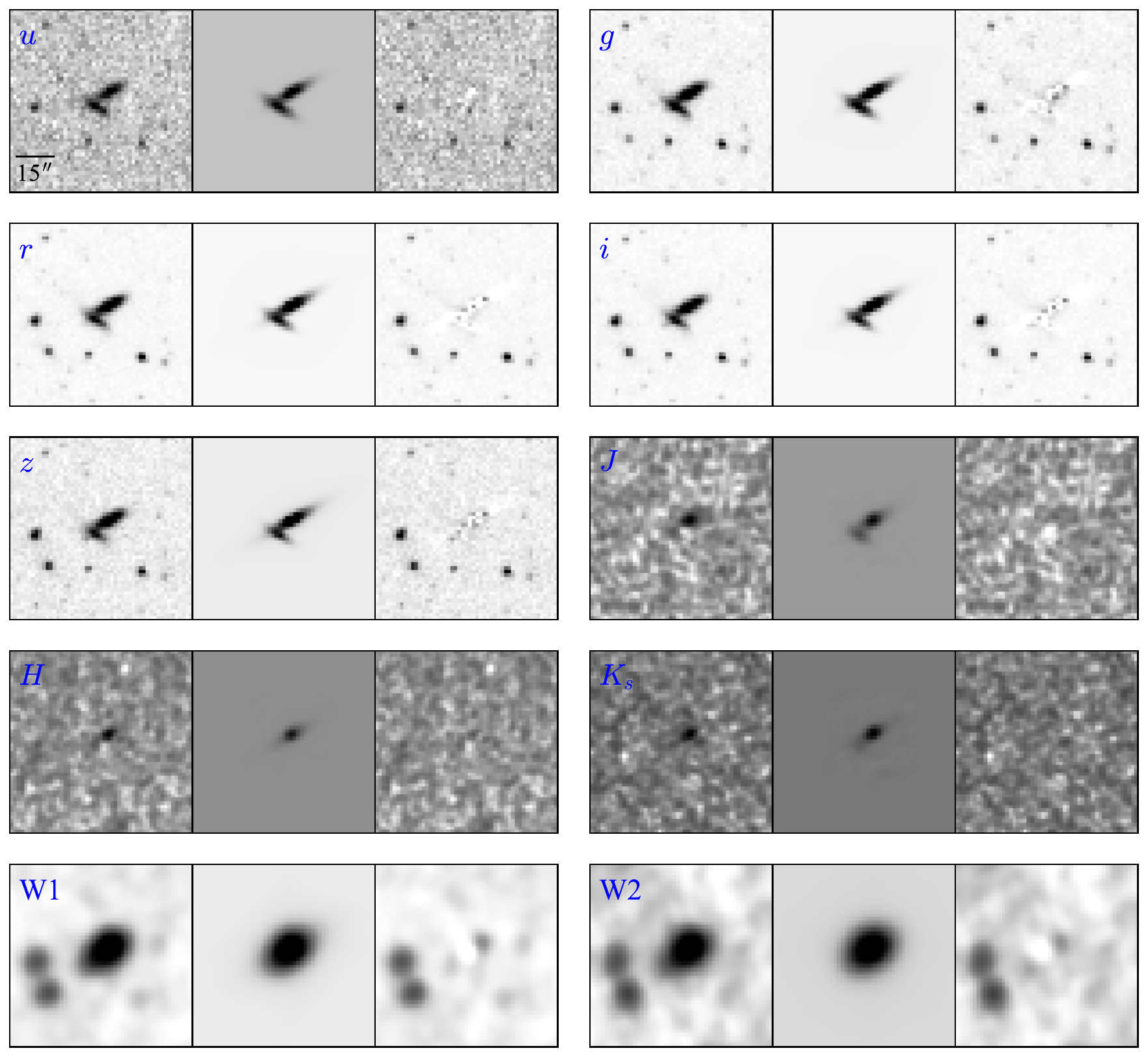}
\caption{Multiband model fitting for a galaxy with a close interacting companion. The original data, best-fit model, and model-subtracted residual map are presented from left to right for each of the 10 bands from $u$ to W2.}
\label{fig:galfitm_fit}
\end{figure*}

\subsection{Model-fitting Photometry for Heavily Blended Galaxies}
\label{sec_galfitm}

A minority ($\sim 340$) of the galaxies in our sample cannot be deblended using our standard procedure (Section~3.1). These largely consist of advanced mergers and close, projected pairs, as well as a handful of galaxies in very dense environments, such as the centers of rich clusters where multiple galaxies may be crowded together. Under these conditions, aperture photometry (Section~3.2) cannot yield meaningful results, and generally, none of the photometry given in conventional survey catalogs can be trusted either. We perform 2D image decomposition to deblend the galaxies using \texttt{GALFITM} \citep{Haussler2013, Vika2013SHAPEGALFIT}, which is a multiwavelength extension of {\tt{GALFIT}} \citep{Peng2002GALFIT, Peng2010GALFIT}. The code fits several bands simultaneously, leveraging the structural parameters of the well-resolved images in some bands against those that are more poorly resolved in others. Our fits incorporate 10 of the 14 filters, spanning from the $u$ band to the W2 band. We use the \texttt{Python} package \texttt{reproject}\footnote{\url{https://reproject.readthedocs.io/en/stable/celestial.html}} to register all the bands to the world coordinate system of the W1 image and rebin them to the lowest common pixel scale of 1\farcs375, which is dictated by WISE. Each galaxy is modeled with a single-component \sersic\ function. Model parameters across the different bands can be mutually constrained by a Chebyshev polynomial. Consistent with the experience of \citet{Haussler2013}, we find that a second-order function describes well the wavelength dependence of $R_e$ and $n$, with initial guesses taken from the catalog of \cite{Bottrell2019}. We keep the axis ratio and PA constant with wavelength, using as initial input $a_{25}$, $b_{25}$, and $\rm PA_{25}$. The magnitude in each band is allowed to vary freely. An example fit is given in Figure~\ref{fig:galfitm_fit}. 

The FUV and NUV bands are excluded from the simultaneous multi-band fit because they consistently give unstable results, possibly a consequence of the extreme variations in galaxy substructure due to the young stellar population and the effects of internal dust extinction. We opt, instead, to fit the two UV bands separately, holding constant the structure parameters ($R_e$, $n$, axis ratio, PA) to the values derived for the $g$ band from the simultaneous 10-band fit. Our 2D decomposition also cannot be applied to the W3 and W4 bands, or to any of the three Herschel bands, because the galaxy pairs are too severely blended in these long-wavelength filters. Moreover, we cannot guarantee that the distribution of the dust emission is identical to that of the stellar emission. However, as mentioned in Section~\ref{sec_data}, the HELP forced-photometry catalog provides deblending results based on the optical/NIR prior coordinates, using the full Bayesian posterior probability distribution. We use the HELP results for the blended pairs if the coordinates match. For pairs that can be matched only in the HELP blind-photometry catalog, we visually check their SPIRE 250 $\mu$m images and SDSS composite-color images. If the SPIRE source is contaminated by any neighboring objects visible in the SDSS image, we regard the SPIRE photometry as an upper limit for the target. We measure the total flux in W3 and W4 for the blended system within a master aperture, as described below. The total MIR flux is used as a flux upper limit for each component object when fitting its SED (Section~\ref{sec_SEDfit}). 

To quantify possible systematic differences between the photometry derived from the model-fitting approach compared to that measured from the aperture-matched technique used for the majority of the sample, we apply both methods to a calibration sample of 50 randomly selected, isolated galaxies that roughly spans the range of $r$-band magnitudes in the parent sample. The two sets of measurements correlate tightly, although small systematic offsets between them exist. The model-fitting photometry is on average brighter than the aperture-matched photometry by $\leq 0.1$~mag in $g$, $r$, $i$, $z$, W1, and W2. The offset is larger in the $u$ band ($\sim 0.3$~mag), a possible manifestation of Runge’s phenomenon \citep{Haussler2022}. The $J$, $H$, and $K_s$ bands have larger discrepancies of 0.1--0.3~mag owing to their lower sensitivity, with a dispersion comparable to the systematical offsets. We apply these average offsets in the optical and IR bands to the \texttt{GALFITM}-derived multi-band photometry for the heavily blended galaxies to mitigate systematic biases with respect to the rest of the sample.

Treating each of the blended galaxies as a single S\'ersic component obviously vastly oversimplifies the intricate, complex substructure often displayed by mergers and strongly interacting galaxies. Our approach, albeit crude, suffices to provide an effective first-order, relatively accurate partitioning of the individual fluxes of the constituent members in the blended system. We find that the sum of the fluxes of the individual model components matches the total flux of the blended system to better than $\sim 0.05$~mag in all bands studied. The total, integrated flux of the blended system is computed within a common, master aperture based on the surface brightness sensitivity of the image. We choose the largest $a_\mathrm{profile}$ and $b_\mathrm{profile}$ (Section~3.1) among all the bands as the final master aperture (see also \citealt{Shangguan2019}).

\begin{figure*} [ht]
\centering
\includegraphics[width=0.85\textwidth]{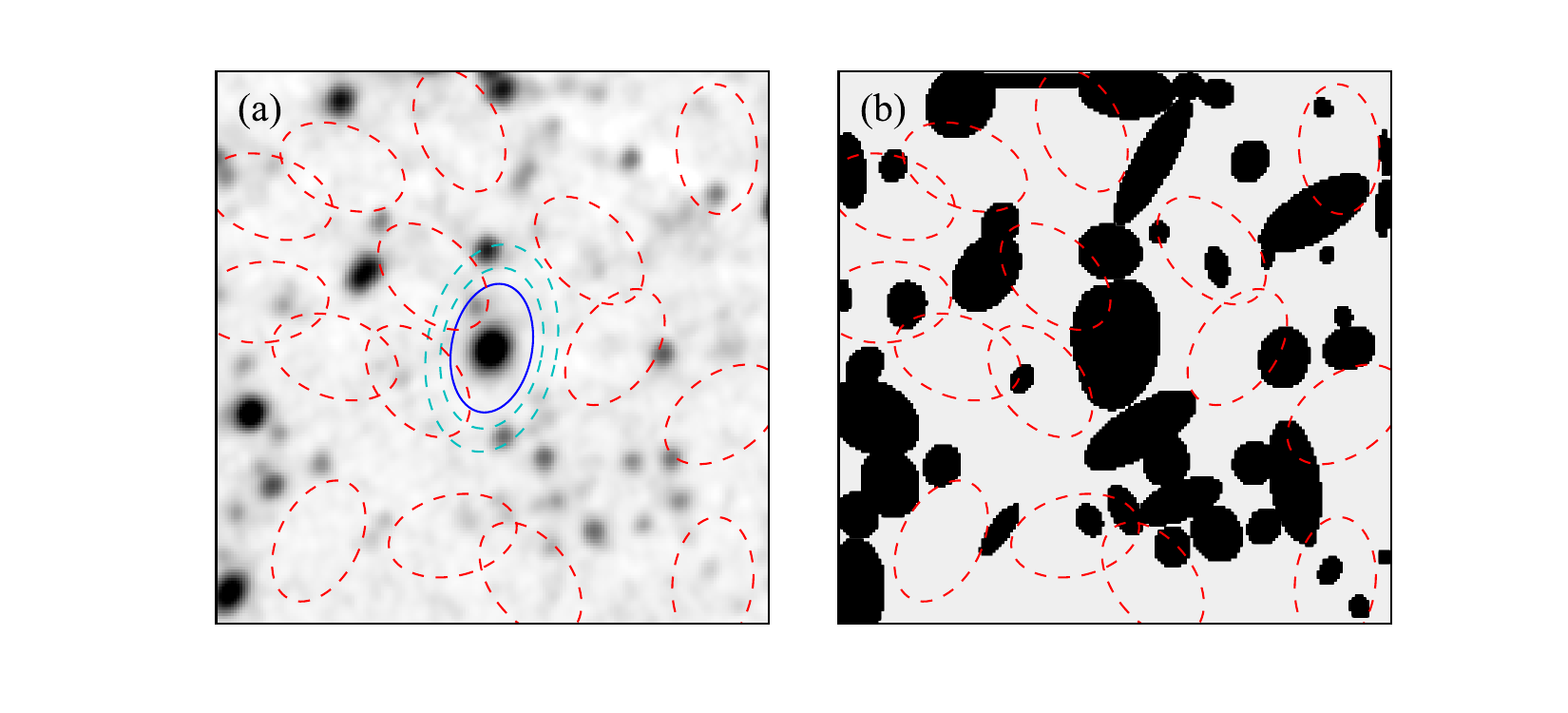}
\caption{Example uncertainty estimation for the aperture photometry. (a) The target is enclosed by the aperture represented by the blue ellipse. The cyan dashed annulus indicates the area used to estimate the background noise. The red dashed ellipses, which are required to overlap with each other less than 20\%, mark the footprints that randomly sample the confusion noise. (b) The mask instead of the image is displayed. We require that the red ellipses contain $<20\%$ of the masked pixels.}
\label{fig_uncertainty}
\end{figure*}

\vspace{0.3cm}
\subsection{Estimating Uncertainties and Flux Upper Limits}
\label{sec_uncertainty}

The uncertainty of the aperture photometry is
 
\begin{equation}
\label{equ_uncertainty}
\sigma^2_{\rm total} = \sigma^2_{\rm source} + \sigma^2_{\rm bkg} + \sigma^2_{\rm conf} + \sigma^2_{\rm star} + \sigma_\mathrm{sys}^2, 
\end{equation}

\noindent
where $\sigma_{\rm source}$ is the Poisson noise of the photons, $\sigma_{\rm bkg}$ is the background noise, $\sigma_{\rm conf}$ is the confusion noise, $\sigma_{\rm star}$ is introduced by blending stars (Equation~\ref{eqn:star}), and $\sigma_\mathrm{sys}$ is the systematic uncertainty of the aperture photometry. We explain each term below.

Assuming that the total flux from the aperture photometry is $I$, $\sigma^2_{\rm source} = I/G$, with $G$ specifying the gain of the image. The background noise can be estimated from an annulus around the target aperture, $\sigma^2_{\rm bkg}= N\,\sigma^2_{\rm ann}$, where $\sigma_{\rm ann}$ is the standard deviation of the background pixels in the annulus and $N$ is the number of pixels in the aperture of the galaxy. As shown in Figure~\ref{fig_uncertainty}a, the annulus has the same ellipticity and PA as the target's aperture. The inner and outer semi-major axes of the annulus are 1.25 and 1.60 times the semi-major axis of the aperture, so that the area of the annulus is the same as that of the aperture.

As the 2MASS and WISE images are resampled and interpolated, we need to consider their pixel-correlated noise \citep{Jarrett2MASS2000}. We adopt
\noindent
\begin{align}
&\sigma^2_{\rm source}\, (\mathrm{2MASS}) =  I/G + F_{\rm corrA}\,N^2\,\sigma^2_{\rm ann}, \\
&\sigma^2_{\rm source}\, (\mathrm{WISE}) =  F_{\mathrm{corrA}}\,\sum_{i}^{N}\sigma^2_{i},\\
&\sigma^2_{\rm bkg} = F_{\rm corrB}\,N\,\sigma^2_{\rm ann},
\end{align}

\noindent
where $F_{\rm corrA}=11.56$ and $F_{\rm corrB}=1$ for 2MASS images\footnote{\url{https://wise2.ipac.caltech.edu/staff/fmasci/ApPhotUncert_corr.pdf}}, accounting for the image coadding and smoothing. For WISE, $F_{\rm corrA}=F_{\rm corrB}$ depending on the aperture size \citep{Cutri2012wise}. The variance of individual pixels ($\sigma^2_{i}$) for the WISE coadded image can be calculated directly as the quadrature sum of the pixels in the uncertainty map in the same aperture.

We only consider the confusion noise ($\sigma_\mathrm{conf}$) for WISE images because their coadded images are very deep but poorly resolved ($>6\arcsec$). We randomly sample the background on the image with the same aperture size as that used to measure the target (Figure~\ref{fig_uncertainty}), estimating $\sigma_\mathrm{conf}$ as the standard deviation of the sampled median flux. To reduce the correlation of the sampling, we require that the samples overlap with each other by less than $20\%$ and contain fewer than $20\%$ of the masked pixels. Finally, the mock tests in Section~\ref{sec_stellar_photo} show that our aperture photometry can measure at least $80\%$ of the total flux of the galaxy. Based on the results of these mock tests,  we adopt $\sigma_\mathrm{sys} = 10\%-20\%$ as the systematic uncertainty of each band. As discussed in Section~\ref{sec_stellar_photo}, the $10\%-20\%$ flux loss is minor when compared to the typical uncertainties of stellar mass and SFR, particularly for faint galaxies that have even larger uncertainties in their physical properties. It suffices to include a systematic uncertainty in our final flux measurements to account for the scatter and offset in the flux recovery rate.

The full error budget of 2D image decomposition described in Section~\ref{sec_galfitm} is difficult to estimate. Apart from formal uncertainties, systematic uncertainties can arise from the influence of S/N and the PSF \citep{Guo2009,Yoon2011}, especially when bright nuclei are involved (e.g., \citealt{Kim2008, Zhuang2022}), and background estimation \citep{Huang2013ApJ, Gao2017}. Because bright AGNs are rare in our sample, we focus on the uncertainties induced by background estimation. We approach this problem empirically, by measuring the parameters for the blended system with different background levels, and calculating the scatter. For each band, we generate an idealized galaxy model using the best-fit parameters given by the 2D multi-band decomposition. After adding the Poisson noise of the source and the Gaussian noise of the background to mimic a realistic mock image of a blended, interacting system, we insert the mock image to different locations on a simulated background image specifically designed for each band. The simulated background images for SDSS, 2MASS, and WISE were created by mosaicing relatively empty sky regions. The background for GALEX is more complicated because its depth depends on the diverse exposure times of the different image tiles. For galaxies that are observed in the MIS or DIS, we simulate the background image of GALEX with the \texttt{Python} package \texttt{make\_noise\_image}, assuming that the large-scale, random noise of the background is dominated by Gaussian noise that can be estimated from the standard deviation of the pixel values in the galaxy-size apertures calculated from the source-masked regions of the FUV and NUV images. The background of the AIS data is primarily influenced by Poisson noise; we combine the real AIS background areas in the same tile to create GALEX background images. We repeat the decomposition multiple times, during each iteration placing the target in a different location in the background. The standard deviation of the multiple trials is used as the final uncertainty.

As discussed throughout the paper, we adopt different methods to measure the flux ($F$) and uncertainty ($\sigma$) of the galaxy in different bands. We follow the same method to estimate the flux upper limit. We consider a source undetected if its flux measurement has $\rm S/N <2 $. As in \cite{Cutri2012wise}, we estimate the upper limit as $2\sigma + F$, setting $F=0$ if the measured $F$ is negative. The only exception comes from the Herschel data. If no measurement is provided by the HELP catalog, we adopt $2\,\sigma_{250}$ as the 250$\,\mu$m upper limit (Section~2.1; \citealt{VieroHerS2014}).

\begin{deluxetable*}{l  l }
\tablecaption{{\tt CIGALE} Model Parameters for SED Fitting} 
\label{tab_CIGALE}
\tablehead{                                                                     
\colhead{Parameter} &                                                    
\colhead{Values}
}
\startdata
\multicolumn{2}{c}{Simple stellar population: \cite{BC03} }  \\ %\hline
                    Initial mass function     &  Chabrier \\
                    Metallicity        &  0.004, 0.02  \\ \hline
\multicolumn{2}{c}{Star formation history: double exponential function}   \\
                    e-folding time of the main stellar population: $\tau_{\rm main}$ (Myr)
                    & 600, 1000, 1500, 2000, 3000, 4500, 6000, 8000, 12000, 18000 \\
                    e-folding time of the latest starburst: $\tau_{\rm burst}$ (Gyr) 
                    &  20   \\ 
                    Fraction of stellar mass formed in starburst: $f_{\rm burst}$ 
                    &  0, 0.001, 0.003, 0.005, 0.01, 0.03, 0.05, 0.1, 0.15   \\ 
                    Age of the main stellar population: $t_{\rm main}$ (Gyr)         
                    &  12   \\ 
                    Age of the latest starburst: $t_{\rm burst}$ (Myr)
                    &  10, 100, 500, 1000, 1500, 2000, 3000, 4000, 5000 \\ \hline
\multicolumn{2}{c}{Nebular emission}    \\ %\hline
                    Ionization parameter: $\log U$                   
                    &  $-3.4$     \\
                    Fraction of Lyman continuum photons absorbed by dust %$f_{\rm dust}$ 
                    &  0.3     \\
                    \hline
\multicolumn{2}{c}{Dust attenuation: \cite{Calzetti2000}}   \\ %\hline 
                    Colour excess of the nebular lines: $E(B-V)$ (mag)
                    &  0.01, 0.15, 0.3, 0.45, 0.6 \\
                    Reduction factor to apply 
                    &  0.25, 0.75  \\
                    Amplitude of the UV bump (Milky Way)
                    &  3.0  \\
                    Slope of the power-law modification: $\delta$
                    &  $-1.2,\, -0.8, \,-0.4,\, 0$ \\ \hline
\multicolumn{2}{c}{Dust emission: \cite{Draine2014}} \\ %\hline 
                    Mass fraction of PAHs: $q_{\rm PAH}$ (\%)      &  2.5 \\
                    Minimum radiation field: $U_{\rm min}$     &  0.1, 0.5, 2.5, 10, 25   \\
                    Power-law slope of the dust radiation: $\alpha$           & 2   \\
                    Fraction illuminated from $U_{\rm min}$ to $U_{\rm max}$: $\gamma$          & 0.001, 0.006, 0.05, 0.2   \\ \hline
\multicolumn{2}{c}{AGN emission: \cite{Fritz2006}} \\
                    Ratio between outer and inner radius of the torus: $R_{\rm max}/R_{\rm min}$ & 60 \\
                    Optical depth at 9.7~$\mu$m: $\tau_{9.7}$ & 1 \\
                    Slope of the radial coordinate: $\beta$  &  $-0.5$ \\ 
                    Angle between equatorial axis and line-of-sight: $\psi$ (deg)  & 0 (type~2), 60 (type~1)  \\
                    Full opening angle of the torus: $\theta$ (deg) & 100 \\
                    Contribution of the AGN to the total $L_{\rm IR}$: $f_{\rm AGN}$ & 0, 0.001, 0.02, 0.1, 0.3, 0.6 \\
                    \hline\hline
\enddata
\end{deluxetable*}

\section{SED Fitting}
\label{sec_SEDfit}

\subsection{{\tt CIGALE} Model}
\label{ssec:model}

We analyze the photometric measurements of our galaxies with the widely used {\tt CIGALE} code \citep{Boquien2019CIGALE}. {\tt CIGALE} models the panchromatic SED from the X-rays to the radio with a flexible collection of model components, including the stellar continuum, nebular emission, dust attenuation and emission, and the AGN continuum \citep{Noll2009,Ciesla2015,Boquien2019CIGALE,Yang2020,Yang2022}. The user specifies the model components and a grid of parameters as prior information. The code calculates the likelihood, ${\cal{L}}=\exp(-\chi^2/2)$, for each model by comparing the model fluxes with the observed fluxes. {\tt CIGALE} uses a Bayesian approach to calculate the marginalized probability distribution function using the $\cal{L}$ values of all models. Based on the probability distribution function, {\tt CIGALE} provides the probability-weighted mean and standard deviation of physical parameters such as stellar mass (\Ms), SFR, dust mass ($M_d$), and AGN fraction ($f_{\rm AGN}$). {\tt CIGALE} adopts an energy conservation algorithm \citep{Burgarella2005, Boquien2019CIGALE}, which requires that the UV and optical emission attenuated by dust reradiates in the MIR and FIR.  In accordance with the tutorial provided by {\tt CIGALE}\footnote{\url{https://cigale.lam.fr/faq/}}, we set the input data and errors as $(2\sigma + F)/2 \pm (2\sigma + F)/2$. This approach allows us to conservatively include all available models.

Table~\ref{tab_CIGALE} summarizes the parameters of the SED model. The stellar emission is represented by simple stellar populations from \cite{BC03} of metallicity 0.004 to 0.02 \citep{Brown2014,Mountrichas2021}, with the metal-poor end suitable for the low-mass galaxies in the sample. We adopt the \citet{Chabrier2003} stellar initial mass function and a double-decaying exponential function to describe the star formation history. This model reproduces the SEDs of both star-forming and quenched galaxies with a modest number of free parameters \citep{Ciesla2015,Ciesla2016,Salim2016}. We mainly follow \cite{Salim2016} to choose the parameter range of the model. The first decaying exponential function reflects the long-term star formation of the primary stellar component of the galaxy. It has an e-folding time ranging from $\tau_{\rm main} = 600$\,Myr to 18\,Gyr, and the age of the main stellar population is set to $t_{\rm main} = 12$\,Gyr. The second exponential function depicts the most recent burst of star formation, which can be achieved by setting $\tau_{\rm burst} = 20$\,Gyr. The age of the starburst varies from $t_{\rm burst} = 10$\,Myr to 5\,Gyr, and the fraction of the stellar mass formed in the starburst spans $f_{\rm burst} = 0$ to 0.15.

For the nebular emission component, we set the ionization parameter to $\log\,U=-3.4$ and the fraction of Lyman continuum photons absorbed by dust to 0.3, which have been found to match the observed emission-line equivalent widths of SDSS spectra (e.g., \citealt{Inoue2001}). The color excess of the nebular lines for both the young and old population ranges from 0.01 to 0.6, with reduction factors of 0.25 or 0.75 compared to the attenuation of the stellar continuum. As in \cite{Brown2014}, we use the starburst attenuation curve of \citet{Calzetti2000}, modified by a power-law term with exponent $\delta = -1.2$ to 0 to steepen it \citep{Salim2018}. The impact of the different attenuation laws will be discussed later. The amplitude of the UV bump \citep{Fitzpatrick1986} is fixed to the Milky Way value of 3.

The MIR and FIR emission of star-forming galaxies is based on the dust emission model of \cite{DL07}, as updated by \cite{Draine2014}. \cite{Draine2014} constrain the dust composition and size distribution based on Spitzer observations. The emission templates are calculated assuming that the dust is exposed to two different environments: (1) a diffuse radiation field described by a constant minimum radiation field intensity $U_\mathrm{min}$; and (2) a photodissociation region with the radiation field intensity ranging from $U_\mathrm{min}$ to $U_\mathrm{max}$. The probability distribution of the dust mass in the radiation field is $dU/dM\propto U^\alpha$, and the fraction of dust mass in the photodissociation region is $\gamma$. Following \cite{Draine2014}, we fix $U_\mathrm{max}=10^7$ and $\alpha=2$, which have a minimal effect on the SED fit \citep{DraineDale2007, Aniano2012}, and set $U_\mathrm{min} = 0.1-25$ and $\gamma = 0.001-0.2$.  The dust composition and the mass fraction of the dust in polycyclic aromatic hydrocarbons (PAHs) is given by the parameter $q_\mathrm{PAH}$. We set $q_\mathrm{PAH}=2.5\%$ because our broadband SED cannot constrain it (see also \citealt{Shangguan2018}).

We include the AGN module of \cite{Fritz2006} for the AGNs and composites in the sample. The module consists of a power-law continuum from the accretion disk in the UV/optical and thermal radiation from the dusty torus in the NIR and MIR. Many parameters of the model cannot be fully constrained by the fit. Similar to \citet{Ciesla2015}, we fix most of the parameters to commonly adopted values, varying only the amplitude because our main goal is to extract SFR, \Ms, and $M_d$ for the host galaxy. We fix the angle between the equatorial axis and the line-of-sight to $\psi=0^{\circ}$ for the type~2 AGNs and composites so that the torus is edge-on and there is no contribution to the UV/optical bands from the AGN; for the type~1 sources, we choose $\psi=60^{\circ}$, which is consistent with the distribution of torus inclination angles obtained for nearby, optically bright type~1 AGNs \citep{Zhuang2018}. We fix the radial and angular dust distribution with $\beta=-0.5$ and $\gamma=0$, as typically found by \citet{Fritz2006}. The optical depth at 9.7$\,\mu$m is set to $\tau_{9.7} = 1$, because it cannot be constrained by the broad SED. We only vary the amplitude of the dust torus model, such that the fractional contribution of the AGN emission to the total IR luminosity, $f_{\rm AGN} = L^{\rm AGN}_{\rm IR}/L^{\rm total}_{\rm IR} = 0-0.6$, where $L^{\rm AGN}_{\rm IR}$ and $L^{\rm total}_{\rm IR}$ are integrated from 1 to 1000\,\um.

\begin{figure*} 
\centering
\includegraphics[width=0.325\textwidth]{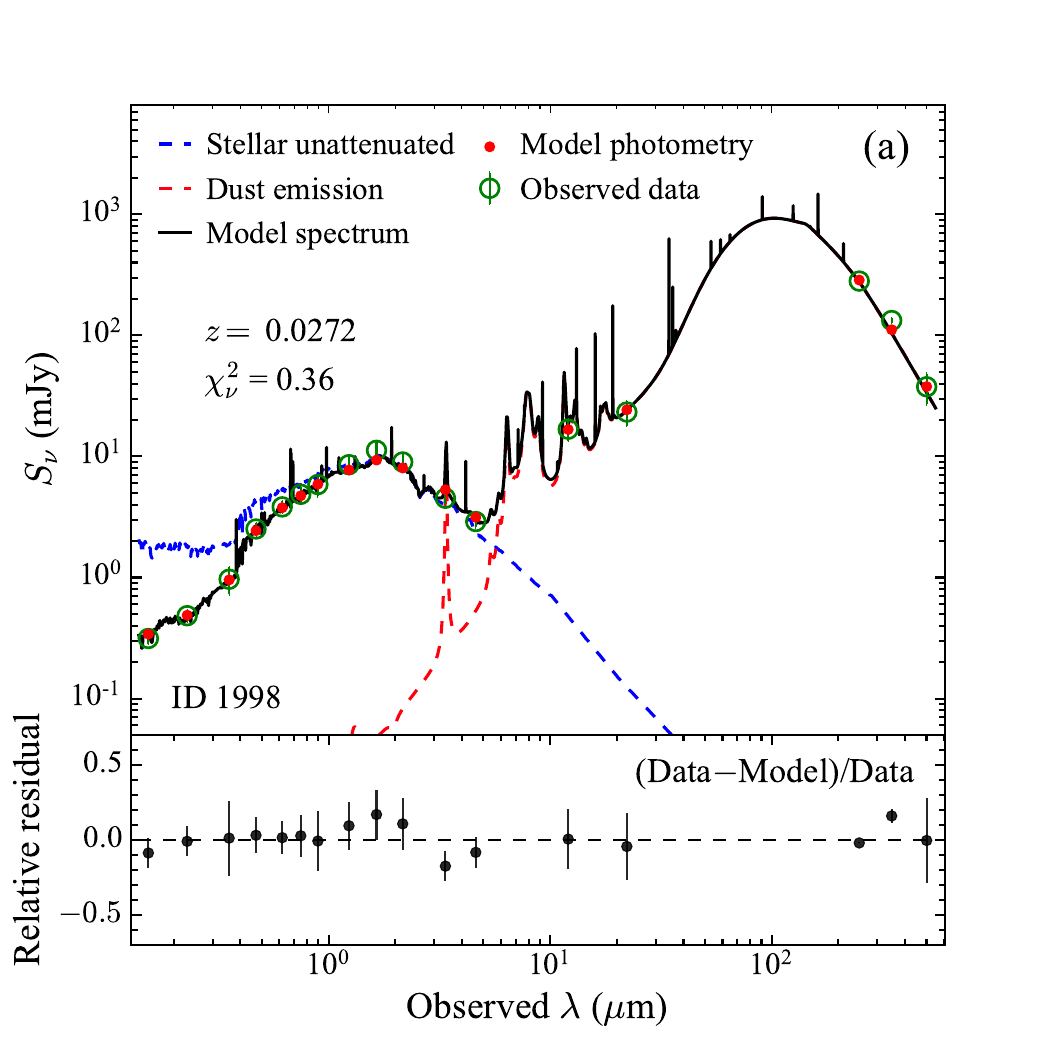}
\includegraphics[width=0.325\textwidth]{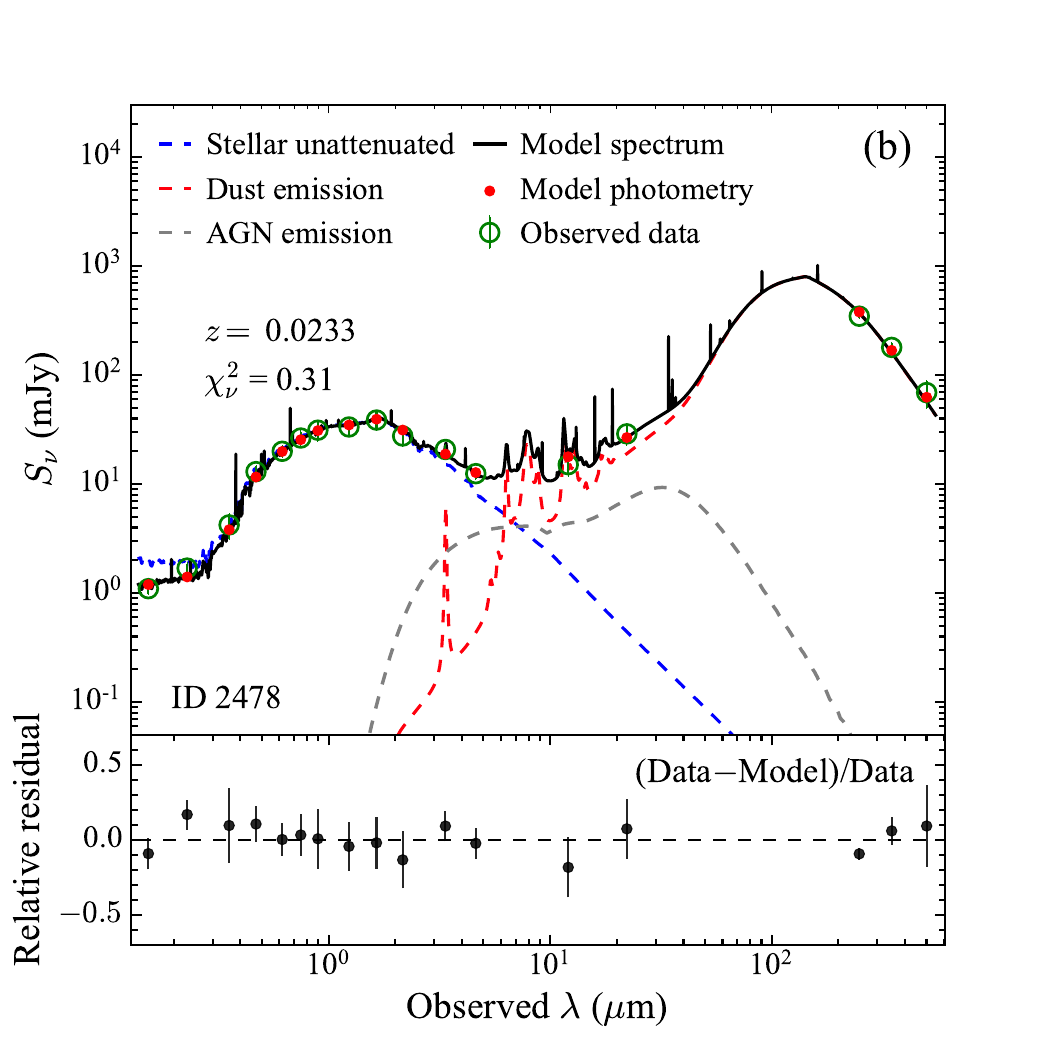}
\includegraphics[width=0.325\textwidth]{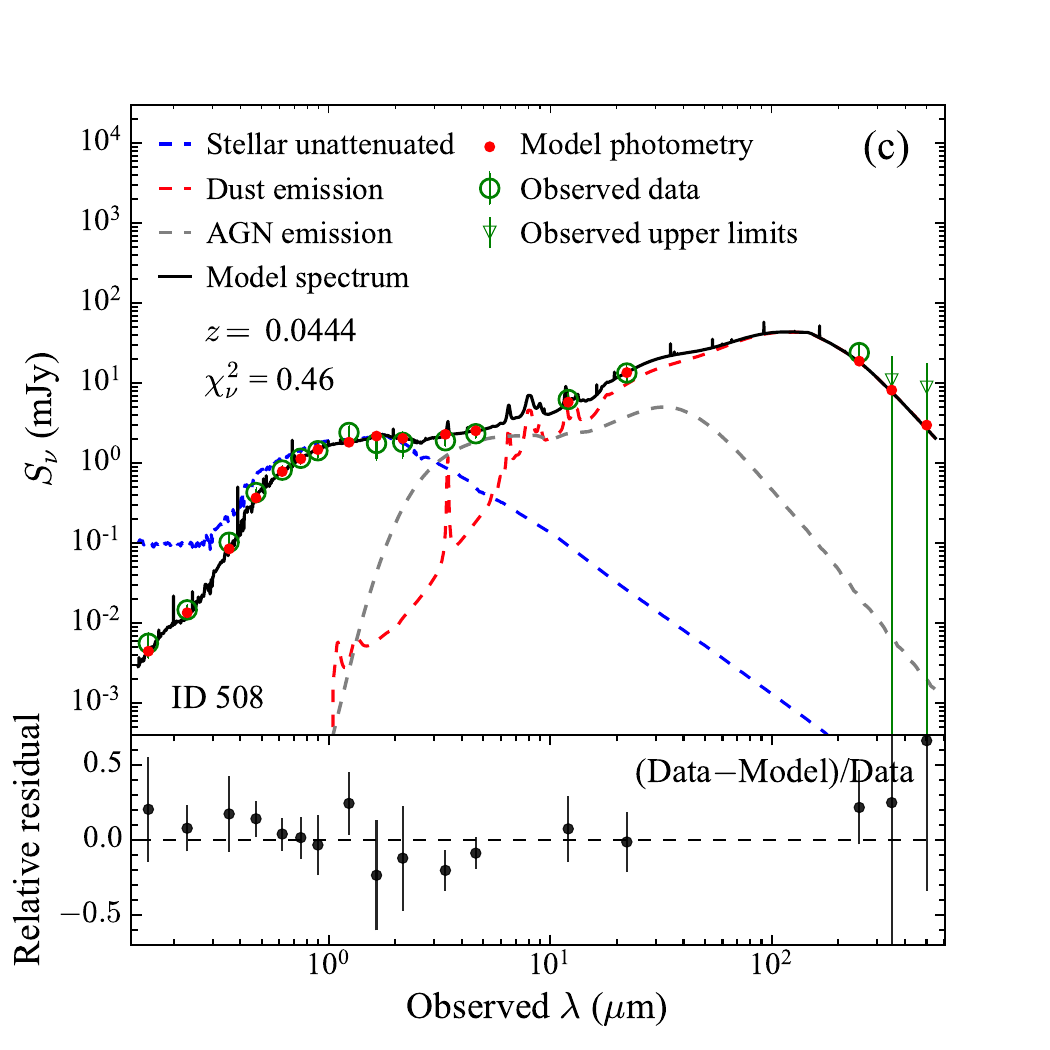}
\caption{Examples of SED fitting for (a) a star-forming galaxy, (b) a typical active galaxy with a moderate AGN fraction ($f_{\rm AGN} = 7.1\%$)}, and (c) an active galaxy with a larger AGN fraction ($f_{\rm AGN} = 26.1\%$). The observed data (open green circles) are compared with model photometry calculated from the model SED (black curve), which consists of an unattenuated stellar continuum (dashed blue curve), dust emission (dashed red curve), and AGN emission (dashed magenta curve). The nebular emission model is not shown for clarity. The $\chi^2_\nu$ of the fit is shown in the upper-left corner. The lower panels give the relative residuals between the data and the model.
\label{fig:SEDfit}
\end{figure*}

\begin{deluxetable*}{r  c  R  c  c c r}
    \tablecaption{Physical Properties of the Sample} 
    \tablewidth{12pc}
    \label{tab_phy}
    \tablehead{                                                                    
    \colhead{ID} &   
    \colhead{$\chi_{\nu}^2$} &                    
    \colhead{$\rm log\,SFR$}&
    \colhead{$\mathrm{log}\,M_*$}&
    \colhead{$\mathrm{log}\,M_d$}&
    \colhead{$f_{\rm AGN}$}&
    \colhead{$\mathrm{log}\,L_{\rm \OIII}$}
    \\
    \colhead{ } & 
    \colhead{ } & 
    \colhead{ ($M_{\odot}\,\mathrm{yr}^{-1}$) } & 
    \colhead{ ($M_{\odot}$) } &
    \colhead{ ($M_{\odot}$) } &
    \colhead{ (\%) } &
    \colhead{ ($\rm erg\,s^{-1}$) } \\
    \colhead{(1) } & 
    \colhead{(2) } & 
    \colhead{(3) } & 
    \colhead{(4) } & 
    \colhead{(5) } & 
    \colhead{(6) } & 
    \colhead{(7) }  
    }
    \startdata
    1 & 2.35 & -1.70$\pm$0.28 & 10.61$\pm$0.03 & 6.76$\pm$0.99 & 0 & 38.94$\pm$0.20\\
    2 & 3.72 & -0.92$\pm$0.19 & 10.84$\pm$0.03 & 6.53$\pm$0.47 & 0 & 38.97$\pm$0.34\\
    3 & 0.76 & -0.10$\pm$0.12 & 11.26$\pm$0.03 & 8.21$\pm$0.29 & 0 & 39.53$\pm$0.07\\
    4 & 0.44 & -2.23$\pm$0.78 & 10.37$\pm$0.03 & 6.41$\pm$0.60 & 0 & 38.67$\pm$0.33\\
    5 & 0.52 & -1.43$\pm$0.32 & 10.27$\pm$0.04 & 6.23$\pm$0.53 & 0 & 39.05$\pm$0.14\\
    6 & 0.69 & -1.90$\pm$0.55 & 10.28$\pm$0.03 & 6.75$\pm$0.64 & 0 & 39.15$\pm$0.11\\
    7 & 0.45 & -1.76$\pm$0.54 & 10.54$\pm$0.04 & 6.42$\pm$0.53 & 0 & 38.76$\pm$0.26\\
    8 & 0.68 & -1.10$\pm$0.22 & 10.63$\pm$0.05 & 6.59$\pm$0.54 & 0 & 39.35$\pm$0.17\\
    9 & 0.74 & -1.42$\pm$0.39 & 10.37$\pm$0.05 & 6.74$\pm$1.17 & 0 & 39.03$\pm$0.20\\
    10 & 0.39 & -1.59$\pm$0.62 & 10.59$\pm$0.02 & 7.56$\pm$0.37 & 0 & 38.56$\pm$1.00\\
    \enddata
\tablecomments{
Col. (1): Index number.
Col. (2): Reduced $\chi^2$ of SED fit.
Col. (3): SFR.
Col. (4): Stellar mass.
Col. (5): Dust mass.
Col. (6): Fractional contribution of the AGN emission to the total IR luminosity; $f_{\rm AGN}=0$ if the galaxy is inactive (see Section~\ref{sec_SEDfit} for details).
Col. (7): Luminosity of \OIII\ $\lambda 5007$, not corrected for extinction.
(The full machine-readable table can be found in the online version.)
}
\end{deluxetable*}

\subsection{Results of SED Fitting}
\label{ssec:res}

The targets can be well fitted by {\tt CIGALE}, achieving $\chi_{\nu}^2 < 2$ for $\gtrsim 97\%$ of the cases (Table~5). This fraction rises to $\sim 99\%$ if the blended merger pairs are excluded, owing to the relatively smaller flux uncertainties derived by \texttt{GALFITM} modeling. The derived physical properties, as well as the $\chi_{\nu}^2$ for the sample, are presented in Table~\ref{tab_phy}. Figure~\ref{fig:SEDfit} shows SED fits for a star-forming galaxy, a typical active galaxy with a low level of AGN activity ($f_{\rm AGN}=7.1\%$), and an exceptionally strong active galaxy ($f_{\rm AGN}=26.1\%$). 

\begin{figure*} 
\centering
\includegraphics[width=\textwidth]{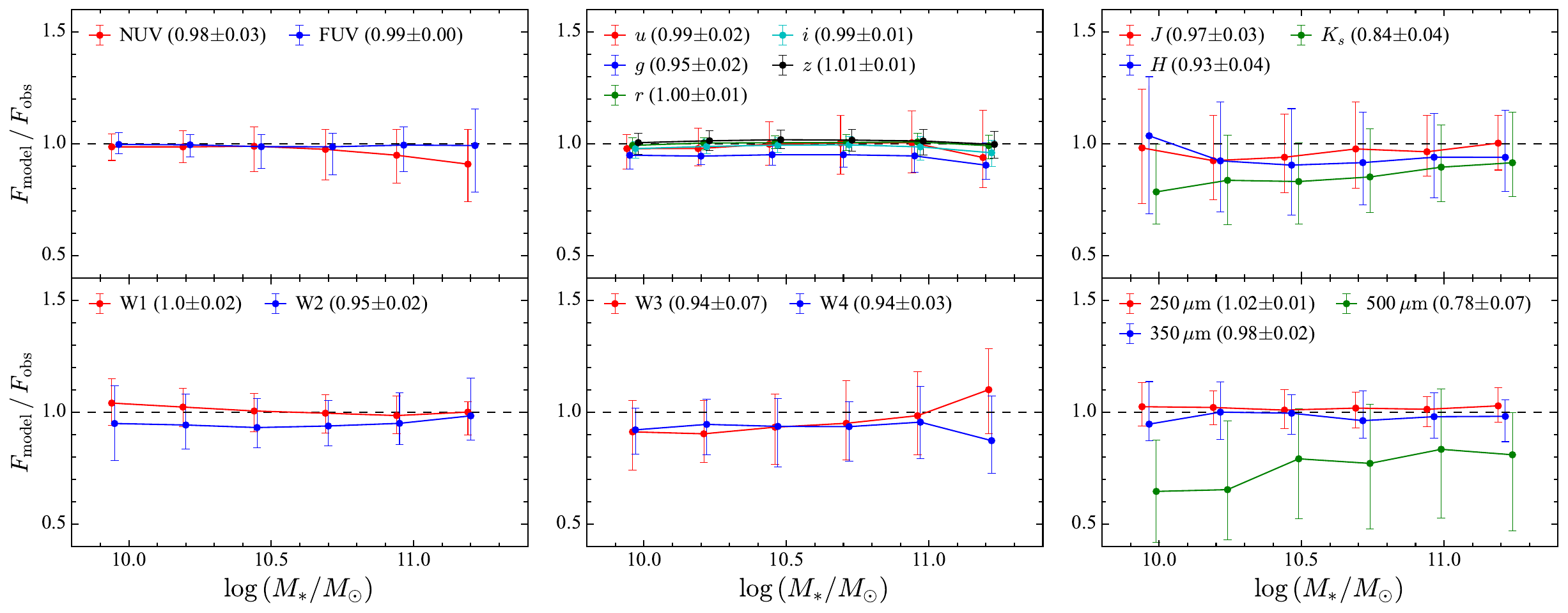}
\caption{Ratio of flux densities between the model and the data, as a function of \Ms. The median and standard deviation in each mass bin are indicated by circles and error bars. Different bands are shown in different colors.}
\label{fig_fitfluxratio}
\end{figure*}

Figure~\ref{fig_fitfluxratio} quantifies the deviation between the data and the best-fitting SED model for each of the bands, for clarity focusing only on sources detected with $\rm S/N>2$ in each band. The data and model agree well for most of the bands across the full range of stellar masses ($\Ms \approx 10^{10}-10^{11.2}\,M_\odot$). The scatter is relatively larger in $H$, $K_s$, and W4 because they have larger measurement uncertainties. Only the 500\,\um\ band shows substantial systematic deviation as a function of \Ms. This is likely related to the ``500\,\um\ excess'' \citep{Galliano2003, Boselli2012, Smith2012, Ciesla2014, Nersesian2019}. The \cite{Draine2014} model used in our fitting corresponds to an effective graybody emissivity index of $\beta \approx 2$, but a flatter index of $\beta=1.5$ or even less is necessary to reproduce flatter slopes of the FIR SED, especially in low-metallicity galaxies \citep{Boselli2012, Smith2012,Smith2019,Lamperti2019}. This may explain the tendency for the lower mass galaxies in our sample to exhibit systematically stronger 500\,\um\ emission.

\subsection{AGN Contribution}\label{AGNcontribution}

Our sample has a median $f_{\rm AGN} = 10\%$ for the type~1 AGNs, $f_{\rm AGN} = 4\%$ for the type~2 AGNs, and $f_{\rm AGN} = 2\%$ for the composites. To avoid the uncertainty of applying the AGN model to objects with very low AGN-contribution \citep{Ciesla2015}, we ultimately exclude the AGN component from the fits when $f_{\rm AGN} < 10\%$. The AGN component in such weakly active systems can be overestimated by up to a factor of 2 from degeneracy with low dust emission due to moderate star formation.
For the overall sample, galaxies with $f_{\rm AGN} \geq 10\%$ comprise 59\% of the type~1 AGNs, 22\% of the type~2 AGNs, and 10\% of the composites; the respective median values of $f_{\rm AGN}$ for these classes are 15\%, 13\%, and 13\%. 

The main goal of our SED analysis is to derive SFR, \Ms, and \Md\ for the sample galaxies. Emissions from the AGN dusty torus may contaminate the NIR and MIR portions of the SED. To evaluate the magnitude of this effect, we compare the SED fitting results of the AGN and composite targets, with and without the AGN torus module included (see Section~\ref{sec_SEDfit} for details). Among the galaxies with $f_{\rm AGN} \geq 10\%$, the mean difference of their SFR, $M_*$, and $M_d$ before and after adding the AGN module are 0.09, 0.01, and $-0.01$ dex, respectively, with a maximum deviation of 0.5, 0.13, and $-0.6$ dex. In contrast, for the galaxies with $f_{\rm AGN}<10\%$, we find little difference in the resulting SFRs ($< 0.03$ dex), $M_*$ ($<0.01$ dex), or $M_d$ ($<0.02$ dex) compared to the uncertainties. This is not surprising given the weakness of the AGNs in our sample. To illustrate this point, we estimate the bolometric luminosity of the AGNs and composites from the observed \OIII\ $\lambda 5007$ luminosity, using the line measurements from the MPA-JHU catalog for the main sample and from the catalog of \citet{Liu2019} for the broad-line AGNs. We adopt a bolometric correction of 3500 as recommended by \citet{Heckman2004}. Other \OIII-based bolometric corrections can be contemplated (see discussion in \citealt{Kong2018}), but they do not alter our basic conclusion: the type~2 AGNs and composite sources in our sample have $\lbol \approx 10^{40}-10^{44}\,\rm erg~s^{-1}$, which is much lower than the threshold for quasars ($\lbol \geqslant 10^{45}\,\rm erg~s^{-1}$; \citealt{Reyes2008}). Only the handful of supplementary type~1 AGNs have $\lbol \approx 10^{44}-10^{45}\,\rm erg~s^{-1}$.

Last, but not the least, the optical emission-line diagnostics may miss heavily obscured AGNs, which, nevertheless, may be identifiable by their MIR colors (e.g., ${\rm W1-W2}>0.8$~mag; \citealt{Stern2012}). This color criterion is met 23 galaxies in our sample, although their optical emission-line ratios classify them as inactive. We fit the SEDs of these 23 sources with and without the AGN module of CIGALE. Most show negligible differences. 

We conclude that black hole accretion is energetically negligible in galaxies with $f_{\rm AGN} < 10\%$ and that it has essentially no impact on the physical parameters of primary interest to this study (SFR, $M_*$, and $M_d$) derived from SED fitting.

\subsection{Reliability of SED Fitting with Nondetections}
\label{SEDuplim}

To investigate the effects of nondetections (upper limits) in different bands on our results, we fit mock SEDs generated from the SEDs of real targets and study the deviation of the derived physical parameters from the true input values, as the number of upper limits increases. We select as the test sample galaxies detected in all the bands and take their best-fit parameters as the true input values. We randomly scale down the amplitude of the observed SED so that their $r$-band flux densities fall between 0.15 to 2\,mJy, while preserving the observed uncertainty of each band. The flux density becomes an upper limit if it is below 2 times the uncertainty. This scaling guarantees that the mock SED does not have upper limits in $u$, $g$, $r$, $i$, $z$, and W1, as is the case in most of our observed data. In total, we randomly generate $\sim 2000$ mock SEDs with nondetections in the FUV, NUV, $J$, $H$, $K_s$, W2, W3, W4, 250\,\um, 350\,\um, and 500\,\um\ bands, closely mimicking the actual observations. We require that there be at least one detection in the other wavelengths to isolate the effect of nondetection on the bands of interest, since the flux densities of an SED scale together.
We also create 100 mock SEDs with all bands detected after rescaling the flux, to serve as a control sample. We then use {\tt CIGALE} to fit the mock SEDs with the same model discussed in Section~\ref{ssec:model}. The input parameters are very well recovered if there are no nondetections in the mock SEDs in the control sample.

\begin{figure*} 
\centering
\includegraphics[width=0.85\textwidth]{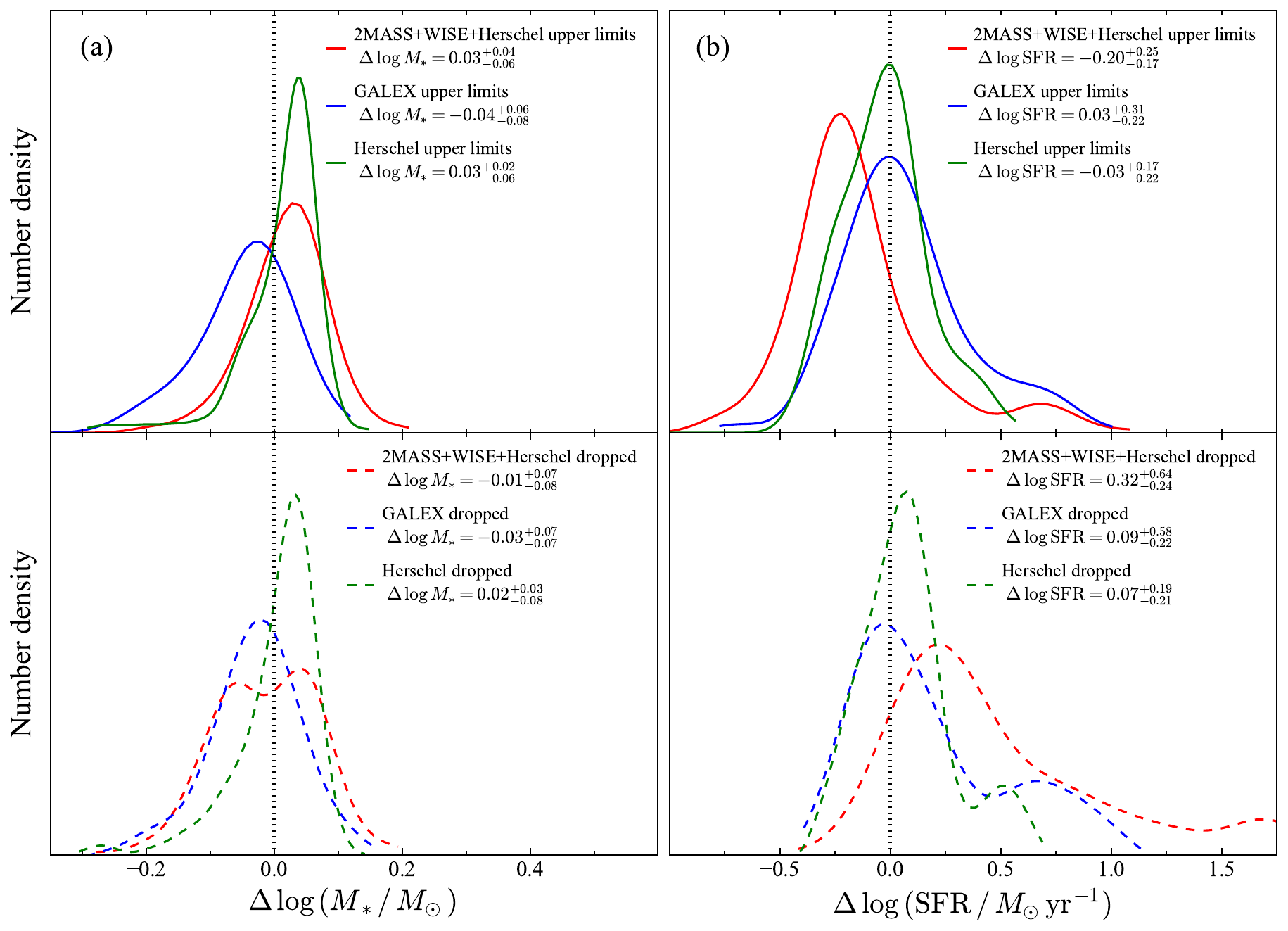}
\caption{The impact on estimates of (a) \Ms\ and (b) SFR from fitting the SED by (top) using upper limits and (bottom) completely dropping them in 2MASS+WISE+Herschel/SPIRE (red), GALEX (blue), and Herschel/SPIRE (green). The X-axis shows output $-$ input; the black dotted line denotes zero deviation. The legend gives the median and standard deviation of each distribution.}
\label{fig:SEDfit_dropband}
\end{figure*}

\begin{figure*}
\centering
\includegraphics[width=0.95\textwidth]{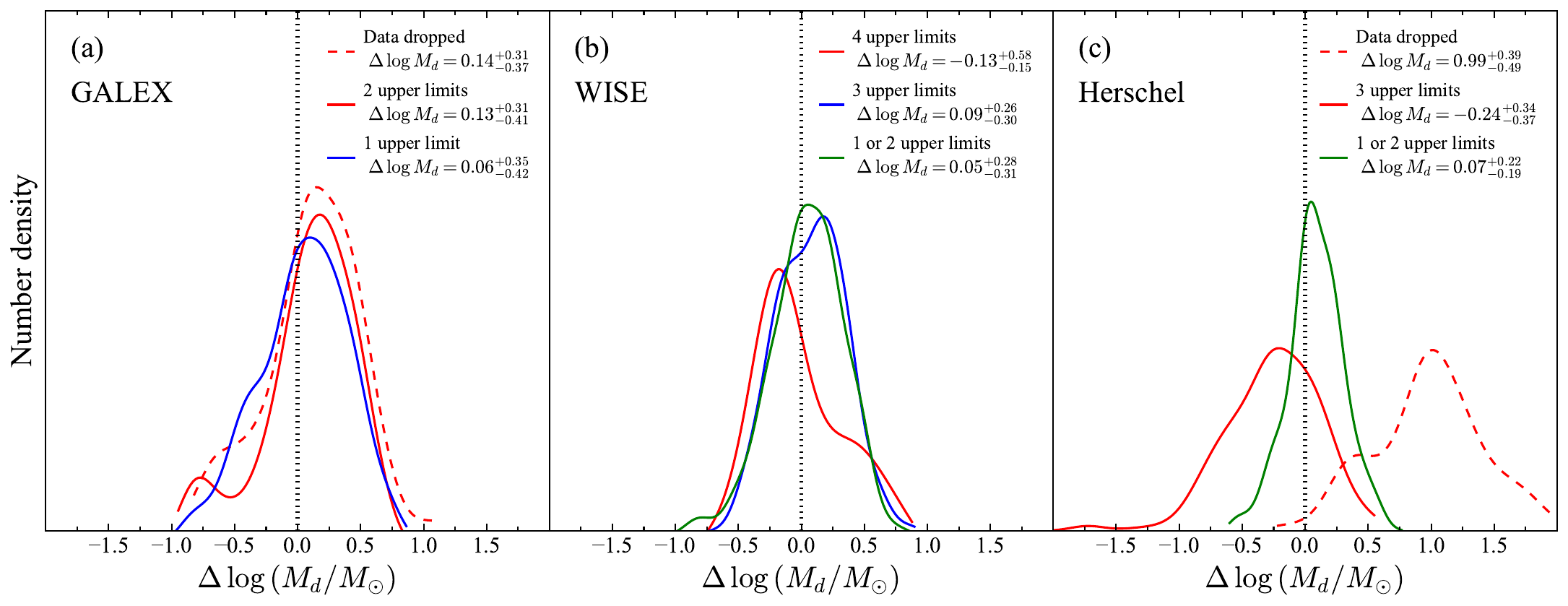}
\caption{The impact on estimates of the dust mass ($M_d$) from fitting the SED with a different number of upper limits in (solid lines) or by completely dropping the data (dashed lines), for (a) GALEX, (b) WISE, and (c) Herschel. The X-axis shows output $-$ input; the black dotted line shows zero deviation. The legend gives the median and standard deviation of each distribution. The effect on dust mass is moderate for GALEX and WISE but substantial for Herschel. Combining WISE detections in the W3 and W4 bands, together with upper limits in Herschel/SPIRE, can provide useful dust masses after correcting for systematic bias.}
\label{Nuplim_band}
\end{figure*}

Figures~\ref{fig:SEDfit_dropband} and \ref{Nuplim_band} summarize the deviation (output $-$ input) of $M_*$, SFR, and $M_d$ from our tests. These physical parameters are not significantly affected by the 2MASS bands, and for the sake of clarity we do not include their results in the plots. The stellar mass is very well recovered ($\Delta\,\mathrm{log\,}M_*\approx 0.03\,$dex) even when the GALEX, Herschel, or even WISE bands are upper limits (Figure~\ref{fig:SEDfit_dropband}a). This is because all galaxies are detected in the five SDSS bands, which play the key role in constraining the stellar mass from the SED fitting. The SFR can be estimated fairly accurately using upper limits in the GALEX or Herschel bands individually ($\Delta\,\mathrm{log\, SFR}\approx 0.03\,$dex), which is small compared to the typical uncertainty of $\sim 0.18$\,dex. SFR is moderately underestimated (0.2~dex) if all of the MIR and FIR data are upper limits. These mock galaxies typically have relatively low input SFRs ($\lesssim 0.3\,M_{\odot}\,\rm{yr^{-1}}$), and the uncertainties of their output SFRs are large ($\sim 0.6$ dex), such that CIGALE derives SFRs consistent with the inputs, although the overall distribution is slightly underestimated. This situation affects $\sim 400$ objects in the actual sample, and for these objects we apply a systematic correction to their SFRs according to this test.

In order to understand the impact of the upper limits on \Ms\ and SFR, we use the mock SEDs to simulate the consequences of dropping the upper limits altogether, successively testing the effect of each band. Dropping the upper limits in the UV or IR bands makes little difference to \Ms\ (Figure~\ref{fig:SEDfit_dropband}a, bottom panel). The situation, however, differs for the SFRs, which can be significantly overestimated when we neglect the upper limits (Figure~\ref{fig:SEDfit_dropband}b, bottom panel). All else being equal, the SFRs are better constrained by including upper limits than not. Even though the median bias is moderate, $\sim 0.1$~dex if GALEX or Herschel is omitted and $\sim 0.3$~dex even if all IR (2MASS, WISE, and Herschel) bands are excluded, neglecting upper limits leaves a long tail of positive residuals: $\sim 0.7$~dex for GALEX, $\gtrsim 1$ dex for Herschel, and $\gtrsim 1.5$ dex for all IR bands. Our results agree with those of \citet{Noll2009} and \citet{Ciesla2015}, who concluded that SFRs in star-forming and AGN host galaxies can be biased too high by more than $0.8$~dex when the IR filters are not used, on account of overestimation of the FUV and visual attenuation factors. Many works (e.g., \citealt{Salim2007,Hoversten2011,Jeong2012,Brown2014,Salim2016}) derive SFRs from panchromatic (UV to FIR) SEDs, but undetected bands often are not included because upper limits are usually not available. For instance, the SED fitting of \cite{Salim2016} only utilizes data from GALEX and SDSS data when W3 and W4 are not detected. The bottom panel of Figure~\ref{fig:SEDfit_dropband}b, which mimics the assumptions of \cite{Salim2016}, illustrates the degree to which SFRs can be overestimated, and that the biases can be mitigated by incorporating upper limits into the analysis.

Upper limits in GALEX (Figure~\ref{Nuplim_band}a) or WISE (Figure~\ref{Nuplim_band}b) have little influence on the accuracy of the dust masses from SED fitting, compared to the typical 0.42~dex uncertainty. The dust mass can be reasonably well constrained if there is a detection in at least one Herschel/SPIRE band (Figure~\ref{Nuplim_band}c). However, if all three SPIRE bands are upper limits, the resulting dust masses are $\sim 0.24$~dex lower than the input values. In contrast, removing all the SPIRE data significantly overpredicts $M_d$ by $\sim 1$~dex or more (red dashed line in Figure~\ref{Nuplim_band}c). This is not unexpected. The FIR data are needed to constrain the peak of the cold dust SED \citep{DraineDale2007}, which not only helps to derive the correct FIR luminosity but also to prevent unphysically large attenuation through the energy-balance constraint. For galaxies that are not detected in any of the Herschel bands, using our mock tests as a guide, we apply a systematic correction of 0.24~dex to their dust mass, although we stress that the correction is small compared to the statistical uncertainty of $M_d$ for individual galaxies (0.42~dex).

We note that our use of galaxies whose SED has been detected in all bands to mimic galaxies with nondetections in some bands may not accurately represent all galaxy types. For example, quiescent galaxies may have systematically different SEDs compared to the mock sample.  However, the main purpose of this series of mock tests is to demonstrate the importance of including upper limits in SED fitting instead of simply disregarding them. This point remains valid regardless of possible sample selection effects. While the empirical corrections derived from our mock tests may not be accurate if the SED of the measured galaxy differs significantly from that of our SED-detected sample, this is likely a higher-order effect, which is beyond the scope of this work. We advise caution when using dust mass and SFR corrections for individual quiescent galaxies.

\subsection{Parameter Uncertainties}
\label{ssec:unc}

As mentioned in Section~\ref{ssec:model}, we adopt Bayesian estimates of $M_*$, SFR, and $M_{\rm d}$ based on the likelihood of each SED model. This method takes into account the degeneracy of parameters among different models if they provide indistinguishable likelihoods. The uncertainty of the likelihood-weighted physical properties is calculated with consideration of the upper limit in the data \citep{Boquien2019CIGALE}. Our choice of parameter ranges covers typical values of nearby galaxies whose SEDs have been well fit \citep[e.g.,][]{Brown2014, Ciesla2015}. Moreover, we test the systematics and scatter introduced by upper limits; manually converting detections to upper limits (Section~\ref{SEDuplim}) demonstrates that our physical parameters of interest are minimally affected by upper limits. Our optical photometry from SDSS delivers robust \Ms\ for all of the targets, with typical uncertainties of 0.05~dex. The simulated values of SFR and dust mass are consistent within $2\,\sigma$ with 92\% and 99\% of the true values, respectively. We show that including upper limits in SED fitting reduces the bias of SFR and dust mass measurements. Although the derived quantities are usually consistent with the input values, considering the uncertainties, the sample-averaged SFRs and dust masses can be underestimated at the level of $\sim 0.2$~dex when the target is not detected in the WISE and Herschel bands (Section~\ref{SEDuplim}). We correct for this source of systematic bias for the faint sources. These galaxies usually have SFR and/or dust mass uncertainties $>30\%$ of the best-fit values from the SED fitting. We note that the sample distribution of SFR and $M_d$ may be biased in the region of low values.

We stress, however, that our tests do not account for uncertainties stemming from prior model assumptions. For example, at fixed stellar initial mass function, the stellar population model and star formation history can introduce 0.3~dex uncertainty to the stellar mass and SFR \citep{Conroy2013}. It is difficult to push the sensitivity of sSFR derived from integrated SEDs down to values $\la10^{-12}$ yr$^{-1}$ because of uncertainties associated with the contribution of UV emission from evolved stars and dust heating \citep{O'Connell1999, Conroy2013}. Nearly 30\% of the galaxies in our sample have sSFR close to or below $10^{-12}$ yr$^{-1}$. One should regard their SFRs with some caution.

%\vspace{0.5cm}
\section{Comparison with Previous Work}
\label{sec_comparison}

This section compares the stellar masses and SFRs derived from our SED fitting with those from two widely used catalogs, the MPA-JHU database \citep{Kauffmann2003, Brinchmann2004, Salim2007} and GSWLC-2 \citep{Salim2018}. The MPA-JHU catalog derives \Ms\ from SED fitting of optical photometry, using model magnitudes (\texttt{ModelMag}) of the five bands of SDSS \citep{Salim2007}. The total SFR of the galaxy comes from a combination of the fiber-based SFR of the central region ($\sim 1-6$\,kpc for the 3\arcsec-diameter fiber in the redshift range $z<0.11$ of our sample) of the galaxy \citep{Brinchmann2004} and the SED-based SFR for the galaxy outskirts \citep{Salim2007}. SFR inside the fiber is estimated from the extinction-corrected \Ha\ emission if the galaxy is purely star-forming according to the \cite{BPT1981} \OIII\,$\lambda 5007$/\Hb\ versus \NII\,$\lambda 6584$/\Ha\ diagram. If, on the other hand, the target shows nuclear activity or has weak \Ha\ emission, the SFR is estimated from the strength of the 4000\,\AA\ break inside the fiber, and from SED fitting outside. \cite{Salim2018} collect broadband (FUV to MIR) photometric data from preexisting catalog measurements provided by GALEX, SDSS, and WISE to derive \Ms\ and SFR from SED fitting. With the aid of a spectral template from \citet{CharyElbaz2001}, they convert the MIR flux densities from W3 or W4 to total IR ($8-1000$\,\um) luminosity, which is used in the SED fitting with a customized version of {\tt CIGALE}. A correction for AGN contribution to the IR luminosity is applied based on the strength of the \OIII\,$\lambda 5007$ line \citep{Salim2016}. 

We derive the dust mass of low-redshift galaxies in Stripe~82 for all the sources using panchromatic (UV to FIR) SED fitting, primarily relying on data from the HerS survey \citep{VieroHerS2014}. Our tests in Section~\ref{SEDuplim} suggest that the upper limits in undetected bands usually can give useful constraints on the dust mass. 
%We, therefore, provide dust masses for all but 41 galaxies, whose dust emission model is barely constrained because they were undetected from W3 to 500~$\mu$m. 
\citet{Bertemes2018} also estimated dust masses for 78 massive galaxies in the Stripe~82 region with CO(1--0) measurements, all of which had $3\sigma$ detections in WISE W4 and Herschel 250 and 350 $\mu$m. Our dust mass measurements agree with those of \citet{Bertemes2018} to $0.05\pm0.19$~dex.

\begin{figure*}
\centering
\includegraphics[width=0.99\textwidth]{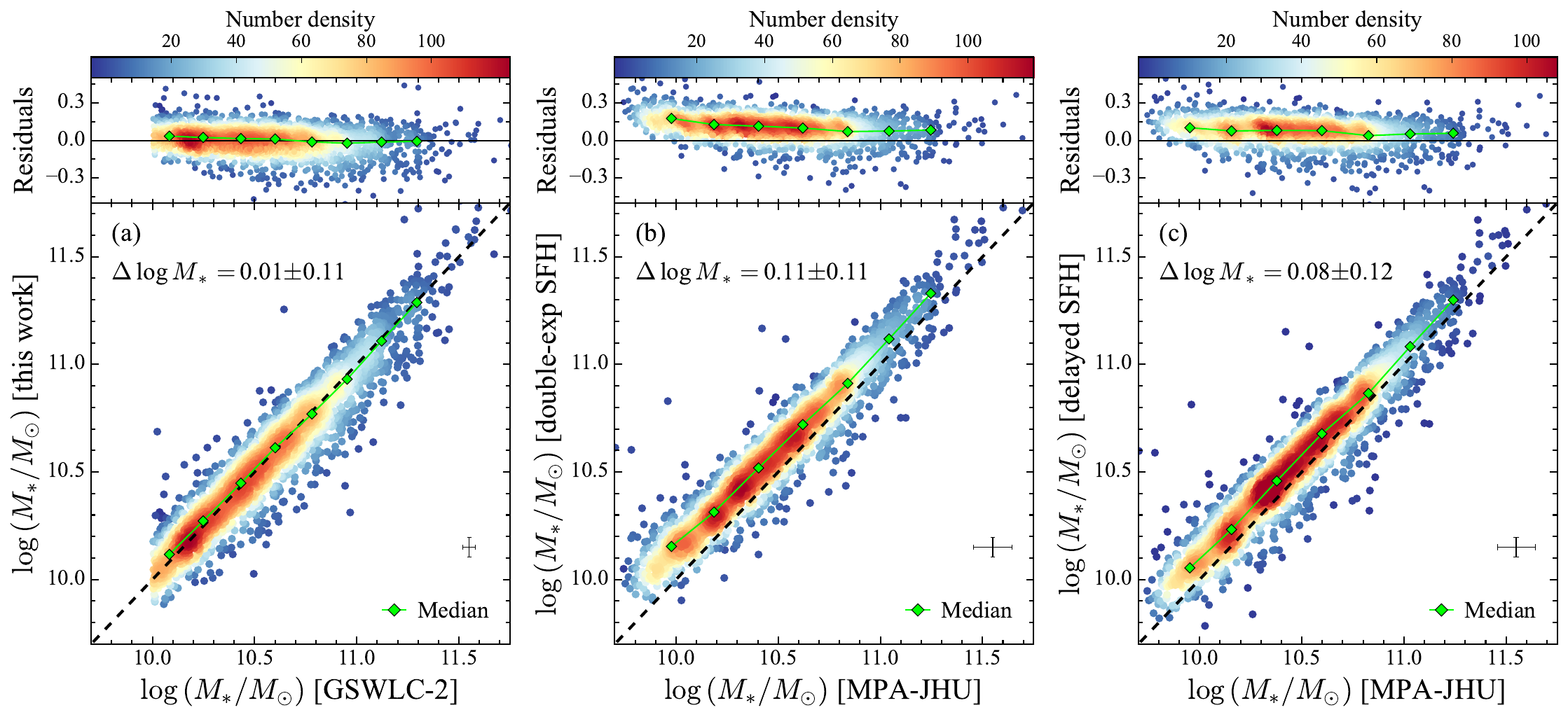}
\caption{Comparison of the stellar mass derived in this work with those from (a) the GSWLC-2 catalog, (b) the MPA-JHU catalog, and (c) the MPA-JHU catalog, for which our comparison stellar masses were calculated assuming the same star formation history (delayed exponential) as that used in the MPA-JHU catalog. The typical uncertainties are given in the lower-right corner. The residuals (this work $-$ other catalogs) are shown in the upper panels; the medians and standard deviations of the residuals are given in the upper left corners in the bottom panels. The color code indicates the number density of objects.}
\label{compareMs_S18_MPA}
\end{figure*}

\subsection{Comparison of Stellar Masses}
\label{ssec:ms}

The stellar masses from our work agree well with those of GSWLC-2 with negligible zero point offset and scatter (median and standard deviation $\Delta \log M_* = 0.01\pm0.11$\,dex; Figure~\ref{compareMs_S18_MPA}a). The $0.1$\,dex scatter is consistent with the measurement dispersion of \Ms\ induced by nondetecions in UV or other bands (Section~\ref{SEDuplim}).  In comparison with the values in the MPA-JHU catalog, our stellar masses are on average slightly larger ($\Delta \log M_* = 0.11\pm0.11$\,dex; Figure~\ref{compareMs_S18_MPA}b), but we note that a systematic trend is visible in the residuals, such that toward the low-mass end our masses tend to be higher. This discrepancy likely arises from the difference in the star formation history used \citep{Salim2016}. Adopting the delayed exponential star formation history assumed in the MPA-JHU catalog slightly reduces the systematic differences between their stellar masses and ours ($\Delta \log M_* = 0.08\pm0.12$\,dex; Figure~\ref{compareMs_S18_MPA}c), but cannot totally remove the trend. The residual offset of 0.08~dex owes to to the fact that our optical fluxes are slightly brighter (e.g., 0.08~mag in $u$ and 0.05~mag in $g$) than the \texttt{modelmag} values used in MPA-JHU catalog (Appendix~A; Figure~\ref{appdix_fluxcompare}2). This effect influences less massive galaxies more significantly because they have a higher fraction of young stars \citep{Sextl2023}, resulting in a larger stellar mass offset toward lower mass. The inclusion of UV photometry, as implemented in GSWLC-2, significantly improves the stellar mass measurements, which otherwise may be biased by only fitting the SDSS photometry, particularly for galaxies with a young stellar population.

The heavily blended, merging galaxies in our sample deserve closer scrutiny. For these sources, which comprise 7\% of the sample, our stellar masses show minor systematic differences compared to the values given in GSWLC-2 ($\Delta \log M_* = 0.03\pm0.19$\,dex), and MPA-JHU ($\Delta \log M_* = 0.11\pm0.23$\,dex). This is not surprising because the stellar mass estimates are mainly driven by the optical photometry based on SDSS images (Section~\ref{SEDuplim}), and compared to the uncertainties, there are no significant statistical differences between the photometry derived from our 2D deblending (Section~\ref{sec_galfitm}) and the SDSS model magnitudes used in other catalogs.

\subsection{Comparison of Star Formation Rates}
\label{ssec:sfr}
\begin{figure*}  [t]
\centering
\includegraphics[width=0.82\textwidth]{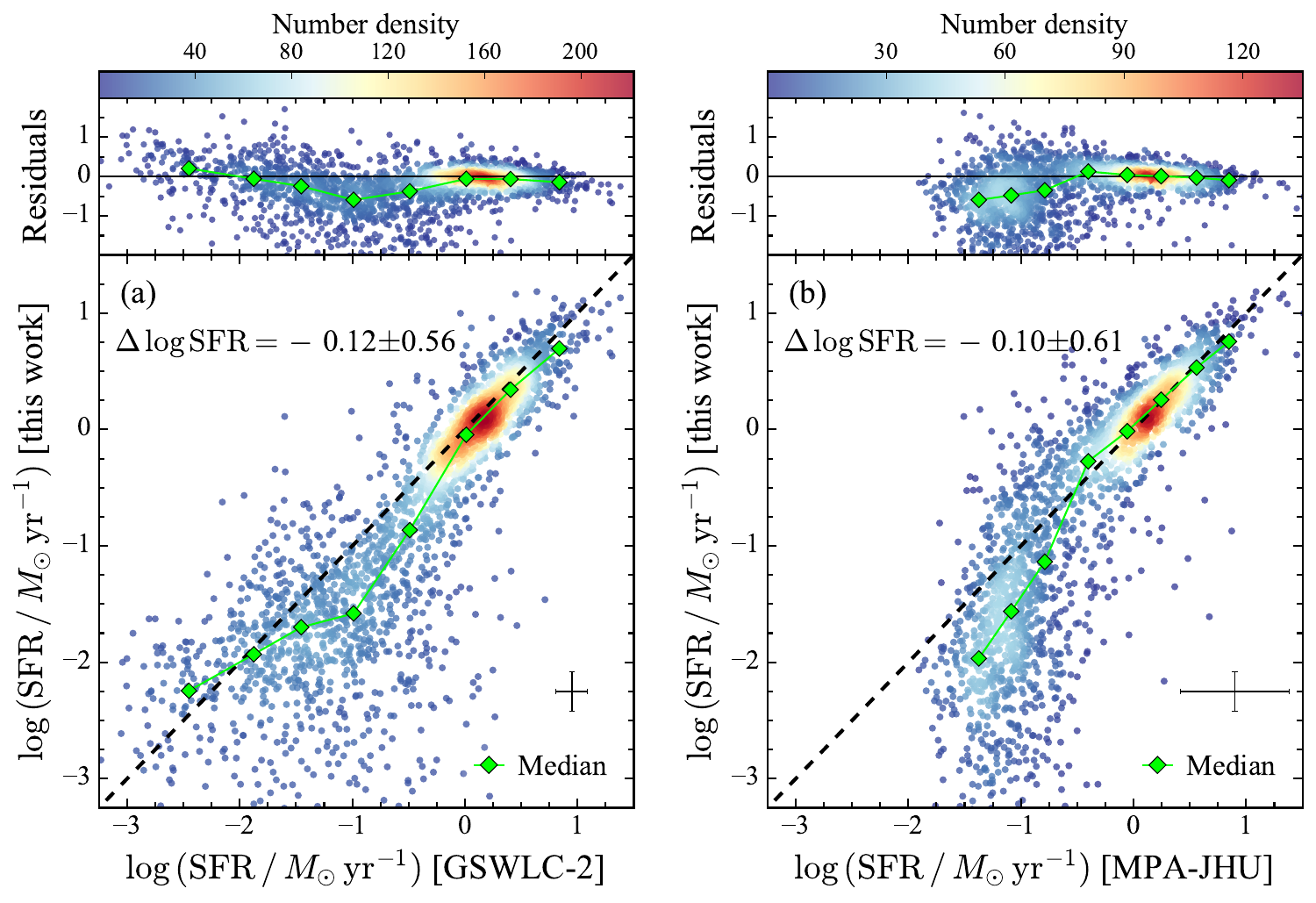}
\caption{Comparison of SFR derived in this work with those from the (a) GSWLC-2 and (b) MPA-JHU catalog. The typical uncertainties are plotted in the lower-right corner. The residuals (this work $-$ other catalogs) are shown in the upper panels; the medians and standard deviations of the residuals are given in the upper left corners in the bottom panels. The color code indicates the number density of objects. }
\label{SFRcompare}
\end{figure*}

For the sample as a whole, our SFRs show a relatively large scatter and modest zero point offset with respect to GSWLC-2 ($\Delta \log \rm SFR = -0.12\pm0.56$\,dex; Figure~\ref{SFRcompare}a). Closer inspection reveals that most of the discrepancy is confined to the 45\% of objects that have low star formation activity: while $\Delta \log \rm SFR = -0.05\pm0.23$\,dex at ${\rm SFR} > 0.3\,M_\odot\,{\rm yr}^{-1}$, $\Delta \log \rm SFR = -0.40\pm0.71$\,dex at ${\rm SFR} \lesssim 0.3\,M_\odot\,{\rm yr}^{-1}$. Now, distinct from our work, which strives to provide full spectral coverage from FUV to FIR for the SED fits, GSWLC-2 did not incorporate the $J$, $H$, $K_s$, W1, and W2 bands into their analysis. To evaluate the extent to which this difference in spectral coverage contributes to the discrepancies between the two studies, we repeat our SED fits without these NIR bands. The resulting SFRs are not strongly affected, for the difference between the results for the full and partial SEDs is only $\Delta \log \rm SFR = 0.002\pm0.06$\,dex. We suspect that the systematic deviation between our results and those of GSWLC-2 stems from the manner in which the two studies treat the SED longward of $\sim 20$\,\um. On the one hand, we rigorously include in the SED fit all (14) bands from the FUV to 500\,\um, using new, customized, more accurate photometric measurements and their associated uncertainties. Upper limits are rigorously included in our sample. \cite{Salim2018} did not have direct access to FIR data and instead extrapolated the total IR emission from MIR flux densities. On the other hand, the unWISE photometry they used for the W4 or W3 (if W4 is unavailable) band is $\sim 0.3$~mag brighter than the profile-fitting photometry we use for the unresolved galaxies that dominate the low-SFR population. For the resolved objects, the unWISE W4 and W3 photometry are $\sim 0.2$~mag brighter than our aperture photometry, as explained in Appendix~A (Figure~\ref{appdix_fluxcompare}5). GSWLC-2 predicted the total IR luminosity from unWISE W3 photometry for $> 40\%$ and W4 photometry for $>20\%$ of the low-SFR galaxies, which accounts, at least in part, for the systematic difference in SFR. Further comparisons are not straightforward because GSWLC-2 amalgamated photometry is derived using multiple methods. For instance, the unWISE photometry was measured by model fitting \citep{Lang2016Tractor}, which differs from the method used for treating the UV bands. Among the population of galaxies with ${\rm SFR} \lesssim 0.3\,M_\odot\,{\rm yr}^{-1}$ for which our two studies show the most glaring inconsistency, $\sim 36\%$ of the sources are not detected in W3 or W4 by unWISE. Without the MIR data, the SFRs of GSWLC-2 default to the UV/optical SED fitting from \citet{Salim2016}, which likely overestimated the SFR. To demonstrate this point, in Section~\ref{SEDuplim}, we fit the mock SEDs with nondetections in WISE, while intentionally dropping all of the data at wavelengths longer than the SDSS $z$ band. Under these circumstances, which mimic the procedure of \citet{Salim2016}, we confirm that the median value of SFR indeed gets systematically overestimated, even up to $\sim 1.5$\,dex. Therefore, we caution that the SFRs from GSWLC-2 may be significantly overestimated in the low-SFR regime (${\rm SFR} \lesssim 0.3\,M_\odot\,{\rm yr}^{-1}$). Section~\ref{SEDuplim} emphasizes the importance of properly including upper limits for the WISE bands and bands at longer wavelengths to avoid overestimating the SFR.

An even more complex situation emerges when we contrast our measurements with those from the MPA-JHU database. First, recall that the total SFRs from the MPA-JHU catalog come from the combination of the contribution measured from the fiber spectrum plus that obtained from SED fitting of broadband UV and optical photometry outside the fiber. This method gives SFRs of considerably large uncertainty. As shown in Figure~\ref{SFRcompare}b, sources with ${\rm SFR} \gtrsim 0.5\,M_\odot\,{\rm yr}^{-1}$ track our measurements fairly well, with $\Delta \log \rm SFR = -0.00\pm0.29$\,dex. The scatter is comparable to the uncertainties of the SFRs in the MPA-JHU catalog in this SFR range. At the highest SFRs, the MPA-JHU values have a mild tendency to be higher than ours. This trend can be explained by our choice of dust attenuation. Although we can achieve better agreement between the two sets of measurements by using the original attenuation law of \cite{Calzetti2000}, a modified attenuation law with a steeper slope has been found to be preferable by many investigators (e.g., \citealt{Buat2011, Salim2018, Qin2022}). The departures between our measurements and those of the MPA-JHU catalog become alarmingly severe ($\Delta \log \rm SFR = -0.47\pm0.71$\,dex) once ${\rm SFR} \lesssim 0.2\,M_\odot\,{\rm yr}^{-1}$. In the range ${\rm SFR} \approx 0.001-0.1\,M_\odot\,{\rm yr}^{-1}$, the SFRs in the MPA-JHU catalog remain more or less fixed at ${\rm SFR} \approx 0.1\,M_\odot\,{\rm yr}^{-1}$. The low-SFR sources also have vastly larger uncertainties in the MPA-JHU catalog (median 0.70\,dex) than in ours (median 0.45\,dex). Within the MPA-JHU catalog, the fiber SFRs for AGNs, composites, and unclassifiable objects (with low-S/N spectra) derive from a relation between the 4000~\AA\ break (D4000) and the specific SFR (sSFR) calibrated from emission-line galaxies for which both quantities could be measured. The objects in our sample with low SFR and large offsets belong to this category. As shown in Figure~11 of \cite{Brinchmann2004}, when D4000 is large (${\rm D4000}\ga1.6$), it starts to become less sensitive to sSFR and exhibits a large spread. In particular, the mean relation exhibits an unphysical rise in sSFR at ${\rm D4000}>2$. The large uncertainty and potential bias of the D4000-sSFR relation may boost the SFR in quiescent galaxies and account for the systematical underestimation of SFRs observed at ${\rm SFR}<0.2\,M_{\odot}$ yr$^{-1}$. 

Lastly, we again examine the subset of heavily blended galaxies that we analyzed by 2D decomposition (Section~\ref{sec_galfitm}). The deviation of SFRs for the blended components is slightly larger than that for isolated objects. Our measurements of the SFRs of the individual galaxies in the blended pairs are $0.14\pm0.69$~dex lower than those in the GSWLC-2 catalog and $0.19\pm0.60$~dex lower than those in the MPA-JHU catalog.

\section{Discussion}
\label{sec_discussion} 

We study the panchromatic (UV to FIR) SEDs of massive galaxies in the SDSS Stripe~82 region using a set of newly measured aperture photometry. For galaxies that are spatially resolved, as is the case for the majority of our low-redshift sample, it is important to adopt a common, matched aperture for all bands to measure the total flux of the source. Careful consideration is given to deblending physical or projected neighboring galaxies, as well as to removing contaminating foreground stars. We pay close attention to the calculation of rigorous uncertainties, both statistical and systematic, and to the derivation of upper limits for nondetections. It is critical to include upper limits into the SED analysis in order to obtain unbiased measurements of physical parameters, especially the SFR and dust mass.

We derive stellar masses that are largely consistent with those reported in the GSWLC-2 and MPA-JHU catalogs. We observe a mild deviation between the MPA-JHU stellar masses and ours among low-mass galaxies, which can be attributed to differences in the choice of star formation history. The SFRs measured by our SED fitting are consistent with those provided by the GSWLC-2 and MPA-JHU catalogs for galaxies with relatively strong star formation activity (${\rm SFR} \gtrsim 0.3\,M_\odot \, \mathrm{yr^{-1}}$). Both catalogs deviate from our measurements for galaxies with ${\rm SFR} \lesssim 0.3\,M_\odot \, \mathrm{yr^{-1}}$. It is not easy to fully explain the deviations between our SFRs and those from GSWLC-2 because the latter combines a set of heterogeneous photometric measurements, which at wavelengths longer than $\sim 20~\mathrm{\mu m}$ are based on extrapolation from an average spectral template. By contrast, our SEDs are constructed from optimized measurements of actual observations up to $500~\mathrm{\mu m}$. The systematic deviations between our results and those from the MPA-JHU catalog can be traced to several factors. Apart from the increased uncertainty incurred from the hybrid method of deriving total SFRs by combining central fiber measurements with (incomplete) photometric measurements in the galaxy outskirts, the reliance on the relation between D4000 and sSFR to estimate SFRs for galaxies with weak emission lines and AGNs introduces not only significant scatter but also bias.

\begin{figure}
\centering
\includegraphics[width=0.46\textwidth]{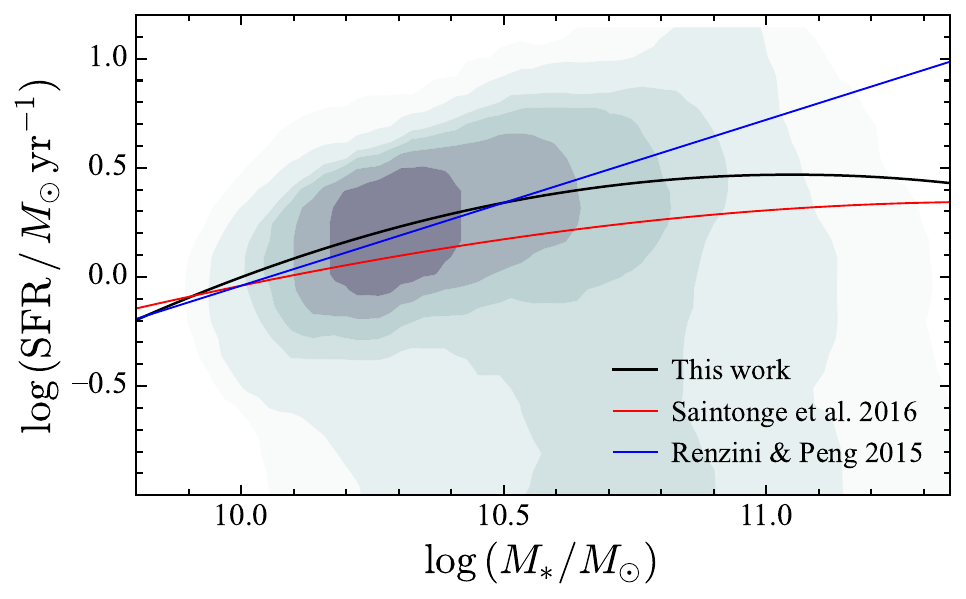}
\caption{Comparison of the main sequence derived in this work with those from \citet{Saintonge2016} and \citet{Renzini_Peng2015}. Our main sequence is derived following the method in \citet{Saintonge2016}. }
\label{compare_MS}
\end{figure}

As a first exploratory application of our database, we examine the distribution of the Stripe~82 galaxy sample in the ${\rm SFR} - M_*$ plane (Figure~\ref{compare_MS}). We construct the main sequence using the star-forming galaxies that have ${\rm sSFR} \equiv {\rm SFR}/M_* > 10^{-11}\,{\rm yr}^{-1}$ \citep{Wetzel2012,Kukstas2020}, calculating the median SFR in stellar mass bins of 0.15~dex, following \citet{Saintonge2016}. We fit the ${\rm SFR} - M_*$ relation with a second-order polynomial, which suffices to describe the data because of the limited number of galaxies at $\Ms \gtrsim 10^{11.3}\,M_\odot$. The best-fit relation is

\begin{equation}\label{equ_MS}
\log\,\left(\frac{\rm SFR}{M_\odot\:\mathrm{yr}^{-1}}\right)= 0.8955x - 0.4268 x^2 - 0.001, 
\end{equation}
 
\noindent
where $x=\log\,(\Ms/10^{10}\,M_\odot)$. Our main sequence is overall close to, but slightly above that reported by \citet{Saintonge2016}: at $M_* = 10^{10.5}\,M_{\odot}$, $\Delta\,\rm log\,SFR\,\approx 0.1$~dex (Figure~\ref{compare_MS}). \citet{Saintonge2016} derive SFRs using a combination of GALEX and AllWISE (W3 or W4) photometry. The W3 and W4 photometry, obtained using a \cite{Kron1980} aperture, can miss $10\%-50\%$ of the total flux of highly concentrated galaxies \citep{Graham2005}. This may explain their lower SFRs relative to ours because galaxies with $\Ms \gtrsim 10^{10}\,M_\odot$ usually have high concentration \citep{Wuyts2011, Popesso2019a}.

We confirm the flattening of the slope of the main sequence relation, which is particularly notable for galaxies more massive than $\sim 10^{10.5}\,M_{\odot}$. Our main sequence lies substantially below the relation reported by \citet{Renzini_Peng2015}, by $\Delta\,\rm log\,SFR\,\approx -0.3$~dex at $M_* = 10^{11}\,M_{\odot}$, and is in better agreement with that of \citet{Saintonge2016}. We suspect that part of the discrepancy, apart from differences resulting from the SFR estimator used, stems from the sample definition. The study of \citet{Renzini_Peng2015} excluded galaxies classified as AGNs and composites. Since nearby AGNs preferentially inhabit earlier type, massive galaxies with relatively evolved stellar populations \citep{Ho1997,Ho2003}, these sources mainly occupy the high-mass region with SFRs lower than those of star-forming galaxies \citep{Leslie2016}. Additional factors may contribute to the flattening of the main sequence at the high-mass end, including the growth of bulges \citep{Wuyts2011, Abramson2014, Feldmann2017, Belfiore2018}, underestimation of SFR in edge-on disks \citep{Morselli2016}, and starvation of cold gas in massive halos \citep{Popesso2019a}. We will explore these issues more fully in a forthcoming work.

\begin{figure*}
\centering
\includegraphics[width=0.45\textwidth]{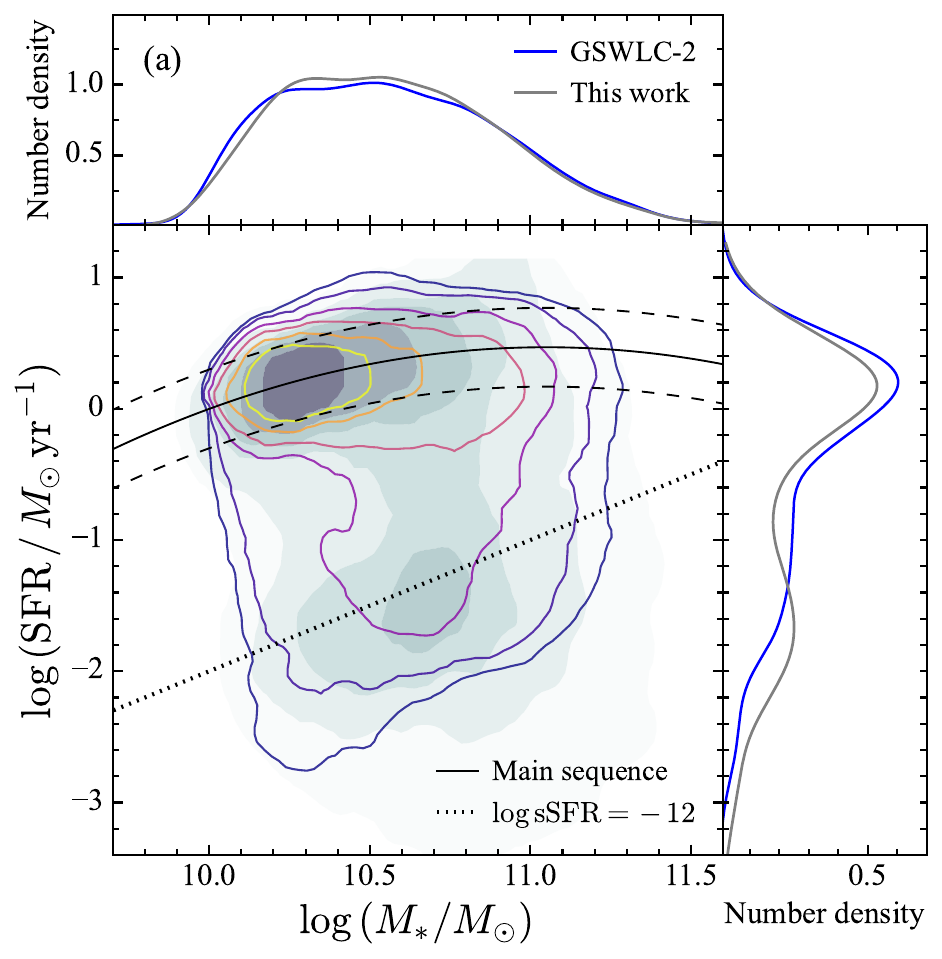}
\includegraphics[width=0.45\textwidth]{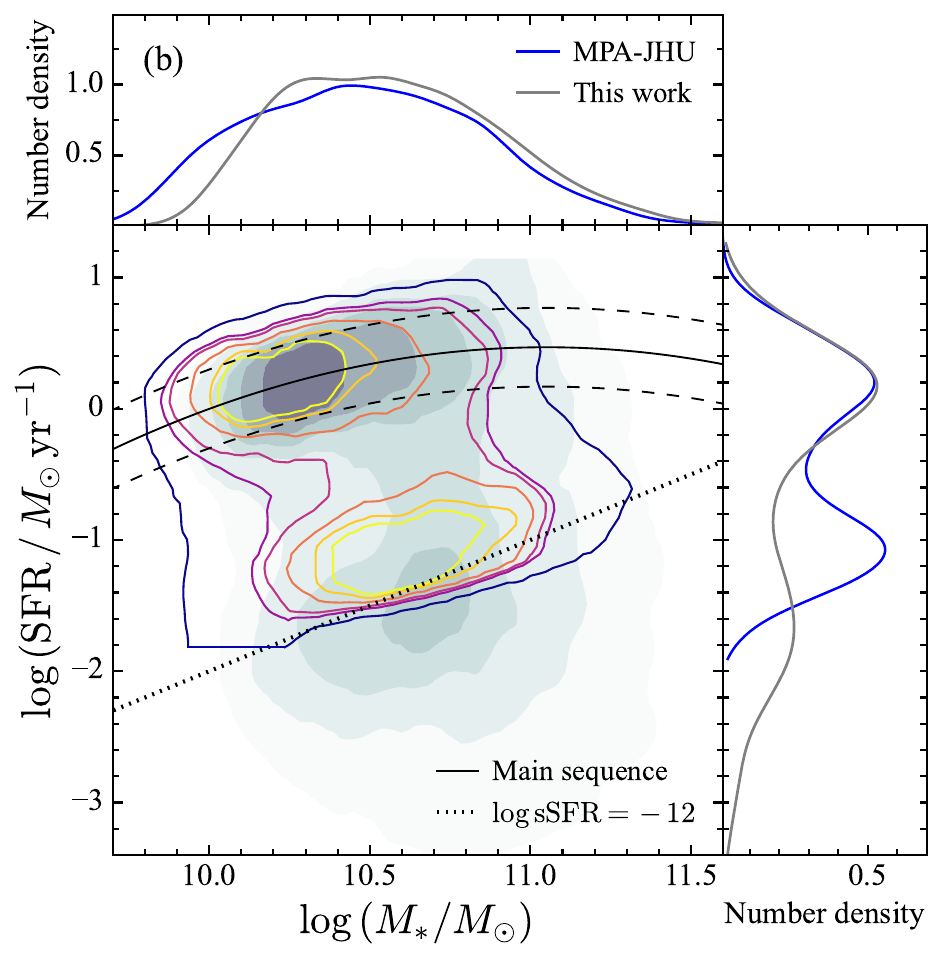}
\caption{Comparison of the ${\rm SFR}-M_*$ diagram derived from this work with that from (a) the GSWLC-2 catalog and (b) the MPA-JHU catalog. The distribution of sources from this work is presented as shadowed background contours, overlaid by a colorful contour showing the distribution from the comparison catalogs. The black solid and dashed curves denote the star-forming main sequence and its $1\sigma$ scatter (Equation~\ref{equ_MS}). The grey dotted line marks $\rm log\,(sSFR/yr^{-1})=-12$. The top and right panels show the histograms of stellar masses and SFRs, respectively.}
\label{fig:SFR_Ms_compare}
\end{figure*}

To illustrate the importance of estimating proper SFRs for investigating the statistical physical properties of galaxies, in Figure~\ref{fig:SFR_Ms_compare} we show the distribution of Stripe~82 galaxies on the ${\rm SFR}-M_*$ plane using the parameters derived from this study, compared to those based on the GSWLC-2 and MPA-JHU catalogs. While the distribution of the star-forming galaxies around the main sequence is largely consistent among the three sets of measurements, in detail there are subtle differences. The ridge line of star-forming galaxies at the massive end of GSWLC-2 is slightly lower than our main sequence. This is a consequence of the fact that GSWLC-2 systematically overestimates SFR over the range $\sim 0.1-1\,M_{\odot}\,\mathrm{yr}^{-1}$ (Figure~\ref{SFRcompare}a), whereby galaxies that are located in the so-called green valley according to our measurements become artificially elevated to the star-forming sequence in GSWLC-2. The contours of star-forming and green valley galaxies from our measurements are consistent with the contours derived from the MPA-JHU catalog (Figure~\ref{fig:SFR_Ms_compare}b). In contrast, the distribution of quenched galaxies (e.g., ${\rm SFR} \lesssim 0.1\,M_\odot\,\rm yr^{-1}$) show dramatic differences between the GSWLC-2 and MPA-JHU catalogs, and between these two catalogs and ours. The discrepancies stem from the different methods and systematic biases incurred when measuring the SFR, as discussed in Section~\ref{ssec:sfr}. The grey dotted 
diagonal line marks $\rm log\,(sSFR/yr^{-1})=-12$, below which the SFRs may not be trustworthy (see Section~4.5).

\section{Summary}
\label{sec_summary} 

We construct panchromatic SEDs of 2781 low-redshift ($z= 0.01-0.11$) galaxies in the SDSS Stripe~82 region, among them 2668 that have stellar masses $M_* > 10^{10}\, M_{\odot}$, that have been observed with the Herschel/SPIRE instrument at 250, 300, and 500 $\mu$m. In combination with other survey data from GALEX, SDSS, 2MASS, and WISE, we develop a hybrid approach of matched-aperture photometry, which can be used effectively for the 14 shorter wavelength bands (FUV to 4.6~$\mu$m), along with profile-fitting photometry for the longer wavelength bands. This approach is optimized for relatively bright, nearby galaxies that are partly or fully resolved. We apply simultaneous, multi-band, 2D image fitting to properly decompose interacting/merging galaxies and sources in crowded environments. A new method, based on random forest regression, is introduced to decontaminate foreground stars from images. Apart from obtaining reliable and physically consistent total galaxy fluxes and their associated uncertainties across multiple bands, we estimate robust upper limits for the nondetections and fully incorporate them into the final SED analysis. Mock tests evaluate the influence of certain critical bands and their respective upper limits. We derive stellar masses, SFRs, dust masses, and AGN luminosity fractions, which will be the subject of a series of forthcoming works.

The main results are as follows:

\begin{enumerate}

\item
Our method of galaxy photometry yields self-consistent, panchromatic total fluxes with systematic uncertainty less than $\sim 20\%$.

\item
Although 40\% of our sample with strong optical emission lines have some level of nonstellar nuclear activity according to their spectral classification, in the vast majority of the cases the AGN contributes negligibly to the broadband SED fits and can be neglected.

\item
The stellar mass can be well constrained solely using photometry from the five SDSS bands. By contrast, estimating accurate SFRs requires the detection of or, at the very minimal, upper limits in the MIR or even bands of longer wavelengths. Measuring the dust mass requires detection in ar least one Herschel/SPIRE band, and whenever possible, even upper limits in the FIR bands should be used to avoid excessively large systematic biases.

\item
Comparison of the stellar masses from our SED fitting reveals good agreement with those from the GSWLC-2 and MPA-JHU catalogs. The same is true for the SFRs of galaxies that occupy the star-forming main sequence, which, for our relatively massive galaxies correspond to ${\rm SFR} \gtrsim 0.3\,M_\odot \, \mathrm{yr^{-1}}$. However, for galaxies with weaker star formation activity, including those on and below the green valley, the SFRs from the GSWLC-2 and MPA-JHU catalogs disagree with each other, and both sets of measurements are systematically overestimated relative to ours. SED fitting-based SFRs of galaxies with moderate to low star formation activity should be used with considerable caution.

\end{enumerate}

\begin{acknowledgments}
LCH was supported by National Key R\&D Program of China (2022YFF0503401), the National Science Foundation of China (11721303, 11991052, 12011540375, 12233001) and the China Manned Space Project (CMS-CSST-2021-A04, CMS-CSST-2021-A06). We acknowledge Yingjie Peng, Linhua Jiang, Chao Ma, and Kai Wang for their discussion and contributions to various aspects of this work.
 We thank the referee for a very constructive report that greatly helped to improve the paper.
\end{acknowledgments}

\newpage
\appendix

\section{Comparison of the Photometric Measurements}
\label{appdix_fluxcompare}

In this appendix, we compare our photometric measurements with those provided in the catalogs of GALEX, 2MASS, and WISE. For clarity, only sources that have flux $\rm S/N>2$ as given by the GALEX, 2MASS, and WISE catalogs are presented in the comparison. We use the \texttt{fit\_left\_censoring}\footnote{\url{https://lifelines.readthedocs.io/en/latest/Survival\%20analysis\%20with\%20lifelines.html\#left-censored-data-and-non-detection}} method in the \texttt{KaplanMeierFitter} class of \texttt{Python} to perform the survival analysis for the comparisons, calculating the median and $1\sigma$ of the flux differences after accounting for upper limits.

\begin{enumerate}

\item{\bf GALEX:} Figure~\ref{fig:compGALEX} compares our FUV and NUV measurements with GALEX \texttt{MAG\_AUTO} from GR6+7, which were measured within a Kron-like elliptical aperture. GALEX only provides detected magnitudes with $\rm S/N>2$. For fainter ($\rm S/N<2$) sources, we only measure upper limits. Overall, we find good consistency, with median difference only $\Delta \rm FUV = -0.02\pm0.11$~mag and $\Delta \rm NUV = 0.03\pm0.15$~mag.

 \item{\bf SDSS:} Figure~\ref{fig:compSDSS} compares our measurements with the SDSS \texttt{modelmag} values. The SDSS values are fainter than our measurements by 0.08~mag in $u$ and by 0.05~mag in $g$, although these systematic offsets are small compared to the scatter. In the $r,\,i$, and $z$ bands, our measurements agree well with the \texttt{modelmag} values. The \texttt{modelmag} photometry for the five SDSS bands are based on the better of two model fits in the $r$ band that assume either an exponential or de~Vaucouleurs function. However, the bluer bands, which are dominated by younger stars, may have different light profiles. Our non-parametric measurements overcome this problem and recover the total flux in the SDSS bands.

\item{\bf 2MASS:} We compare our measurements with two sets of 2MASS catalog results. Figure~\ref{fig:comp2MASSXSC} shows good consistency ($\Delta J = -0.04\pm0.20$~mag, $\Delta H = -0.04\pm0.25$~mag, $\Delta K_s = -0.04\pm0.24$~mag) between our measurements and the ``total flux'' from the 2MASS XSC \citep{Jarrett2MASS2000}. The XSC fitted the galaxy light profile with a \sersic\ modified exponential function in the $J$ band, which was then applied to all bands and integrated out to 4 times the disk scale length to measure the total flux of the galaxy. The good consistency confirms that our measurements enclose the total flux of the galaxy. However, only about 60\% of our sample is included in the XSC. While PSF-fitting magnitudes are available for all sources in the 2MASS PSC \citep{Cutri2003}, the large, systematic discrepancy ($\Delta J = -0.95\pm0.45$~mag, $\Delta H = -0.95\pm0.48$~mag, $\Delta K_s = -0.88\pm0.41$~mag) with our measurements (Figure~\ref{fig:comp2MASSPSC}) indicates that most of the galaxies are resolved significantly in 2MASS images.

\begin{figure*}
\centering
\figurenum{A1}
\includegraphics[width=0.63\textwidth]{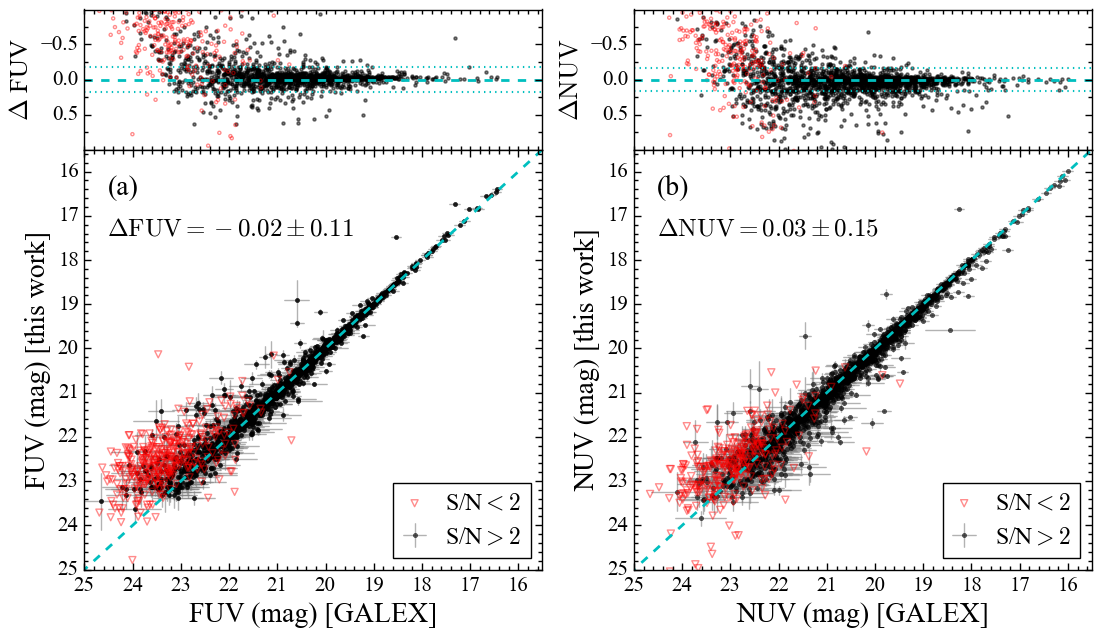}
\caption{Comparison between our measurements and those from the GALEX catalog in the (a) FUV and (b) NUV bands. The upper limits ($\rm S/N<2$) from our measurements are shown in red triangles. The upper panels show the differences between the two sets of measurements (this work $-$ GALEX). The dotted lines give the quadrature sum of the typical uncertainties of the two sets of measurements. The median and standard deviation derived through survival analysis are shown in the upper left of the bottom panels.}
\label{fig:compGALEX}
\end{figure*}

\item{\bf AllWISE:} For resolved galaxies flagged by $\texttt{ext\_flg}>0$ in the AllWISE catalog, our measurements are about 0.5~mag brighter than the profile-fitting results (Figure~\ref{fig:compAllWISE}). Even for galaxies classified as unresolved ($\texttt{ext\_flg}=0$), our measurements are still 0.2~mag brighter in W1 and W2. For W3 and W4 bands, we only compare the measurements of resolved galaxies selected according to the criteria outlined in Appendix~\ref{appdix_psffit}. Our aperture photometry is about 0.4~mag brighter than the profile-fitting results. The elliptical aperture photometry of AllWISE, which is based on the 2MASS extended source aperture, still systematically misses some flux compared to our measurements. A similar result was found by \cite{Cluver2014}.

\item{\bf unWISE:} The unWISE catalog provides forced-photometry measurements based on the galaxy's $r$-band profile from SDSS \citep{Lang2016Tractor}. As shown in Figure~\ref{fig:comp_unWISE}, our measurements agree much better with the unWISE results ($\Delta \rm W_1 = 0.13\pm0.14$~mag, $\Delta \rm W_2 = 0.13\pm0.26$~mag, $\Delta \rm W_3 = 0.19\pm0.23$~mag, $\Delta \rm W_4 = -0.19\pm0.39$~mag) than with those from AllWISE.

\end{enumerate}

\begin{figure*}  
\centering
\figurenum{A2}
\includegraphics[width=0.99\textwidth]{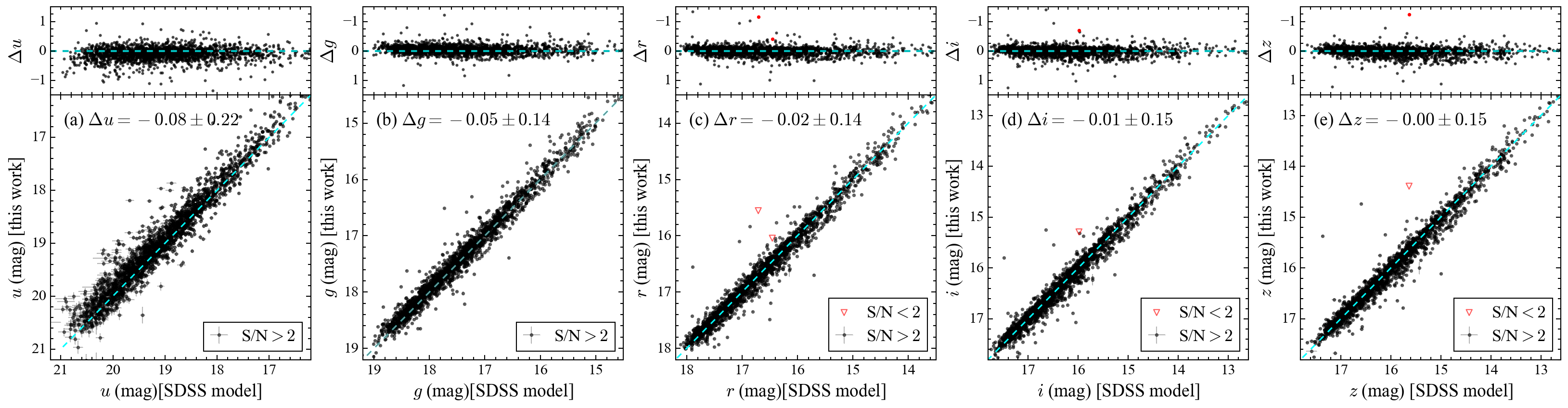}
\caption{ Comparison between our measurements and the SDSS \texttt{modelmag} photometry in the (a) $u$, (b) $g$, (c) $r$, (d) $i$, and (e) $z$ band. The upper limits ($\rm S/N<2$) from our measurements are plotted as red triangles. The upper panels show the differences between the two sets of measurements (this work $-$ \texttt{modelmag}). The dashed line denotes the 1:1 relation. The median and standard deviation derived through survival analysis are shown in the upper left of the bottom panels.} 
\label{fig:compSDSS}
\end{figure*}

\begin{figure*}  
\centering
\figurenum{A3}
\includegraphics[width=0.9\textwidth]{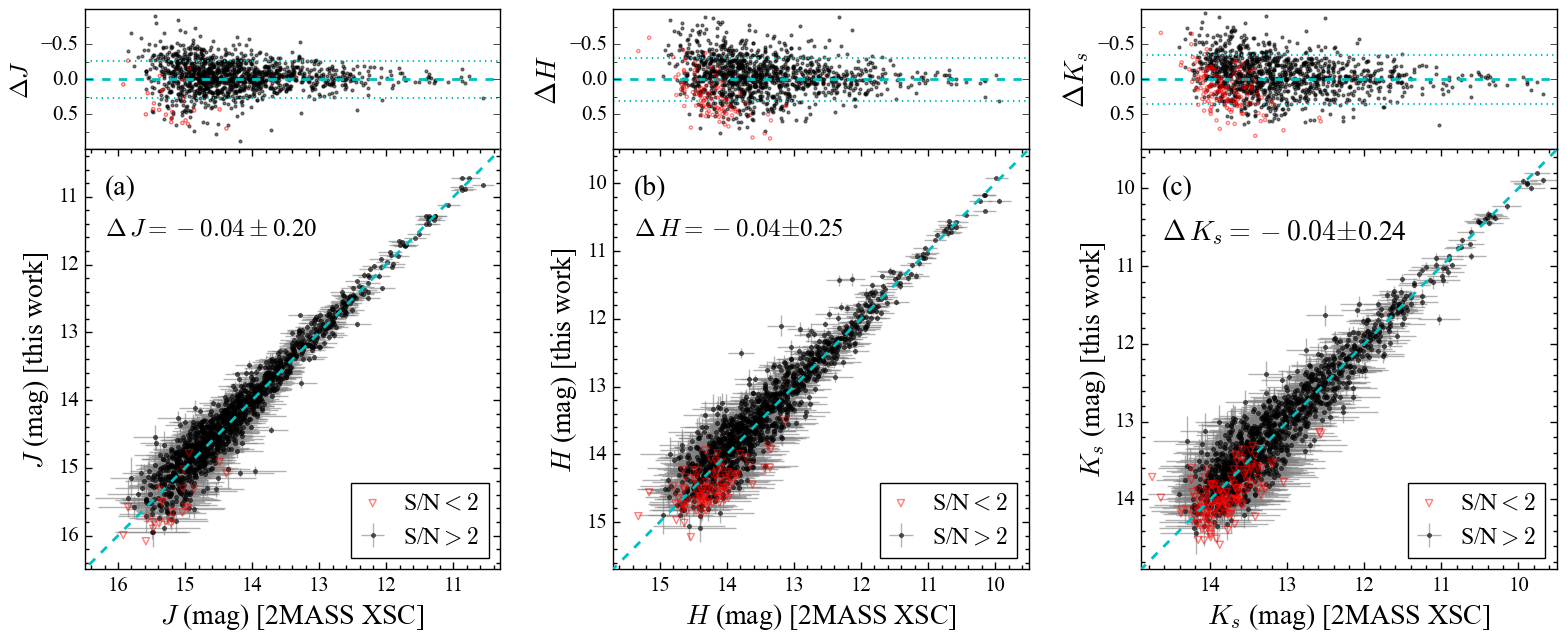}
\caption{Comparison between our measurements and the ``total flux'' measurements of the 2MASS XSC in the (a) $J$, (b) $H$, and (c) $K_s$ bands. The upper limits ($\rm S/N<2$) from our measurements are shown in red triangles. The upper panels show the differences between the two sets of measurements (this work $-$ 2MASS). The dotted lines give the quadrature sum of the typical uncertainties of the two sets of measurements. The median and standard deviation derived through survival analysis are shown in the upper left of the bottom panels.}
\label{fig:comp2MASSXSC}
\end{figure*}

\begin{figure*}  
\centering
\figurenum{A4}
\includegraphics[width=0.88\textwidth]{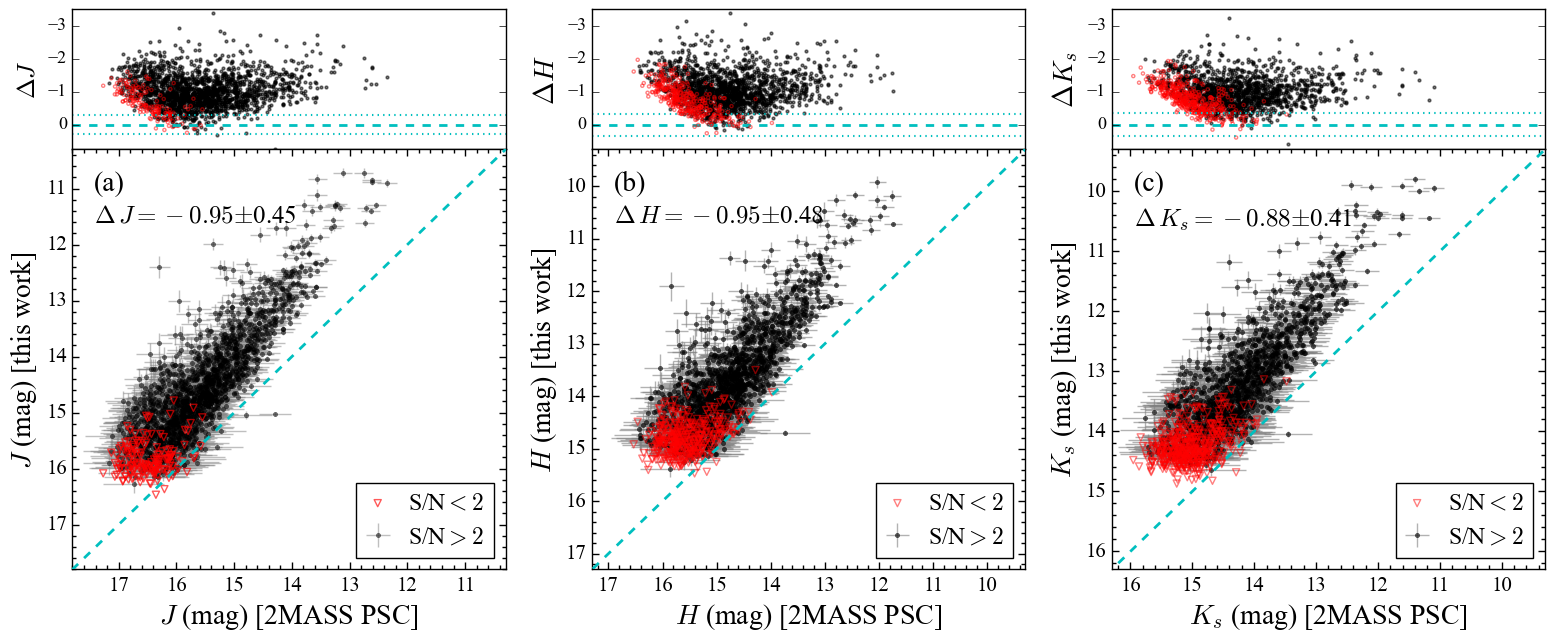}
\caption{Comparison between our measurements and the 2MASS PSF-fitting magnitudes in the (a) $J$, (b) $H$, and (c) $K_s$ bands. The upper limits ($\rm S/N<2$) from our measurements are shown in red triangles. The upper panels show the differences between the two sets of measurements (this work $-$ 2MASS). The dotted lines give the quadrature sum of the typical uncertainties of the two sets of measurements. The median and standard deviation derived through survival analysis are shown in the upper left of the bottom panels.}
\label{fig:comp2MASSPSC}
\end{figure*}

\begin{figure*} [hb]
\figurenum{A5}
\gridline{
\fig{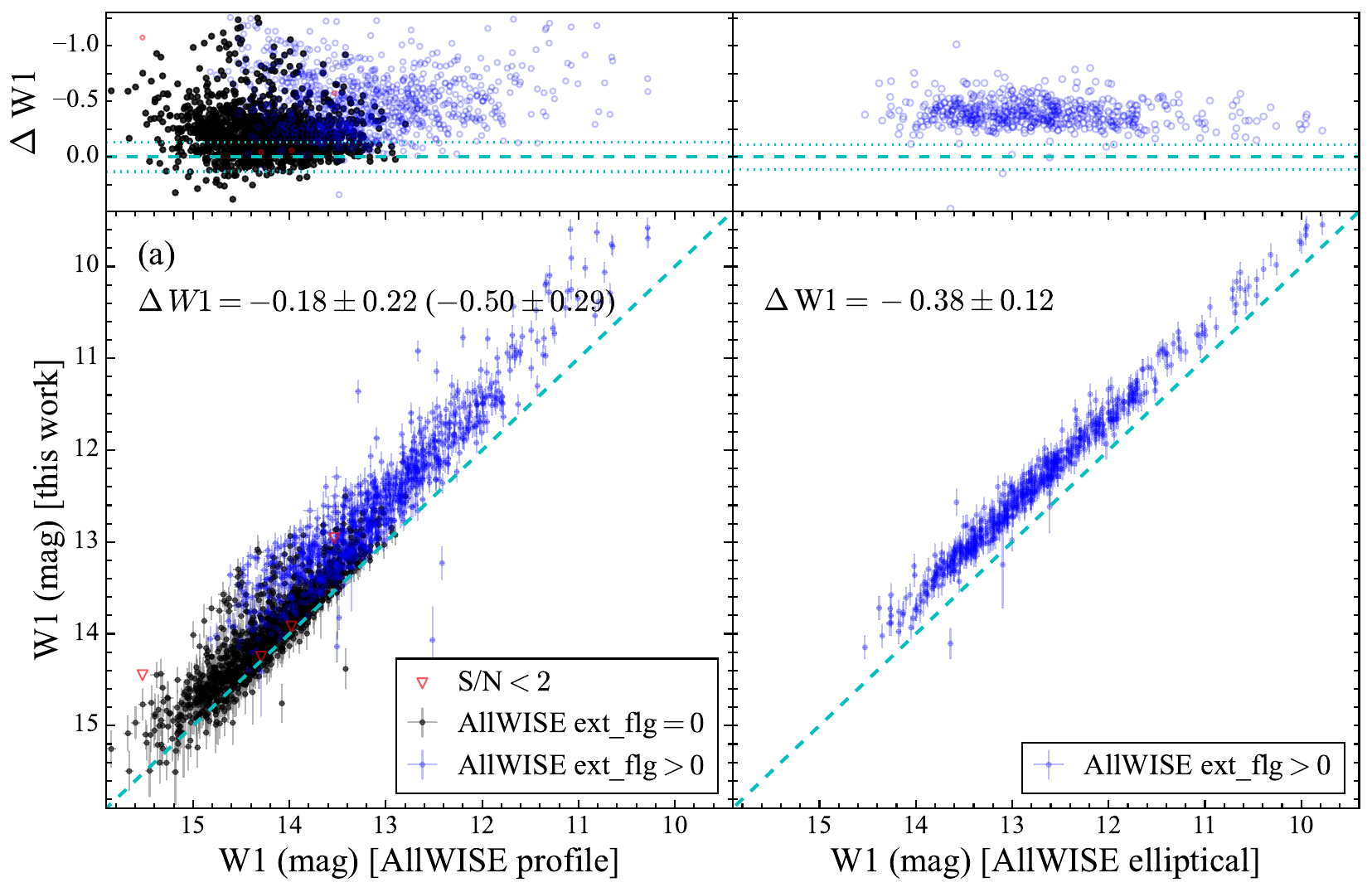}{0.495\textwidth}{}
\fig{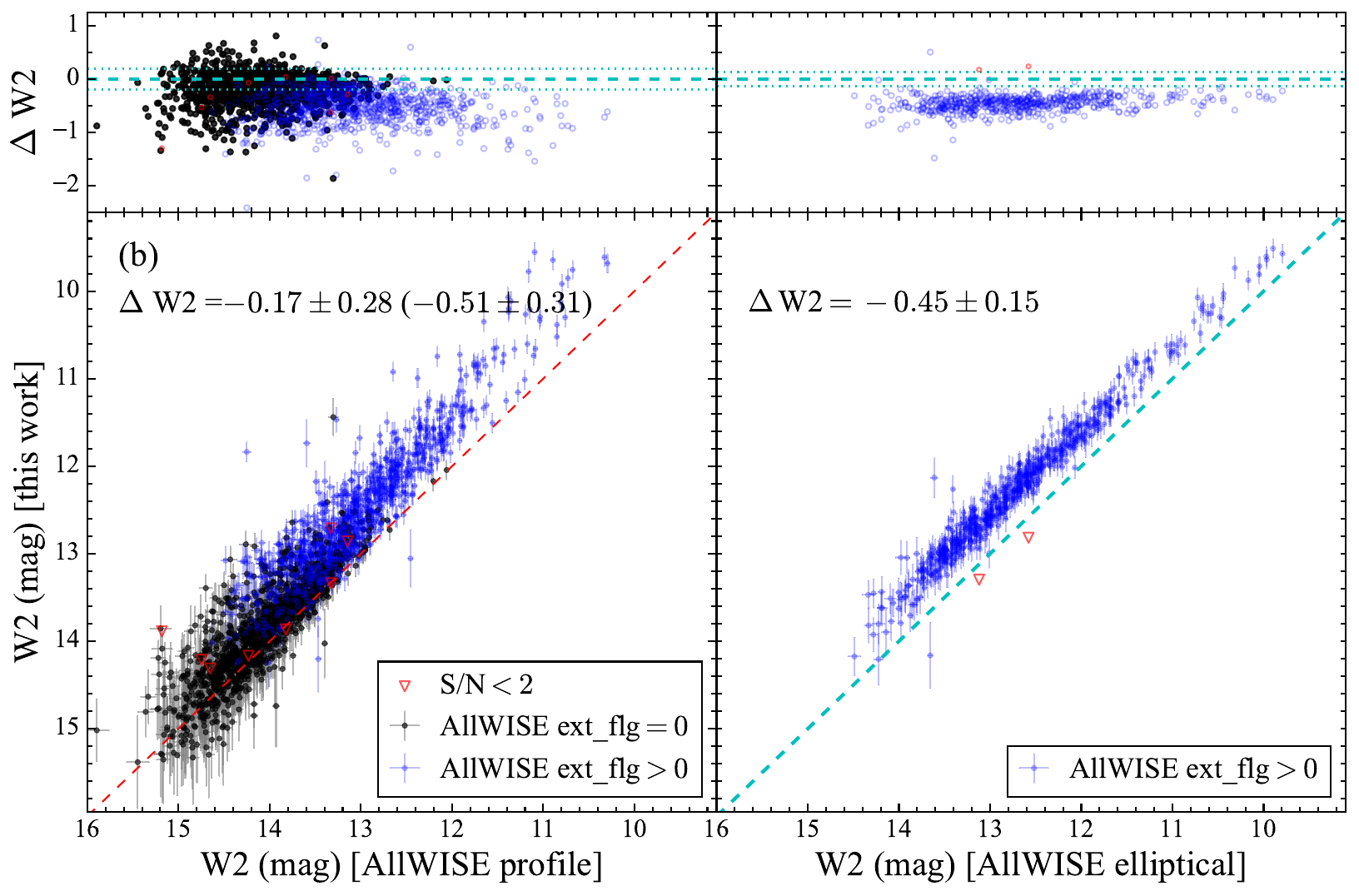}{0.495\textwidth}{}
}
\gridline{
\fig{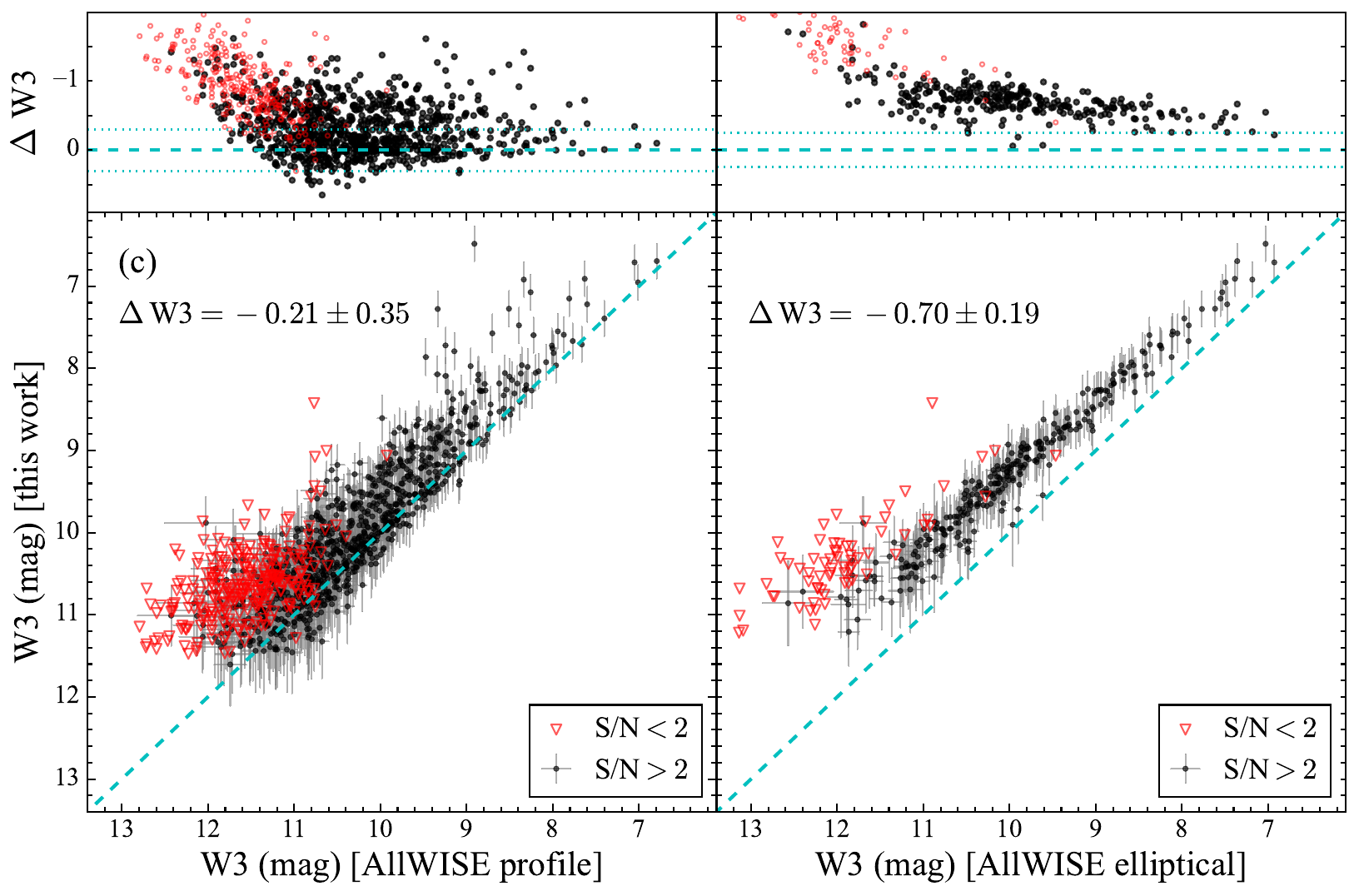}{0.495\textwidth}{}
\fig{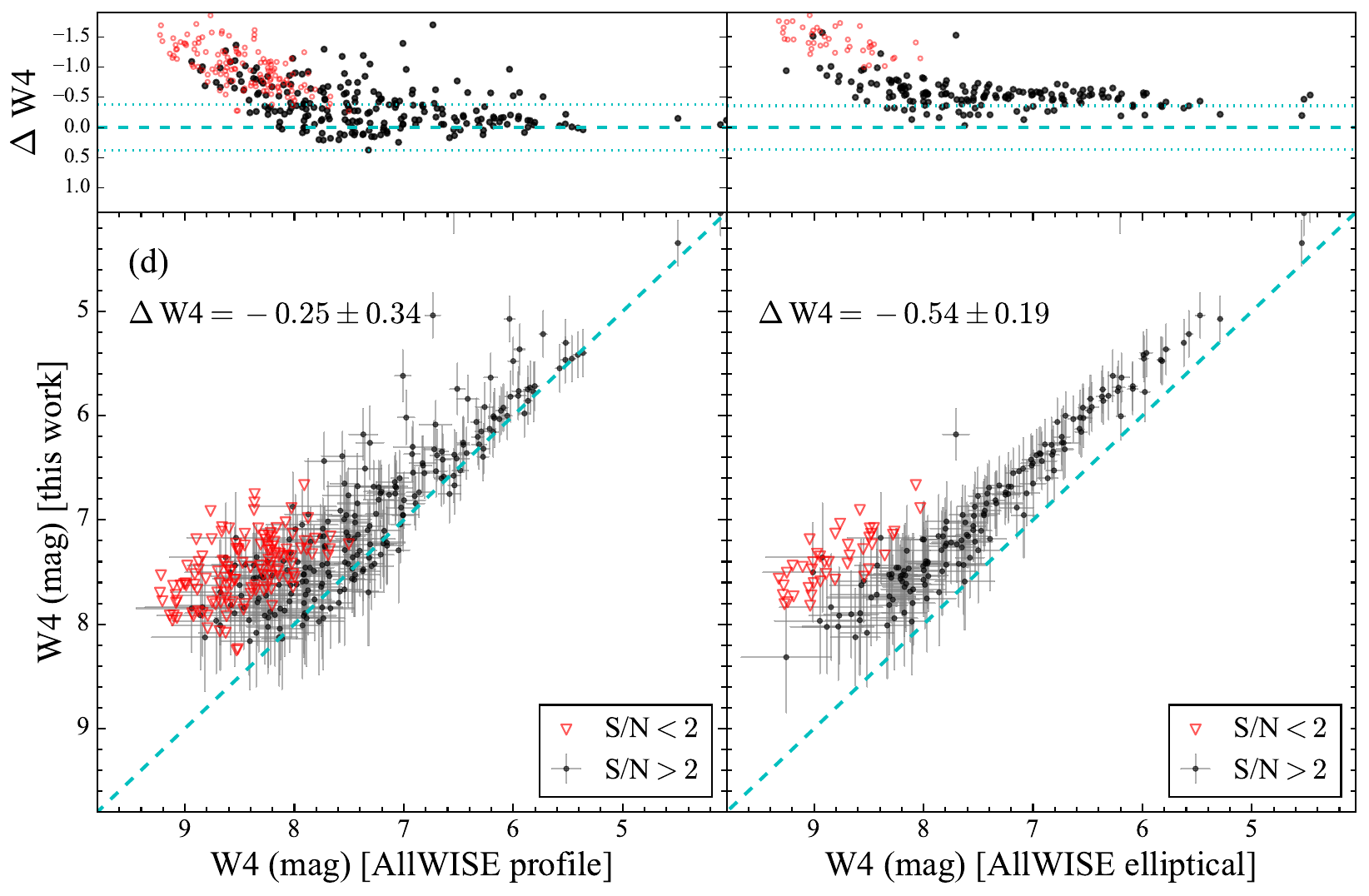}{0.495\textwidth}{}
}
\caption{Comparison between our measurements and the AllWISE measurements obtained from profile fitting (left) and elliptical aperture photometry (right) for (a) W1, (b) W2, (c) W3, and (d) W4. For W1 and W2, unresolved and resolved sources are plotted in black and gray points, respectively. The upper limits ($\rm S/N<2$) from our measurements are shown in red triangles. The upper panels show the differences between the two sets of measurements (this work $-$ WISE). The dotted lines give the quadrature sum of the typical uncertainties of the two sets of measurements. The median and standard deviation derived through survival analysis are shown in the upper left of the bottom panels. For W1 and W2, the numbers in parentheses pertain to the resolved sources.}
\label{fig:compAllWISE}
\end{figure*}

\begin{figure*}  
\centering
\figurenum{A6}
\includegraphics[width=\textwidth]{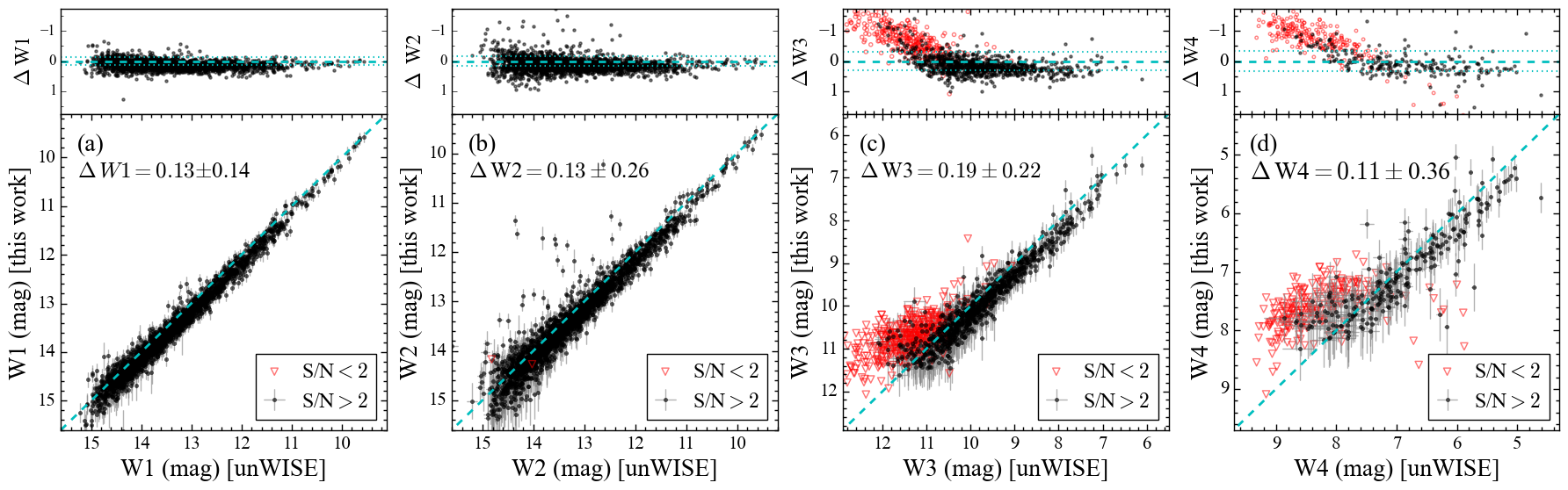}
\caption{Comparison between our measurements and the measurements from the unWISE catalog for (a) W1, (b) W2, (c) W3, and (d) W4. The upper limits ($\rm S/N<2$) from our measurements are shown in red triangles. The upper panels show the differences between the two sets of measurements (this work $-$ unWISE). The dotted lines give the quadrature sum of the typical uncertainties of the two sets of measurements. The median and standard deviation derived through survival analysis are shown in the upper left of the bottom panels.}
\label{fig:comp_unWISE}
\end{figure*}

\begin{figure*} 
\centering
\figurenum{B1}
\includegraphics[width=\textwidth]{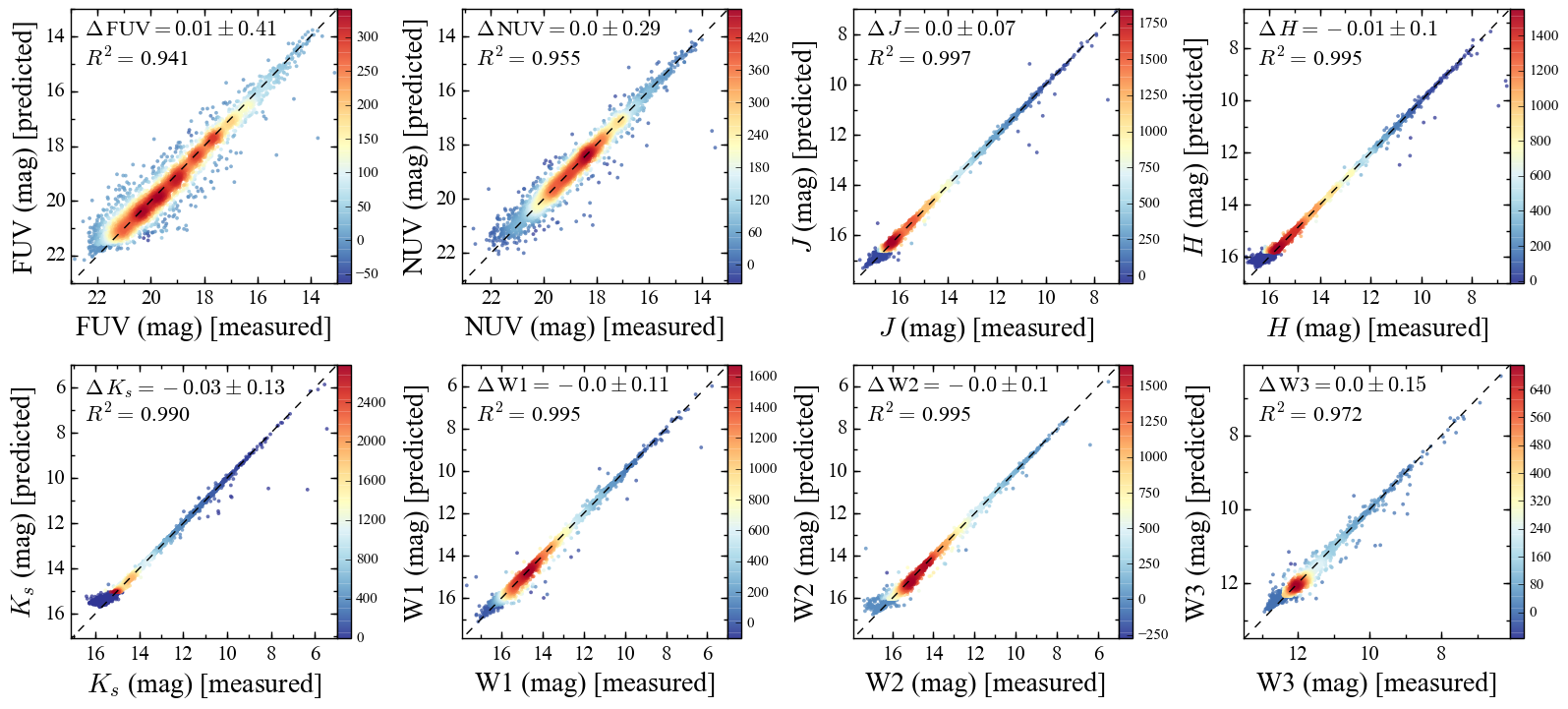}
\caption{Comparison between the measurements and RF-predicted magnitudes of the testing sample of stars. The median and standard deviation between the measurements and the prediction, as well as the $R^2$ score, are displayed in the upper-left corner of each panel. The red dashed line indicates the 1:1 relation. The color code indicates the number density of objects.}
\label{RFpredict}
\end{figure*}

\section{Removing Blended Emission from Foreground Stars}
\label{sec_star_decontam}

We develop a new method to remove the contamination of a foreground star if its emission cannot be deblended from the target galaxy following the standard deblending procedure described in Section~\ref{sec_deblend}. While the galaxy and foreground stars are well resolved and can usually be deblended in SDSS images, they become heavily blended in WISE images on account of its much coarser PSF. Blending can even be problematic in GALEX and 2MASS images, too. As robust measurements exist of the positions and magnitudes of the galaxy and the star(s) from SDSS, we can remove the contamination of the stellar emission if we can accurately estimate the magnitude of the star(s) in the other bands based on the SDSS measurements. 

We use random forest (RF) regression \citep{Breiman2001RandomForest} to predict the magnitude of the star in the UV and IR bands based on their five-band SDSS optical magnitudes. The RF algorithm is an effective machine-learning model for regression problems. It is a supervised learning algorithm that links the input individual subsamples of the data set and outputs predicted values by building multiple decision trees. The decision trees work through minimizing the Gini Impurity \citep{Pedregosa2012sklearn} at each stage of the process. Models are constructed to uncover the highly nonlinear correlation between the targeted output value and the input subsamples that may contribute to it \citep{Bluck2022RF}. We will show that the optical SED from the five bands of SDSS contains enough information to strongly constrain the stellar type and accurately predict the magnitude of the star in the other bands.

We employ the \texttt{Python} function \texttt{RandomForest Regressor}\footnote{\url{https://scikit-learn.org/stable/modules/generated/sklearn.ensemble.RandomForestRegressor.html}} from the package \texttt{scikit-learn} \citep{Pedregosa2012sklearn}. To construct the training samples and test the feasibility of the method, we collect unsaturated ($z>13$ and $u>16$~mag) field stars from the SDSS DR16 database (class=\texttt{STAR}) that do not have companion sources within 20\arcsec, a distance larger than 3 times the FWHM of the PSF. We also exclude stars near the Galactic plane ($|b|<10^{\circ}$) because dust extinction strongly affects the optical SED of the star. We crossmatch the SDSS coordinates of the stars to the GALEX GR6+7, the 2MASS PSC, and the AllWISE catalog. Limiting to measurements with $\mathrm{S/N}>3$, we find a sample of more than 21,000 stars for the UV and 54,000 stars for the IR, which covers the bands $J$ through W3 but not W4 and the FIR, for which stellar emission is typically very faint.

We arbitrarily divide the stars into a training sample and a testing sample, roughly in the number ratio 8:2. For each of the eight bands (FUV, NUV, $J,\,H,\,K_s$, W1, W2, W3), we use the training sample to train an RF model to make predictions based on the SDSS bands. Then we use the testing sample to evaluate the effectiveness of the model. We adopt \texttt{n\_estimators}=300 instead of the default value of 100 when training the FUV and NUV data, because we find that this choice significantly improves the prediction for the UV bands but not for the IR bands. Otherwise, we adopt the default parameters of \texttt{RandomForestRegressor}. Figure~\ref{RFpredict} compares the measured and predicted magnitudes of stars in the testing sample. For each band, we calculate the $R^2$ score,
 
\begin{equation}
R^2 = 1-\frac{\sum_{i=1}^{k}\, (y_i - y_i^{\prime})^2}{\sum_{i=1}^{k}\, (y_i - \hat{y})^2},
\end{equation}

\noindent 
where $k$ is the number of data in the test sample, and $y_i$, $y_i^{\prime}$ and $\hat{y}$ are the truth value, prediction, and mean of the truth value. The best possible $R^2$ score for a model is 1. Our models show very high consistency between data and prediction. We find little systematic uncertainty between the measured and predicted magnitudes for stars with $z<18.5$ and $u<18.5$~mag. As the contamination is negligible if the star is fainter than these limits, we only consider the contamination from stars brighter than these criteria. The scatter between the measured and predicted magnitudes is typically 0.1~mag, except for FUV (0.4~mag) and NUV (0.3~mag). We take the scatter as the uncertainty of the prediction.

\begin{figure}  [b]
\centering
\figurenum{B2}
\includegraphics[width=0.43\textwidth]{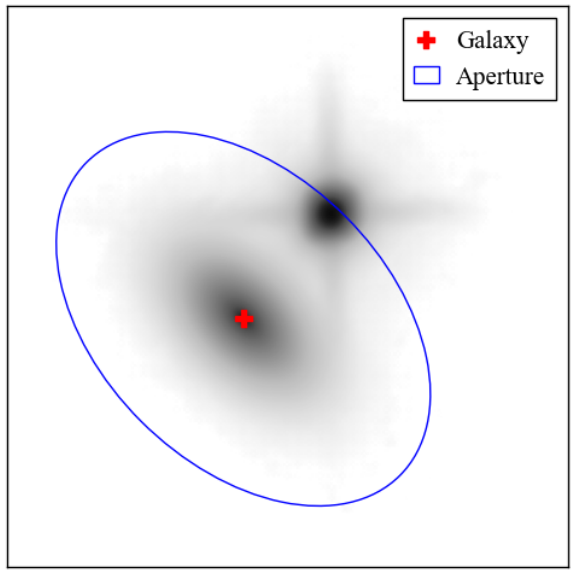}
\caption{Example of a target galaxy (marked by the red cross) blended with a bright star. The aperture of the galaxy (blue ellipse) encloses a significant fraction of the star.}
\label{PSFsubtr}
\end{figure}

For each target, we search for the stars with $z<18.5$ and $u<18.5$~mag inside twice the aperture size from the center of the galaxy. If the star cannot be deblended from the target in the UV and IR images following the procedure of Section~\ref{sec_deblend} (e.g., Figure~\ref{PSFsubtr}), we use the SDSS magnitudes and RF model to predict the magnitude of the star in these images. We calculate the fraction of a star's flux ($f_{\rm star}$) within the aperture of the target in the unmasked region from the PSF models, which are provided by the GALEX and WISE websites\footnote{\url{http://www.galex.caltech.edu/researcher/techdoc-ch5.html} and \url{https://wise2.ipac.caltech.edu/docs/release/allsky/expsup/sec4_4c.html}}. The PSF of 2MASS varies from image to image, and for each image tile we extract stars with $\rm S/N>3$ in an uncrowded field to generate the PSF model using the \texttt{IRAFstarFINDER} function from \texttt{photutils}\footnote{\url{https://photutils.readthedocs.io/en/stable/epsf.html}}. 

The uncertainty introduced by star contamination is
 
\begin{equation}\label{eqn:star}
\sigma_{\rm star}\,=\,\sigma_\mathrm{RF}\times f_{\rm star},
\end{equation}

\noindent 
where $\sigma_\mathrm{RF}$ is the scatter of the RF prediction in the corresponding band. We exclude targets whose SDSS images contain a saturated star ($z<13$ or $u<16$~mag) within twice their aperture size.

\begin{figure*} 
\centering
\figurenum{C1}
\includegraphics[width=0.43\textwidth]{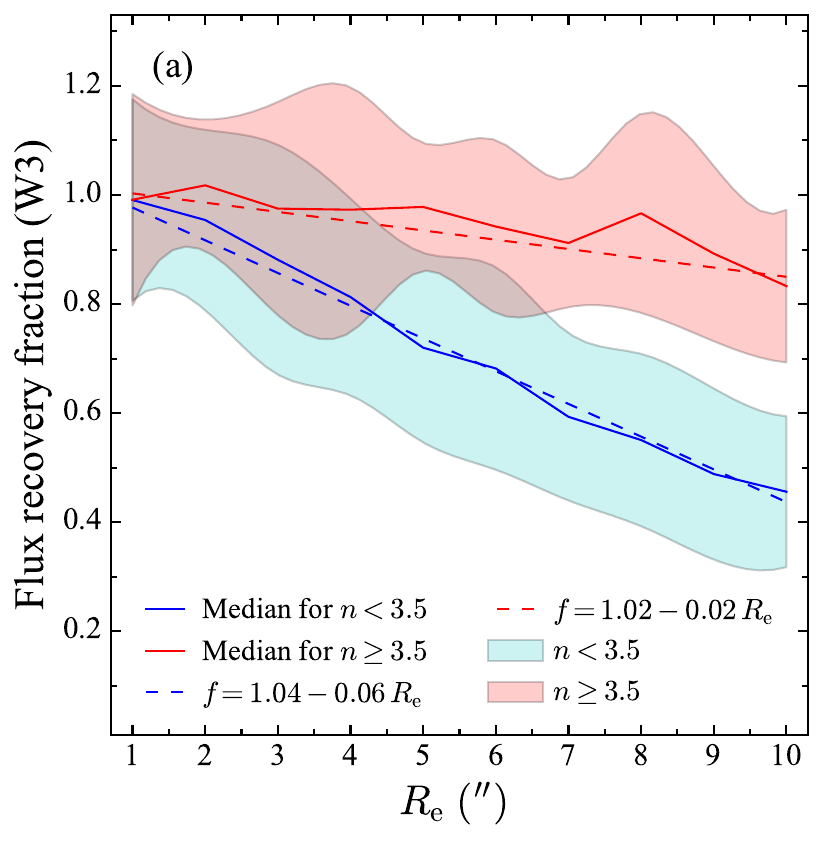}
\includegraphics[width=0.43\textwidth]{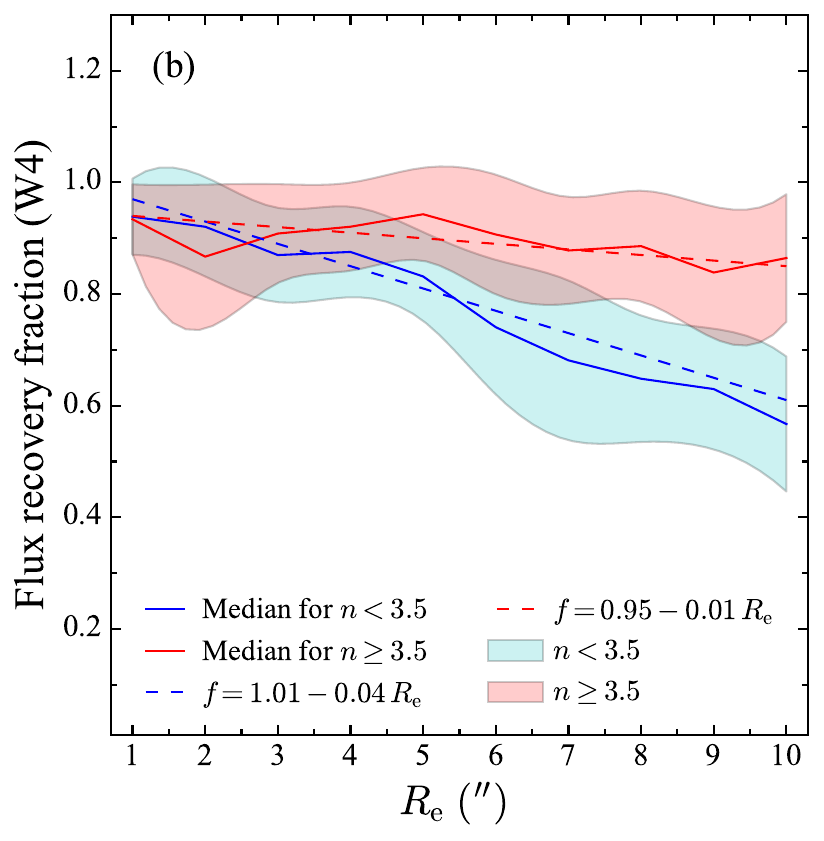}
\caption{Flux recovery fractions from the PSF-fitting test described in Appendix~B on (a) W3 and (b) W4 data. Galaxies with low ($n<3.5$; blue) and high ($n\ge3.5$; red) \sersic\ indices are shown as a function of effective radius $R_e$. We fit the relations with a linear function (dashed lines).}
\label{psffitw34snr2}
\end{figure*}

\begin{figure} [b]
\centering
\figurenum{C2}
\includegraphics[width=0.4\textwidth]{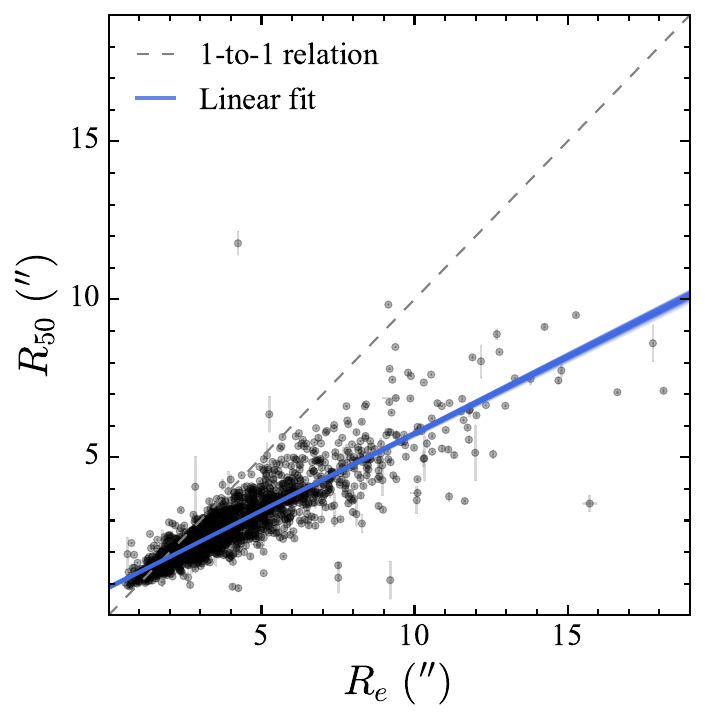}
\caption{Relation between \sersic\ $R_e$ and Petrosian $R_{50}$ for our galaxies. The black dashed line is the 1-to-1 relation. The blue line indicates the best-fit linear relation.}
\label{fig:ReR50}
\end{figure}

\section{Resolved Sources in W3 and W4}
\label{appdix_psffit}

In the AllWISE catalog, sources are flagged as resolved ($\texttt{ext\_flg}>0$) if they are included in the 2MASS extended source catalog or PSF-fitting provides poor results ($\chi^2_\nu > 2$). This subjective flag cannot be used to guide our selection of galaxies that are resolved in W3 and W4. To obtain a more reliable empirical guide, we apply the PSF-fitting method to mock galaxies to determine the critical optical size above which PSF fitting fails to yield a robust measure of galaxy flux. Guided by the measurements of \cite{Bottrell2019}, we construct mock galaxy images using a single-component model with S\'ersic index $n = 1-5$ and effective radius $R_e = 1\arcsec-10\arcsec$. To mimic the actual observations, we set $\rm W3 = 8.5-11.5$~mag and $\rm W4 = 5.5-7.5$~mag, and we convolve the mock images with the corresponding PSFs. Employing the \texttt{Levenberg-Marquardt} algorithm for the PSF fitting, we follow the same approach as AllWISE\footnote{\url{https://wise2.ipac.caltech.edu/docs/release/allsky/expsup/sec4\_4c.html}} and use as an initial guess a circular aperture of radius 8\farcs25 for W3 and 16\farcs5 for W4. The fit considers confusion uncertainty and pixel value uncertainty. 

Figure~\ref{psffitw34snr2} illustrates that PSF fitting can measure $>80\%$ of the galaxy flux in both W3 and W4 if the galaxy has high \sersic\ index ($n\geq 3.5$). However, if the galaxy is less centrally concentrated, the flux measured by PSF fitting drops quickly below 80\% if the galaxy has $R_e > 2\arcsec$ in W3 or $R_e > 4\arcsec$ in W4. Therefore, we consider the galaxy possibly resolved in W3 and W4 if it has $R_e$ larger than 2\arcsec\ and 4\arcsec, respectively, and $n<3.5$ in the SDSS $r$ band. For galaxies without $R_e$ measured by \cite{Bottrell2019}, we establish an empirical correlation between $R_e$ and the Petrosian radius $R_{50}$ in the $r$ band (Figure~\ref{fig:ReR50}), for which a linear regression {\tt LinMix} \citep{Kelly2007Linmix} yields
 
\begin{equation}
R_{50} = 0.49\, R_e + 0.89.
\end{equation}

\noindent
Thus, a galaxy is resolved in W3 if $R_{50} > 1\farcs9$ and in W4 if $R_{50} > 2\farcs9$. To facilitate other investigators who might wish to correct the PSF-fitting fluxes of galaxies from the AllWISE catalog, we fit linear functions to the flux recovery fractions in W3 and W4 as a function of $R_e$, separately for galaxies with $n<3.5$ and $n\ge3.5$. The best-fit functions are shown in Figure~\ref{psffitw34snr2}.

\end{document}